\documentclass[referee,iicol,sn-basic]{sn-jnl}

\usepackage{epstopdf}
\usepackage{graphicx}
\usepackage{amssymb}
\usepackage{tabularx}
\usepackage{subcaption} 
\usepackage{amsmath}
\usepackage{pifont}
\usepackage{supertabular}
\usepackage{eurosym}
\usepackage{xcolor} 
\usepackage{mathtools}
\usepackage{multirow}
\usepackage{appendix}
\usepackage{booktabs}
\usepackage{array}
\usepackage{diagbox}
\newcommand{\coo}{\ensuremath{\mathrm{CO_2}}}
\hypersetup{
     colorlinks=true,
     linkcolor=blue,
     filecolor=blue,
     citecolor=black,      
     urlcolor=blue,
     }

\jyear{2021}%

\begin{document}

\title[Article Title]{The energy return on investment of whole-energy systems: application to Belgium.}


\author*[1]{\fnm{Jonathan} \sur{Dumas}}\email{jdumas@uliege.be}
\author[1]{\fnm{Antoine} \sur{Dubois}}\email{antoine.dubois@uliege.be }
\author[2]{\fnm{Paolo} \sur{Thiran}}\email{paolo.thiran@uclouvain.be}
\author[2]{\fnm{Pierre} \sur{Jacques}}\email{p.jacques@uclouvain.be}
\author[2]{\fnm{Francesco} \sur{Contino}}\email{francesco.contino@uclouvain.be}
\author[1]{\fnm{Bertrand} \sur{Corn\'elusse}}\email{Bertrand.Cornelusse@uliege.be}
\author[2]{\fnm{Gauthier} \sur{Limpens}}\email{gauthier.limpens@uclouvain.be}

\affil*[1]{\orgdiv{Departments of Computer Science and Electrical Engineering}, \orgname{Liege University}, \orgaddress{\street{Place du 20-Août, 7}, \city{Liege}, \postcode{B-4000}, \state{State}, \country{Belgium}}}

\affil[2]{\orgdiv{Institute of Mechanics, Materials and Civil Engineering}, \orgname{Catholic University of Louvain}, \orgaddress{\street{Place de l'Universit\'e 1}, \city{Ottignies-Louvain-la-Neuve}, \postcode{1348}, \country{Belgium}}}



\abstract{
Planning the defossilization of energy systems while maintaining access to abundant primary energy resources is a nontrivial multi-objective problem encompassing economic, technical, environmental, and social aspects.
%
However, most long-term policies consider the cost of the system as the leading indicator in the energy system models to decrease the carbon footprint. 
%
This paper is the first to develop a novel approach by adding a surrogate indicator for the social and economic aspects, the \textit{energy return on investment} (EROI), in a whole-energy system optimization model. In addition, we conducted a global sensitivity analysis to identify the main parameters driving the EROI uncertainty.
%
This method is illustrated in the 2035 Belgian energy system for several greenhouse gas (GHG) emissions targets. Nevertheless, it can be applied to any worldwide or country energy system. 

The main results are threefold when the GHG emissions are reduced by 80\%: (i) the EROI decreases from 8.9 to 3.9; (ii) the imported renewable gas (methane) represents 60 \% of the system primary energy mix; (iii) the sensitivity analysis reveals this fuel drives 67\% of the variation of the EROI.
%
These results raise questions about meeting the climate targets without adverse socio-economic impact, demonstrating the importance of considering the EROI in energy system models. 
}




\maketitle
\section{Introduction}\label{sec:introduction}

To limit climate change and achieve the ambitious targets prescribed by the Intergovernmental Panel on Climate Change \citep{ipccar6spm}, the transition toward a carbon-free society goes through an inevitable increase in the share of renewable generation in the energy mix.
Integrating these new energy resources and technologies will lead to profound structural changes in energy systems, such as an increasing need for storage and radical electrification of the heating and mobility sectors. Therefore, energy planners face the double challenge of transitioning towards more sustainable fossil-free energy systems, including high penetration of renewables, while preserving access to abundant and affordable primary energy resources. 
In the literature, a large variety of energy system models exists. \citet{limpens2019energyscope} has conducted an extensive review of 53 energy system models and tools. They all consider a cost-based objective function with sometimes a greenhouse gas emissions target. However, designing an optimal energy system is a multi-objective problem as it encompasses economic, technical, environmental, and social aspects. 
Thus, new flexible and open-source optimization modeling tools are required to capture the increasing complexity of future energy systems. 

This study addresses this issue by considering a comprehensive indicator: the \textit{energy return on investment} (EROI). It better encompasses the technical and social challenges of the energy transition than the cost.
The field of net energy analysis was first developed following the 1970s oil crises, to assess how much energy is made available to society \citep{Cutler1984}. Then, various metrics have been introduced in the past few decades, including the energy profit ratio, energy gain, energy payback time, and the most well-known, the EROI. 
Expressed as a ratio between energy outputs and energy inputs of a given system or process \citep{hall1979efficiency}, the EROI captures the extent to which helpful energy is yielded from that system or process. The lower the EROI of an energy source, the more input energy is required to produce the output energy, which results in less net energy available for the rest of the economy. Thus, the EROI can be understood as the ease with which the energy system can extract energy sources and transform them into a form beneficial to society. The work of \citet{mulder2008energy} established a theoretical framework for EROI analysis that encompasses the various methodologies in the literature.

Access to abundant and affordable primary energy resources has been recognized as an essential element for the prosperity of human societies, and the concept of EROI is commonly used to measure their quality. The literature about EROI is abundant and, without being exhaustive, concerns three main fields of research, which are reviewed in the following paragraphs: (1) the link between EROI and societal well-being; (2) estimation of the EROI of an energy resource or technology; (3) the estimation of a global EROI at the level of an economy or a society, and a lower bound below which a prosperous lifestyle would not be sustainable.


The potential connections between societal well-being and net energy availability are investigated by \citet{lambert2014energy}. The results for a large sample of countries point out that the estimated societal EROI is correlated with the Human Development Index (HDI), which is a standard of living indicator. However, for a few countries with a high level of development, \textit{i.e.}, HDI above 0.75, there is a saturation point where increasing the EROI above 20 is not associated with further improvement in society. In addition, the relationship between societal EROI and HDI is non‑linear as the HDI increases less and less rapidly with societal EROI.

The characteristics of the primary energy sources, including the standard EROI of each fuel, are investigated by \citet{hall2014eroi}\footnote{The EROI values presented in \citet{hall2014eroi} refer to \textit{standard} EROI, which only considers energy inputs required for the extraction of the energy resource. Other EROI metrics exist, such as the final stage EROI or the societal EROI. The final stage EROI considers not only the energy inputs used for extracting, but also for processing and delivering an energy carrier.}\textsuperscript{,}\footnote{The mean and standard error values of EROI provided are estimated based on several published studies, and the references are listed in \citet{lambert2012eroi}.}. They conclude that: 
(i) overall, the standard EROI of conventional fuels, such as oil and gas, has been declining over the last decades for all nations examined (United States, Canada, Norway, Mexico, and China), reducing from 10\% to 50\%, depending on the location of production. The study of \citet{en20300490} estimates a global standard EROI value for the resources of private oil and gas companies of about 30 in 1995 and 18 in 2006. In addition, alternatives to conventional fossil fuels such as tar sands and oil shale have a lower standard EROI, with mean values of 4 and 7 \citep{hall2014eroi}, respectively;
(ii) various renewable and non-conventional energy alternatives have substantially lower EROI values, such as PV with an average EROI value of 10, than traditional conventional fossil fuels. However, wind power energy seems competitive, with a mean EROI value of 20.
Another more recent study \citep{Brockway2019} calculates the EROI of fossil fuels at both primary and final energy stages. Their results suggest that the current EROI of fossil fuels may not differ from the EROI of renewables when computed at the final stage, which illustrates the difficulty of adequately assessing the EROI of resources or technologies.

Several attempts to determine a lower bound of the societal EROI below which a prosperous lifestyle would not be sustainable have been performed over the past few years. However, the estimation of this threshold differs from one study to the other, as listed in \citet{dupont2021estimate}: (i) the study of \citet{hall2009minimum} provides an educated guess of a minimum societal EROI value of around 5. However, this value is not the result of a calculation; ii) the work of \citet{FIZAINE2016172} focuses on the USA by conducting an econometric equation linking growth rates, energy expenditures, capital formation, and population. It results in an estimated EROI value of 11 to maintain the economic growth of the USA; iii) the paper of \citet{brandt2017does} provides an estimated minimum societal EROI of 5 by using a simple template economy with four sectors and inputs for each sector defined at an order-of-magnitude level using data for the US; iv) the study of \citet{court2019estimation} indicates that this lower bound decreases as technical change improves the conversion efficiency of primary-to-final and final-to-useful exergy processes. They estimated that the minimum sustainable societal EROI of the world has declined from 20 in 1900 to 6 in 1970 to remain constant so far.

Finally, a few studies have attempted to estimate the societal EROI at a country or world level.
The work of \citet{dupont2021estimate} provides an estimate of the societal EROI using a macroeconomic model with two sectors, an energy sector and a final sector aggregating the rest of the economy. In addition, they use the net energy ratio, which is more comprehensive than the EROI, to assess the energy embodied in the intermediate and capital consumptions of the entire economy. The model estimates a net worldwide EROI of 8.5 for 2018, and the sensitivity analysis performed on the model parameters demonstrates the robustness of the model. However, the model of \citet{dupont2021estimate} focuses only on the current EROI and does not assess how it would evolve with a transition towards an energy system based mainly on intermittent renewable energy sources to meet the IPCC targets.

The papers presented above illustrate the difficulty of assessing the EROI of a given resource, technology, or society. However, they depict the key elements which have contributed to the increased attention paid to the EROI and the field of net energy analysis: 
(1) the standard EROI of the main fuels, in particular fossil fuels, has been declining due to depletion of finite resources \citep{hall2009minimum,Lambert2013}; 
(2) the estimated EROI values for renewable energy sources in comparison to conventional fossil fuels are often controversial and vary significantly depending on the adopted methodology \citep{hall2014eroi,Brockway2019}. This matter raises concerns that the renewables-led energy transition required to meet climate targets may have adverse socioeconomic impacts \citep{sers2018energy}; 
(3) the EROI captures the efficiency of energy conversion technologies and provides some macro-economic perspective because of its link to the well-being of society \citep{lambert2014energy}.

We provide two reasons to assess the EROI of a whole-energy system instead of a set of technologies and resources of a given subsystem of the energy sector, such as the electricity grid. 
First, it is not relevant to compare the EROI of renewable resources or technologies independently. For instance, solar and wind energies are intermittent and non-dispatchable. Gas and nuclear power plants are adjustable and can meet fluctuating demand. Thus, comparing the EROI of solar \textit{vs.} nuclear without taking into account storage systems and other assets to balance the system is not pertinent.
Second, a whole-energy system comprises several sectors (mobility, heat, electricity, industry) that use several technologies and resources that can be imported or extracted. These resources are transported, stored, and converted by energy conversion technologies to supply end-use demands such as electricity, transport, heating, and the production of goods. Assessing an energy system as a whole opens the opportunity for the full deployment of synergies and can generate unexpected results \citep{CONTINO2020100872}. Thus, the EROI of the system cannot be simply the sum of the EROIs of each of its components. 

The literature comprises a large variety of energy models (based on optimization and simulation). We refer the reader to the reviews proposed by \citet{limpens2019energyscope,BORASIO2022111730}, which compare models according to the following features: open-source, time resolution (from monthly to hourly), exhaustivity (\textit{i.e.}, considering the electricity sector only, or also mobility and heat supply).
While there are many studies devoted to planning a whole-energy system based on the cost indicator, those that consider the EROI are much rarer.
MEDEAS-World \citep{CAPELLANPEREZ2019100399,C9EE02627D}, a global, one-region energy-economy-environment model, is one of the few models which take into account the evolution of the societal EROI. MEDEAS is a policy-simulation dynamic-recursive model that has been designed using the theory of System Dynamics. The EROI of the system is estimated, based on a detailed review of the life cycle analyses of the different energy sources, including the ancillary structures required to handle the intermittency of renewable energies. For 2015, using aggregated data at the world level, the model estimated a global societal EROI of 12. Then, the results indicate that a fast transition to reach a 100\% renewable electric system by 2060, consistent with the Green Growth narrative, could cause a decrease of the EROI of the system from 12 to 3 by mid-century. 

In the present work, we focus on optimization models that reveal optimal configurations among many available options and degrees of freedom. They are suitable for analyzing complex systems, where many combined alternatives need to be explored.
However, the study of \citet{BORASIO2022111730} illustrates that there is no perfect energy model capable of addressing all case studies and research topics. It is improbable that a single modeling framework will ever be able to capture all the relevant and interlinked dynamics of the energy transition, which is a complex and interdisciplinary challenge \citep{CONTINO2020100872}. Various models can answer different research questions and can be complementary. However, selecting a particular model among the wide range of available energy models is a difficult task. Thus, building on existing and consolidated frameworks can be advantageous rather than developing new case-specific models from scratch.

We decided to use EnergyScope Typical Days (EnergyScope TD) \citep{limpens2019energyscope}, an open-source model for the strategic planning of urban and regional energy systems. Compared to other existing energy models, which are often proprietary, computationally expensive, and primarily focused on the electricity sector, EnergyScope TD optimizes both the investment and operating strategy of an entire energy system, including electricity, heating, mobility, and the non-energy demand (NED)\footnote{The NED comprises energy products used as raw materials in different sectors, not consumed as a fuel or transformed into another fuel \citep{rixhon2021comprehensive}. The NED amounts to 20\% of the total energy demand in the case of Belgium and 10\% of the world final energy consumption.}. 
Therefore, the EnergyScope TD model offers several benefits compared to other modeling approaches and can easily be extended to include new indicators such as the EROI. In the following section, we focus on the recent works related to EnergyScope TD.

\subsection{Related work}

A first attempt to study the Belgian energy system using the EROI metric was conducted by \citet{limpens2018electricity}. The study focuses on the mix of energy storage technologies to allow a high penetration of intermittent renewable energies. A simplified hourly-based model optimizes the renewable energy and storage assets by maximizing the EROI while respecting energy balance constraints. The results indicate that depending on the deployment of renewable energies and on the nuclear share in the energy mix, the EROI of the system ranges from 5 to 10.5. However, one of the main limitations of this study is related to the model, which is not a whole-energy system model.
This issue is addressed with EnergyScope TD \citep{limpens2019energyscope}, a more advanced model of the Belgian energy system.

First, the EnergyScope TD model was applied to analyze the 2035 Belgian energy system for different carbon emissions targets \citep{limpens2020belgian}. Choosing the year 2035 constitutes a trade-off between a long-term horizon shaped by policy choices and a horizon short enough to be able to define the future of the energy sector with a group of known technologies. The results indicate a lack of endogenous renewable resources in Belgium of 275.6 [TWh/y], amounting to 30-40\% of the primary energy demand. Several recommendations are proposed to obtain additional potential such as importing renewable fuels and electricity or deploying geothermal energy. In the study from \cite{limpens2020belgian}, a mix of solutions is the most cost-effective for reaching low carbon emissions.

Second, a further step is achieved by considering the importance of renewable fuels in a low-carbon energy system \citep{rixhon2021role}. This study performs an uncertainty quantification on a whole-energy system model by considering the total annualized cost of the system  (which is the objective function of EnergyScope TD). The polynomial chaos expansion method is implemented to perform the sensitivity analysis and to highlight the influence of the critical parameters on the cost of the system. This approach is applied to study the case of Belgium in 2050\footnote{EU strives to be a climate-neutral continent in 2050 with the European Green Deal \citep{eugreendeal}, which sets a target of zero net GHG emissions.}, and the results indicate: (i) when considering uncertain parameters, the average value of the system cost is 17\% higher at carbon neutrality than in a deterministic setting; (ii) the standard deviation of the cost increases when decreasing the GHG emissions; (iii) the price of renewable fuels is the primary driver of the uncertainty on the system's cost with 53\% of the cost variation.

Finally, a preliminary implementation of a multi-criteria approach in the EnergyScope TD model, which is currently a single objective optimization model, is proposed by \citet{muyldermansmulti}. Given the challenges associated with the energy transition, this work allows to assess an energy system including economic, environmental, technical, and social aspects. The case study is similar to \citet{limpens2020belgian} with the 2035 Belgian energy system. The analysis emphasizes the environmental impacts of the energy system depending on the weights associated with each criterion in the objective function: the total system cost, EROI, and global warming potential. The authors conclude that considering multiple criteria leads to a more nuanced and robust solution than a single criterion approach.
However, this work is introductory, and the results must be consolidated with a more extensive analysis. In addition, it does not: (i) consider several GHG emissions scenarios; (ii) assess the uncertainty of the model input parameters. Nevertheless, it paves the way for this paper.

\subsection{Research gaps and scientific contributions}

To the best of our knowledge, the research gaps motivating this paper are four-fold:
\begin{enumerate}
	\item Many studies using the EROI are focused on specific technologies and resources. There have been several attempts to estimate the EROI of society in the economic and social sciences \citep{court2019estimation,dupont2021estimate}, but none have considered a whole-energy system using a bottom-up, optimization-based model;
	\item While many studies are devoted to planning a whole-energy system based on the cost indicator, those considering the EROI are rarer. MEDEAS-World \citep{CAPELLANPEREZ2019100399}, a global energy-economy-environment system dynamics model, is one of the few that consider the societal EROI evolution but is not a bottom-up, optimization-based model; 
	\item There currently exists no open-access consolidated EROI dataset for all technologies and resources of a whole-energy system;
	\item There is no comparison of the EROI of a whole-energy system, accounting for parameters uncertainties, with the deterministic cost-optimum situation.
\end{enumerate}

\noindent With these research gaps in mind, the main contributions of this paper, built on the previous studies \citep{limpens2018electricity,limpens2019energyscope,rixhon2021role,muyldermansmulti}, are fourfold:
\begin{enumerate}
	\item Develop a novel and open-source approach by adding the EROI in a whole-energy system optimization model. This approach can be applied to an energy system to investigate the evolution of the societal EROI during the energy transition at the worldwide, country, or regional level;
	\item Propose and implement a methodology to assess the impact of uncertain parameters on the EROI and compare the results with those of the deterministic analysis;
	\item Use a real-world case study, the Belgian energy system, for several GHG emissions targets in 2035 to illustrate the novel approach by comparing the results when considering the cost as a leading indicator. In this case study, we emphasize the role of renewable fuels for decreasing the GHG emissions; 
	\item Provide a transparent and collaborative database of the EROI of all technologies and resources of a whole-energy system.
\end{enumerate}

In addition to these contributions, this study also provides open access to the code repository\footnote{\url{https://github.com/energyscope/EnergyScope}} and the latest documentation\footnote{\url{https://energyscope-td.readthedocs.io/en/master/}} to help the community reproduce the experiments.
Table \ref{tab:contributions} presents a comparison of the present study with several state-of-the-art papers analyzing energy transition systems. Appendix \ref{appendix:intro} provides the justifications for the comparison.

The present work provides decision-makers with insightful guidelines to answer the following questions: (i) What are the main changes in the planning of the transition of an energy system when using the EROI as a guiding indicator instead of the total system's cost ?
(ii) To what extent should uncertainties be considered when planning a low-carbon energy system based on the maximization of EROI? Furthermore, which are the key parameters that drive the uncertainty in the system's EROI ?
(iii) Given the limited availability of local renewables in Belgium, what solutions of the Mix scenario presented by \citet{limpens2020belgian}, such as electrification, nuclear energy, and import of synthetic fuels, would most affect the variation of the system's EROI?
\begin{table*}[htbp]
\renewcommand{\arraystretch}{1.25}
\begin{center}
\begin{tabular}{lccccccc}
\hline \hline
Criteria & [1]  &  [2]  & [3] & [4] & [5] & [6] & This study\\ \hline
Authors & EU & FPB  & EV & ELIA & RTE & UCL & UCL-ULG\\
Multi-sectors  & \checkmark & $\times$   & \checkmark & $\times$    & $\times$ & \checkmark & \checkmark \\
Multi-scenario  &  $\times$ & $\sim$   & \checkmark & \checkmark  &\checkmark & \checkmark & \checkmark \\
Model  & PRIMES  &  CSG & TIMES  &  \multicolumn{2}{c}{Antares}   &  \multicolumn{2}{c}{EnergyScope TD}   \\
EROI         &     $\times$   &  $\times$   &  $\times$&  $\times$& $\times$& $\times$& \checkmark \\
Sensitivity analysis        &   $\times$  &  $\times$  & $\times$   & $\times$  & $\sim$ & \checkmark & \checkmark \\
Open dataset             &  $\times$    & $\times$   & $\sim$ & $\sim$ &\checkmark & \checkmark & \checkmark \\
Open-access code         &  $\times$  &  $\times$  &  $\times$  & \checkmark &\checkmark  & \checkmark & \checkmark \\
\hline \hline
\end{tabular}
\caption{The contributions of the present study are compared to several state-of-the-art studies about the transition of the energy system. Justifications are provided in Appendix \ref{appendix:intro}. \\
References: [1] \citep{EU2020}; [2] \citep{devogelaer2021bon};  [3] \citep{meinke2017energy}; [4] \citep{elia2017}; [5] \citep{rte2021}; [6]  \citep{limpens2020belgian,rixhon2021role}\\
References of the models: PRIMES \citep{primeseu}; CSG  \citep{csgartelys}; TIMES \citep{fishbone1981markal}; Antares \citep{doquet2008new,antares}; EnergyScope TD \citep{limpens2019energyscope} \\
\checkmark: criteria fully satisfied, $\sim$: criteria partially satisfied, $\times$: criteria not satisfied. 
Multi-sectors: whole-energy system considered; Multi-scenario: several scenarios of GHG emissions; EROI: EROI-based objective function; Sensitivity analysis: uncertainty analysis of the model parameters; Open dataset: the data used are in open-access; Open-access code: the code used to conduct the experiments is in open-access. \\
Abbreviations: European Commission (EU), Federal Planning Bureau of Belgium (FPB), EnergyVille (EV), France’s transmission system operator (RTE), Belgium’s transmission system operator (ELIA), UCLouvain (UCL), ULi\`ege (ULG), Price-Induced Market Equilibrium System (PRIMES), Crystal Super Grid (CSG), The Integrated MARKAL-EFOM System (TIMES).}
\label{tab:contributions}
\end{center}
\end{table*}

\subsection{Organization}

\begin{figure}[tb]
\centering
\includegraphics[width=1\linewidth]{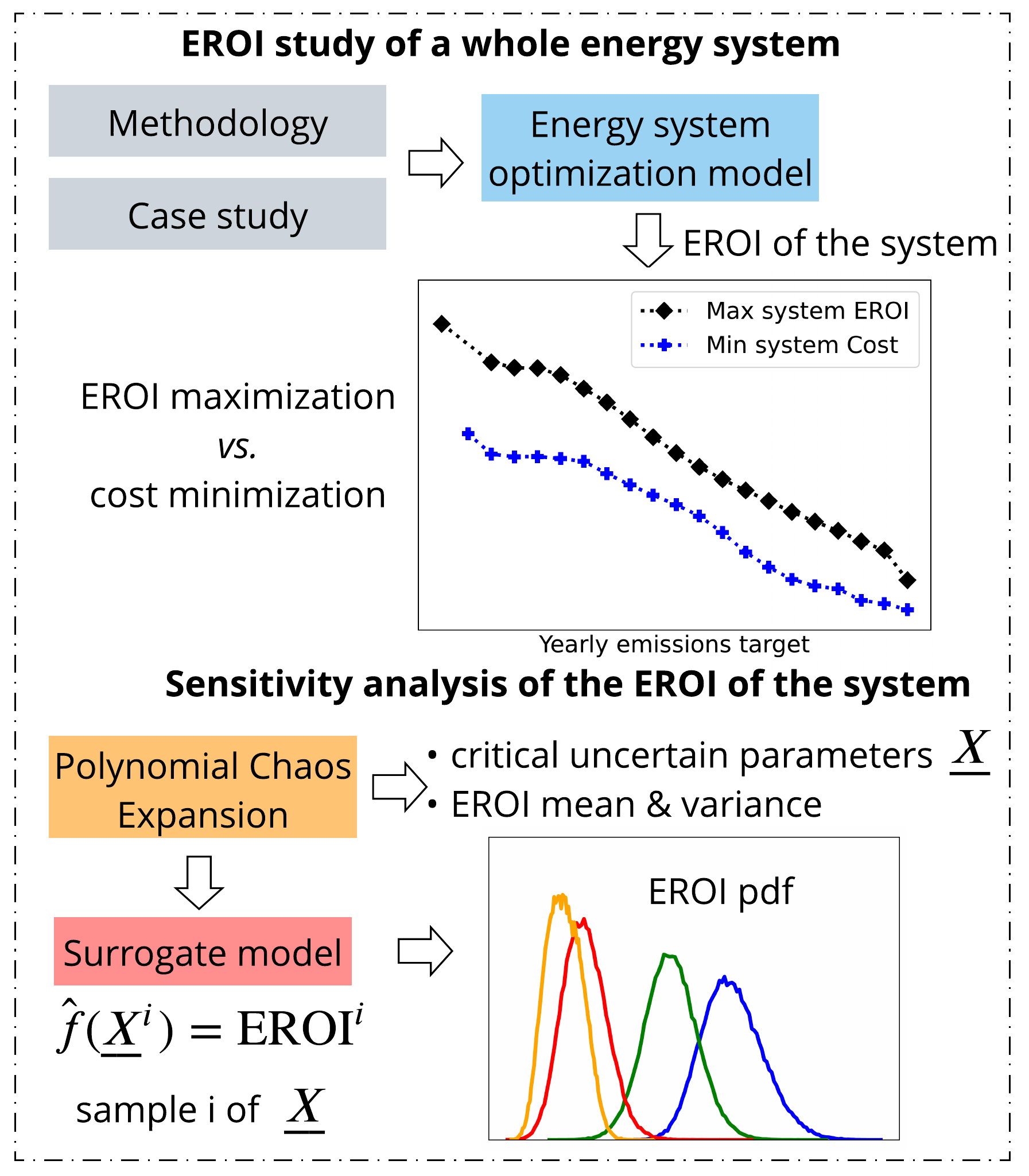}
\caption{Paper skeleton with the main contributions. The energy system optimization model EnergyScope TD is used to assess the EROI of the Belgian energy system for several GHG emissions targets. Then, an EROI sensitivity analysis identifies the critical uncertain parameters $\underline{X} \in \mathbb{R}^N$ and estimates the EROI probability density functions using the surrogate model $\hat{f}$.}
\label{fig:paper-skeleton}
\end{figure}
Figure \ref{fig:paper-skeleton} depicts the paper skeleton, which is organized as follows. Section~\ref{sec:methodology} presents the EROI definition used in this study and provides the succinct formulation of the EnergyScope TD model with the main assumptions. Section~\ref{sec:case-study} presents the real-world case study of the Belgian energy system in 2035, and Section~\ref{sec:section-4} investigates its EROI evolution for several GHG emissions targets. Section~\ref{sec:section-5} presents the results of the EROI sensitivity analysis, and Section~\ref{sec:section-6} points out the model and methodology limitations. Finally, Section~\ref{sec:conclusions} outlines the main findings and proposes ideas for further work.
Appendix \ref{appendix:intro} presents the justifications for the comparison conducted in Table \ref{tab:contributions}. Appendix \ref{appendix:section-3} provides the methodology to derive the final energy consumption from the simulation results in EnergyScope TD. Finally, Appendices \ref{appendix:section-4} and \ref{appendix:section-5} give additional results for the EROI study of the Belgian energy system in 2035 and for the sensitivity analysis, respectively.

\section{Methodology}\label{sec:methodology}

This section presents the EROI definition used in this study and details the main assumptions of the model, including the EROI-based objective function. The EnergyScope TD complete formulation, the documentation of the model, and the input data are described in \citet[Appendix C. Supplementary material]{limpens2019energyscope}, and the code repository with the data is open-access.

\subsection{EROI definition}\label{sec:eroi}

This study considers the EROI defined at the final energy stage
\begin{align}
    \text{EROI}_\text{fin} &= \frac{\text{Gross energy produced}}{\text{Energy invested}} = \frac{\text{E}_\text{out}}{\text{E}_\text{in}}.
\label{eq:eroi-final}
\end{align}
$\text{E}_\text{out}$ is computed at the final stage, \textit{i.e.}, the quantity of gasoline or electricity required by cars and trains, the heat produced for warming buildings, or the electricity delivered to households and companies.
$\text{E}_\text{in}$ is also measured in terms of final energy. It is composed of: (1) the energy required for building and operating all the infrastructure of the energy system, from the cradle to the grave; (2) the energy used for operating the energy system.

\begin{figure*}[tb]
\centering
\includegraphics[width=\linewidth]{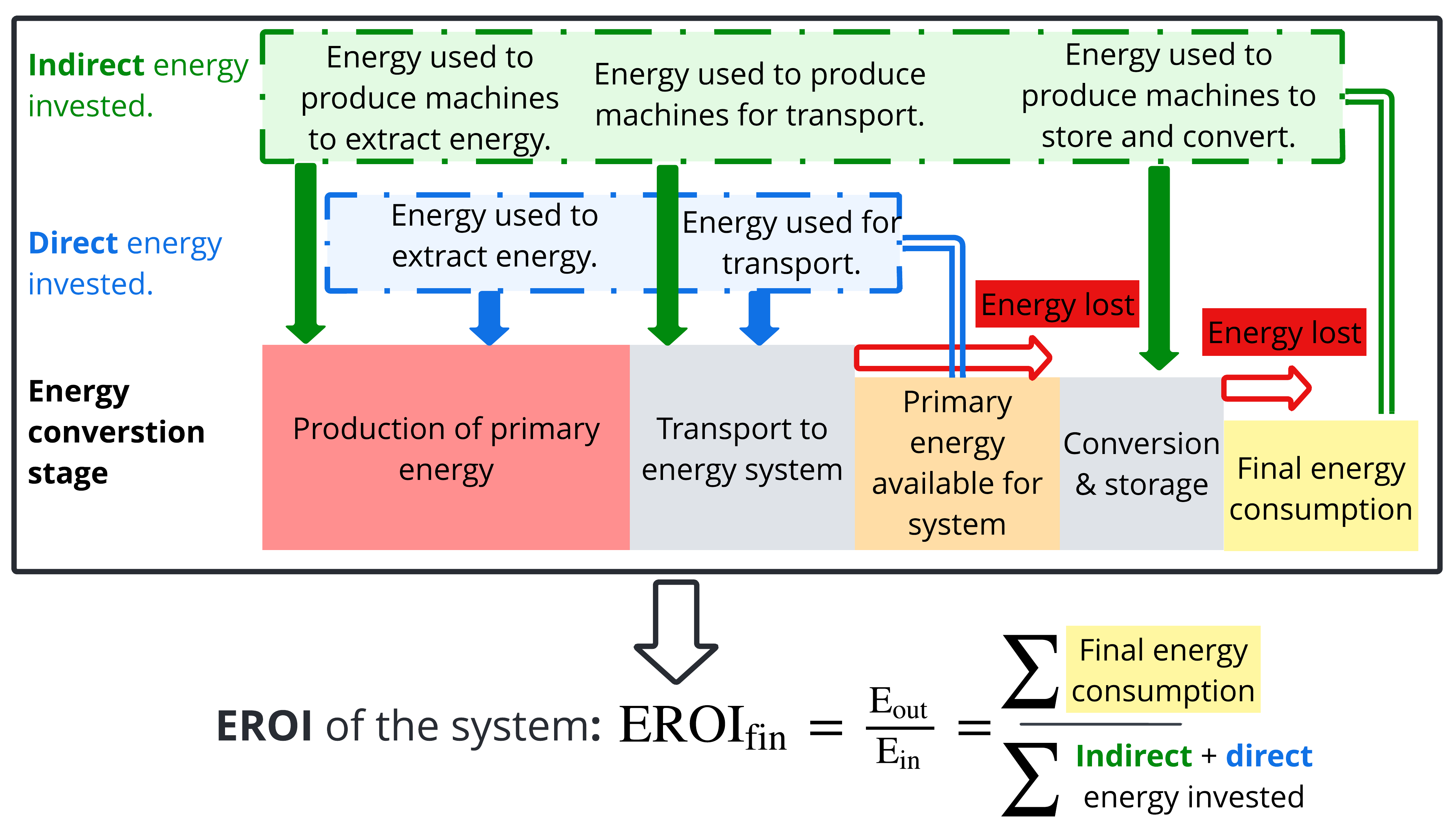}
\caption{The energy cascade illustrates the EROI of the system at the final stage ($\text{EROI}_\text{fin}$), considered in this study, with the direct and indirect energy invested at each step of the energy conversion required to produce final energy and satisfy the final energy consumption. This Figure was adapted from \citet{Brockway2019}.}
\label{fig:energy-cascade}
\end{figure*}

The motivation of this definition is that energy enters the productive economy at the final energy stage. Figure \ref{fig:energy-cascade} depicts the differences between primary energy production and final energy consumed with the concept of energy cascade. It illustrates the EROI of the system at the final stage.
The numerator is the sum of all the types of final energy consumption considered in the system, such as electricity, heat, mobility, and non-energy. The denominator is the sum of all the indirect and direct energy invested for each step of the energy conversion from the primary energy production to the conversion and storage of the final energy into final energy consumption (FEC). 
Direct energy is required for the conversion energy process, such as the energy to extract and produce gas and oil in the field. Indirect energy is related to the products used at each step of the conversion process, such as the infrastructure used to extract and produce gas and oil in the field.
Note that the FEC definition adopted in this study is provided by \citet{eurostat-glossary}: it excludes energy used by the energy sector, including distribution and transformation, and it is the energy that reaches the final consumer's door, such as households or industries. Therefore, $\text{EROI}_\text{fin}$ defined in Eq. (\ref{eq:eroi-final}) corresponds to the net external energy return (NEER) from the nomenclature of \citet{brandt_general_2011}.

\subsection{EROI formulation of EnergyScope TD}

This study uses the open-source energy system optimization model EnergyScope TD \citep{limpens2019energyscope}, built on previous works \citep{Moret231814,girones2015strategic}. EnergyScope TD is a linear programming (LP), multi-sector, and multi-carrier model for a regional whole-energy system such as a country. 
This model has been validated for the 2035 Belgian whole-energy system by \citet{limpens2020belgian}, which is the case study of interest. Furthermore, the model has been used for several regions\footnote{\url{https://energyscope.readthedocs.io/en/master/sections/Releases.html\#case-studies}} including Italy \citep{BORASIO2022111730}, Spain \citep{Martinez2021}, Switzerland \citep{Moret231814}, and Europe-26 \citep{Dommisse2020}. In addition, \citet{thiran2020energyscope} developed a multi-regional version, called EnergyScope Multi-Cell, which was applied to Western Europe \citep{cornet2021energy} and Italy \citep{thiran2021flexibility}.

EnergyScope TD represents the heating, mobility, and electricity sectors with the same level of detail. The main characteristics are: (i) satisfying the system end-use demand (EUD) instead of final energy consumption (FEC). The system EUD is composed of electricity, heat, transport, and non-energy demands. For instance, passenger mobility is defined in passenger kilometers per year rather than in a certain amount of gasoline to fuel cars or electricity to power trains; (ii) optimizing the system design and operation by minimizing its overall cost; (iii) using an hourly resolution which makes the model suitable for analyzing the integration of intermittent renewable energy resources and storage; (iv) modelling the country as a single node where transmissions constraints within the country are not considered. Energy demand is balanced by energy generation without considering the flows between the producers and the consumers. However, the heating and electricity grid costs are considered, including an extra electricity network investment related to the integration of intermittent renewable energies. These costs are proportional to the installed capacity of electricity production and heating technologies; (v) achieving a short computational time, typically a few minutes, due to the use of typical days, and a mapping method to represent the storage over a year with an hourly resolution. 

In this study, the objective function is modified to maximize $\text{EROI}_\text{fin}$ instead of minimizing the total cost of the system.
The model uses fixed EUD as input parameters where the EUD is related to the numerator in the $\text{EROI}_\text{fin}$ definition Eq. (\ref{eq:eroi-final}). 
Maximizing the EROI of the whole-energy system is similar to maximizing the ratio between the energy services, expressed in this work in terms of EUD and the quantity of energy invested. Since the same EUD is considered for each simulation, we only minimize the denominator: the energy invested $\text{E}_\text{in}$.
In practice, $\text{EROI}_\text{fin}$ Eq. (\ref{eq:eroi-final}) is computed ex-post by using the optimization result with 
\begin{subequations}
\label{eq:eroi-f-1}
\begin{align}
    \text{E}_\text{out} &= \text{FEC}, \\
    \text{E}_\text{in} & = \arg \min \mathbf{E_\text{in,tot}},
\end{align}
\end{subequations}
where the final energy consumption (FEC) is derived from the simulation result with the methodology detailed in Appendix \ref{appendix:section-3}.

Thus, the model optimizes the design and the operational strategies to minimize the total annual system energy invested $\mathbf{E_\text{in,tot}}$ given: (1) the exogenous EUD in electricity, heat, non-energy demand, and mobility; (2) the availability and the operating energy required to use the resources  (RES). The resources can be for example natural gas, wind and solar energies, biomass or renewable fuels; (3) the efficiency and the energy invested in the construction of technologies (TECH). Examples of technologies include gas power plants, wind turbines, heat pumps, or cars.
The objective function to minimize is thus
\begin{align}
\label{eq:objective-function}
    \min \mathbf{E_\text{in,tot}} &= \sum_{j\in TECH} \frac{\mathbf{E_\text{constr}}(j)}{\text{lifetime}(j)} + \sum_{i\in RES} \mathbf{E_\text{op}}(i),
\end{align}
with $TECH$ and $RES$ the sets of all the technologies and resources, respectively, and $\text{lifetime}(j)$ the lifetime of the technology $j$. 
$\mathbf{E_\text{constr}}(j)$ is the energy invested by the system in the construction of technology $j$ over its entire lifetime. This total energy invested is then allocated between years based on the technology lifetime $\text{lifetime}(j)$.
$\mathbf{E_\text{op}}(i)$ is the energy invested by the system annually in the operation of resource $i$.
They are defined as follows
\begin{subequations}
\label{eq:einv-def}
\begin{align}
& \mathbf{E_\text{constr}}(j)  = e_\text{constr}(j) \mathbf{F}(j) \quad \forall j \in TECH,\\
& \mathbf{E_\text{op}}(i)  =\sum_{\mathclap{t\in \mathcal{T} \vert\{h,td\}\in T\_H\_TD(t)}} e_\text{op}(i) \mathbf{F_t}(i,h,td)t_\text{op}(h,td) \quad \forall i \in RES, \label{eq:einv-op}
\end{align}
\end{subequations}
with $e_\text{constr}(j)$ [GWh/GW] the specific value of energy invested in construction of technology $j$ which is the cumulative energy demand associated to the construction of one GW of this technology, $e_\text{op}(i)$ [GWh/$\text{GWh}_\text{fuel}$] the specific value of energy invested in operation of resource $i$ which includes energy inputs for extraction/production/transportation and combustion, $\mathbf{F}(j)$ [GW] ([GWh] for storage technologies) the installed capacity of the technology $j$, $\mathbf{F_t}(i,h,td)$ [GWh] the quantity of the resource $i$ that is used at the hour $h$ of the typical day $td$, $t_\text{op}(h,td)$ (1h by default) the time period duration, $\mathcal{T}$ the set of all the periods of the year, \textit{i.e.}, 8760 hours, and $T\_H\_TD(t)$ the hour $h$ and the typical day $td$ associated to the period $t$. In Eq. (\ref{eq:einv-op}) summing over the different typical days and the hours of typical days, using the set $T\_H\_TD(t)$, is equivalent to summing over the 8760 hours of the year.

The choice made in EnergyScope TD to model Climate change is the global warming potential (GWP) [Mt\coo-eq./y]. The annual greenhouse gas emissions of the system $\mathbf{GWP_\text{tot}}$ are defined as the sum of: (1) the emissions related to the construction and end-of-life of the energy conversion technologies $\mathbf{GWP_\text{constr}}$, allocated to one year based on the technology lifetime; (2) the emissions related to the operation of resources $\mathbf{GWP_\text{op}}$ which accounts for extraction, transportation and combustion. They are defined as follows
\begin{subequations}
\begin{align}
\label{eq:GWP-tot-def}
& \mathbf{GWP_\text{tot}} = \sum_{\mathclap{j\in TECH}} \frac{\mathbf{GWP_\text{constr}}(j)}{\text{lifetime}(j)} + \sum_{i\in RES} \mathbf{GWP_\text{op}}(i), \\
& \mathbf{GWP_\text{constr}}(j)  = \text{gwp}_\text{constr}(j) \mathbf{F}(j) \quad \forall j \in TECH,\\
& \mathbf{GWP_\text{op}}(i)  =\sum_{\mathclap{t\in \mathcal{T} \vert\{h,td\}\in T\_H\_TD(t)}} \text{gwp}_\text{op}(i)   \mathbf{F_t}(i,h,td)t_\text{op}(h,td) \notag \\ 
& \quad \quad \quad \quad \quad \quad \forall i \in RES.
\end{align}
\end{subequations}
Similarly to the energy invested, the total emissions related to the construction of technologies are the product of the specific emissions $\text{gwp}_\text{constr}$ and the installed capacity $\mathbf{F}$. The total emissions of resources operation are the emissions associated with fuels from cradle to combustion and imports of electricity $\text{gwp}_\text{op}$ multiplied by the quantities of resources used $\mathbf{F_t}(i,h,td)$ and the period duration $t_\text{op}$.

The GHG emissions scenario or target is defined by setting a limit, $\text{gwp}_\text{limit}$, on the annual system GHG emissions $\mathbf{GWP_\text{tot}}$ as follows
\begin{align}
\label{eq:GWP-tot-target}
\mathbf{GWP_\text{tot}} &  \leq \text{gwp}_\text{limit}.
\end{align}
Therefore, the method relies on a snapshot approach \citep{girones2015strategic} where for two different GHG emissions targets specified in Eq. (\ref{eq:GWP-tot-target}), two different strategies result from the optimization of the system in the year 2035 without any dependence on the state of the system at any previous year. The two obtained strategies are thus totally independent from each other.

GWP data ($\text{gwp}_\text{op}$ and $\text{gwp}_\text{constr}$) are estimated by using a life cycle assessment (LCA) approach taken from the Ecoinvent database v3.2 \citep{wernet2016ecoinvent} using the ``allocation at the point of substitution", \textit{i.e.}, taking into account emissions of technologies and resources ``from the cradle to the grave" and following the indicator ``GWP100a- IPCC2013" developed by the Intergovernmental Panel on Climate Change (IPCC) \citep{change2013physical}. The ``Input Data" section of the online documentation provides the input data to apply the EnergyScope TD model to the Belgian energy system in 2035. 
Table \ref{tab:energy-invested-data} summarizes the specific values of operational energy inputs $e_\text{op}(i)$ and GHG emissions $\text{gwp}_\text{op}(i)$ for each resource in 2035. We refer the reader to the online documentation for the data related to the construction of technologies.
\begin{table}[tb]
\renewcommand{\arraystretch}{1.25}
\centering
\begin{tabular}{lrr} 
 \hline
Resources & $\text{gwp}_\text{op}$ & $e_\text{op}$  \\ \hline
Elec. import    & 206 & 0.123  \\
NG  & 267 & 0.0608   \\ 
Renewable gas  & 0 & 0.269   \\ 
Gasoline  & 345 & 0.281   \\ 
Bio-ethanol  & 0 & 0.316   \\ 
Diesel  & 315 & 0.21   \\ 
Bio-diesel  & 0 & 0.432   \\ 
LFO  & 312 & 0.204   \\ 
H2  & 364 & 0.083   \\ 
Renewable H2  & 0 & 0.579   \\ 
Ammonia  & 285 & 0.065   \\ 
Renewable ammonia  & 0 & 0.579   \\ 
Methanol  & 350 & 0.0798   \\ 
Renewable methanol  & 0 & 0.579   \\ 
Wood  & 11.8 & 0.0491   \\ 
Wet biomass  & 11.8 & 0.0559   \\ 
Waste  & 150 & 0.0577   \\ 
Uranium  & 3.9 & 0.0434   \\ 
Wind & 0 & 0   \\ 
Solar   & 0 & 0   \\ 
Hydro    & 0    & 0     \\   
Geothermal            & 0 & 0  \\ \hline
\end{tabular}
\caption{The specific value of operational energy inputs $e_\text{op}$ [GWh/$\text{GWh}_\text{fuel}$] and GHG emissions $\text{gwp}_\text{op}$ [Mt\coo-eq./$\text{GWh}_\text{fuel}$] for each resource in 2035 used in the case study. 
The methodology of the data collection of $\text{gwp}_\text{op}$ and $e_\text{op}$ relies on \citet{muyldermansmulti} where the data have been collected from the \textit{ecoinvent} database \citep{wernet2016ecoinvent}.
}
\label{tab:energy-invested-data}
\end{table}

Finally, EnergyScope TD has an hourly resolution and a tractable formulation with a few minutes of computational time thanks to a formulation with twelve typical days \citep{limpens2019energyscope}. This number allows to reach a good trade-off with: (i) a limited impact on the resulting energy system strategy, \textit{i.e.}, the installed capacity of the technologies and the use of resources remain in the same order of magnitude as without the use of typical days; (ii) a significant gain in computational time from 20 hours, with no typical day, to a few minutes. This relatively short computation time allows: (1) for a detailed analysis of the high integration of renewable energy resources and storage capacities; (2) to carry out an uncertainty quantification requiring several thousands of runs.

\section{Case study: the 2035 Belgian energy system}\label{sec:case-study}

The model is applied to the 2035 Belgian energy system for several GHG emission targets, capitalizing on previous studies \citep{limpens2020belgian,muyldermansmulti} to define the input parameters.
%


The energy transition relies on renewable energies, making their deployment potential a critical parameter. Tables \ref{table:installed-capacity-2015-vs-availability-2035} and \ref{table:renewable-availability-2015-vs-potential-2035} provide the installed capacity of technologies using renewable energy and the renewable resource potential, respectively. The data for 2015 and 2020 are compared to the predictions for 2035 used as input parameters of the model. In the following paragraphs, we comment on some of the parameters considered in the case study.

Based on the current policies, Belgium intends to phase out coal and nuclear. Hence, we suppose coal and nuclear power plants are shut down in 2035; thus not available. However, in the sensitivity analysis conducted in Section \ref{sec:section-5}, the installed nuclear capacity is considered an uncertain parameter.
A limit of 27.9\% \citep{limpens2020belgian} of the 2035 electricity end-use demand, representing 27.57 [TWh/y], bounds the electricity imports and restricts the Belgian electrical dependence on neighboring countries.
This study assumes an upper limit of 59.2 GWe \citep{limpens2020belgian} of PV installed capacity. This estimation relies on two hypotheses; (i) the actual area of available well-oriented roofs is approximatively 250km$^2$ \citep{devogelaer2012towards}, which is around 1\% of total Belgian lands; (ii) a 23\% PV efficiency expected in 2035 with an average daily total irradiation, similar to historical values, of 2820 Wh/m$^2$. Notice that PV and solar thermal technologies compete with this land availability constraint of 250km$^2$, equivalent to 59.2 GWe of PV or 70 GWth of solar thermal (centralized or decentralized).
\begin{table}[tb]
\renewcommand{\arraystretch}{1.25}
\centering
\begin{tabular}{llrrr} 
 \hline
 & Technology & 2015 & 2020 & 2035 \\ \hline
 \multirow{5}{*}{\begin{tabular}[c]{@{}l@{}}Electricity \\ production\end{tabular}}  & PV  & 3.85 & 4.49 & $59.2^\dagger$  \\
                   & onshore wind  & 1.49 & 2.49 & 10    \\ 
                   & offshore wind & 0.70 & 1.95 & 6     \\ 
                   & hydro river   & 0.11 & 0.115 & 0.115     \\ 
                   & geothermal    & 0    & 0 & 0     \\   \hline
 \multirow{3}{*}{\begin{tabular}[c]{@{}l@{}}Heat \\ production\end{tabular}}  & geothermal & 0 & 0 & 0  \\
                   & cen. solar th.  & 0 & 0 & $70^\dagger$    \\ 
                   & dec. solar th. & 0 & 0 & $70^\dagger$     \\ \hline
\end{tabular}
\caption{Comparison of installed capacity [GWe] or [GWth] of technologies using renewable energy in 2015, 2020 and their maximal potential expected in the model for 2035. $^\dagger$ PV and solar thermal technologies compete with the land availability constraint of 250km$^2$ which is equivalent to 59.2 GWe of PV or 70 GWth of solar thermal (centralized or decentralized). 2015 and 2020 data are based on \citet{EU2020} and the 2035 projection on \citet{limpens2020belgian}.
Abbreviations: centralized (cen.), decentralized (dec.), thermal (th.).
}
\label{table:installed-capacity-2015-vs-availability-2035}
\end{table}
\begin{table}[tb]
\renewcommand{\arraystretch}{1.25}
\centering
\begin{tabular}{llrrr} 
 \hline
 & Resources & 2015 & 2020 & 2035 \\ \hline
 \multirow{6}{*}{\begin{tabular}[c]{@{}l@{}}Imported \\ fuels\end{tabular}}    & bio-ethanol  & 0.48 & ?$^\dagger$ & no limit    \\ 
                   & bio-diesel     & 2.89 & ?$^\dagger$  & no limit     \\ 
                   & gas-RE         & 0    & 0& no limit    \\ 
                   & H2-RE          & 0    & 0& no limit \\
                   & ammonia-RE     & 0    & 0& no limit    \\ 
                   & methanol-RE    & 0    & 0& no limit     \\   \hline
\multirow{2}{*}{Biomass}    & woody  & 13.9 & ?$^\dagger$ & 23.4    \\ 
                   & wet & 11.6 & ?$^\dagger$ & 38.9     \\ \hline
Waste              &  & 7.87 & ?$^\dagger$ & 17.8     \\ \hline
Elec. import &  & 24.54 &-0.6 & 27.57     \\ \hline
\end{tabular}
\caption{Comparison of renewable resources [TWh] in 2015, 2020, and their maximal potential in the model for the year 2035. Waste is a non-renewable resource. $^\dagger$ no consolidated data available. Abbreviations: electricity (elec.), renewable (RE).
}
\label{table:renewable-availability-2015-vs-potential-2035}
\end{table}
This study accounts for two types of biofuels: bio-diesel and bio-ethanol. They can substitute diesel and gasoline, respectively. 
Finally, a new type of renewable fuel, produced from renewable electricity, is considered. This study allocates them a zero-global warming potential, \textit{i.e.}, $\text{gwp}_\text{op}$ = 0 [Mt\coo-eq./GWh]. However, there is always a residual \coo-impact that must be compensated for through biomass or direct air capture. Nevertheless, the analysis of these \coo-compensation approaches is out of the scope of this work. Four fuels are considered: hydrogen, ammonia, methanol, and methane. They can be ``renewable" or fossil with a $\text{gwp}_\text{op} > $ 0 [Mt\coo-eq./GWh].

The 2035 heating, electricity, and mobility EUD projections are based on \citet{limpens2020belgian} and Table \ref{table:EUD-2035} summarizes the differences between the EUD in 2015, 2020 and 2035. It is possible to notice the COVID-19 impact in 2020.
The 2035 passenger transport demand, 194 [Mpass.-km/y], is divided between public and private transport. The lower and upper bounds for the use of public transport are 19.9\% and 50\% of the annual passenger transport demand, respectively.
The freight demand, 98 [Mt-km/y], can be supplied by trucks, trains, or inland boats with corresponding lower and upper bounds: 0\% and 100\%, 10.9\% and 25\% 15.6\% and 30\%, respectively.
\begin{table}[tb]
\renewcommand{\arraystretch}{1.25}
\centering
\begin{tabular}{lrrr} 
 \hline
EUD & 2015 & 2020 & 2035  \\ \hline
Electricity   [TWhe]      & 81.5  & 74.8 & 91.9      \\ 
Heat high T.  [TWh]      & 74.1  & 71.2 & 50.4       \\ 
Heat low T.   [TWh]      & 124.8 & 116.8 & 147.3     \\ 
Non-energy       [TWh]   & 89.0   & 86.3& 53.1         \\   
Mobility pass. [Mpass.-km] & 147   & 110 & 194      \\
Mobility freight   [Mt-km] & 70     & 69 & 98        \\ \hline
\end{tabular}
\caption{Comparison of EUD for 2015, 2020, and the 2035 projection. The 2015 and 2020 heat low/high T. EUD are derived from the FEC of the corresponding sector (residential, service and industry for heat low T. and energy intensive industries for heat high T.) by removing the corresponding electricity FEC. 83.1\% of the electricity FEC of Belgium is allocated to the residential and services sectors, and the remaining is the energy-intensive industrial sector. This ratio is estimated based on \citet{eurostat2018}. 2015 and 2020 data are based on \citet{EU2020} and the 2035 projection on \citet{limpens2021generating}.
Abbreviations: end-use demand (EUD), temperature (T.), passenger (pass.), tons (t).}
\label{table:EUD-2035}
\end{table}

\section{EROI analysis with fixed parameters}\label{sec:section-4}

This section presents the model's results in the real-world case study of the Belgian energy system in 2035.
First, Section \ref{sec:ref-case-study} presents the results of the EROI maximization of the system without imposing a GHG emissions target. Then, Section \ref{sec:eroi-evolution} investigates the evolution of the EROI of the system for several GHG emissions targets. It studies how this impacts the energy and technology mix of the system, the breakdown of the GHG emissions, and the evolution of the primary energy mix. Finally, Section \ref{sec:eroi-vs-cost} compares a solution maximizing the system's EROI to a solution minimizing the system's cost for the same GHG emissions targets.

\subsection{Reference case study results}\label{sec:ref-case-study}

EnergyScope TD reaches an EROI-optimum Belgian energy system in 2035 at around 8.9 and a $\mathbf{GWP_\text{tot}}$ of 100.3 [Mt\coo-eq./y] without limiting the GHG emissions. This case is called ``reference scenario-100\%", where the constraint Eq. (\ref{eq:GWP-tot-target}) is not activated, and the parameters are at nominal values. In comparison, the EU reference scenario 2020 \citep{EU2020} provides the actual 2015 and 2020 values and the 2035 forecast value for the total GHG emissions\footnote{These GHG emissions values do not consider the international intra-EU and international extra-EU and Use, Land-Use Change and Forestry (LULUCF).} of 118.6, 106.6, and 100.0 [Mt\coo-eq./y], respectively. Notice that 2020 is particular due to the COVID-19, and the GHG emissions are expected to increase in the coming years before decreasing to achieve the EU targets.
As explained in the following paragraphs, the 2035 energy system, when maximizing the EROI, relies mainly on natural gas, which is less carbon-intensive than the actual Belgian energy system, which uses oil for mobility. In addition, the share of renewable energies is higher with maximal wind installed capacities. 
\begin{figure*}[htbp]
\centering
\includegraphics[width=\linewidth]{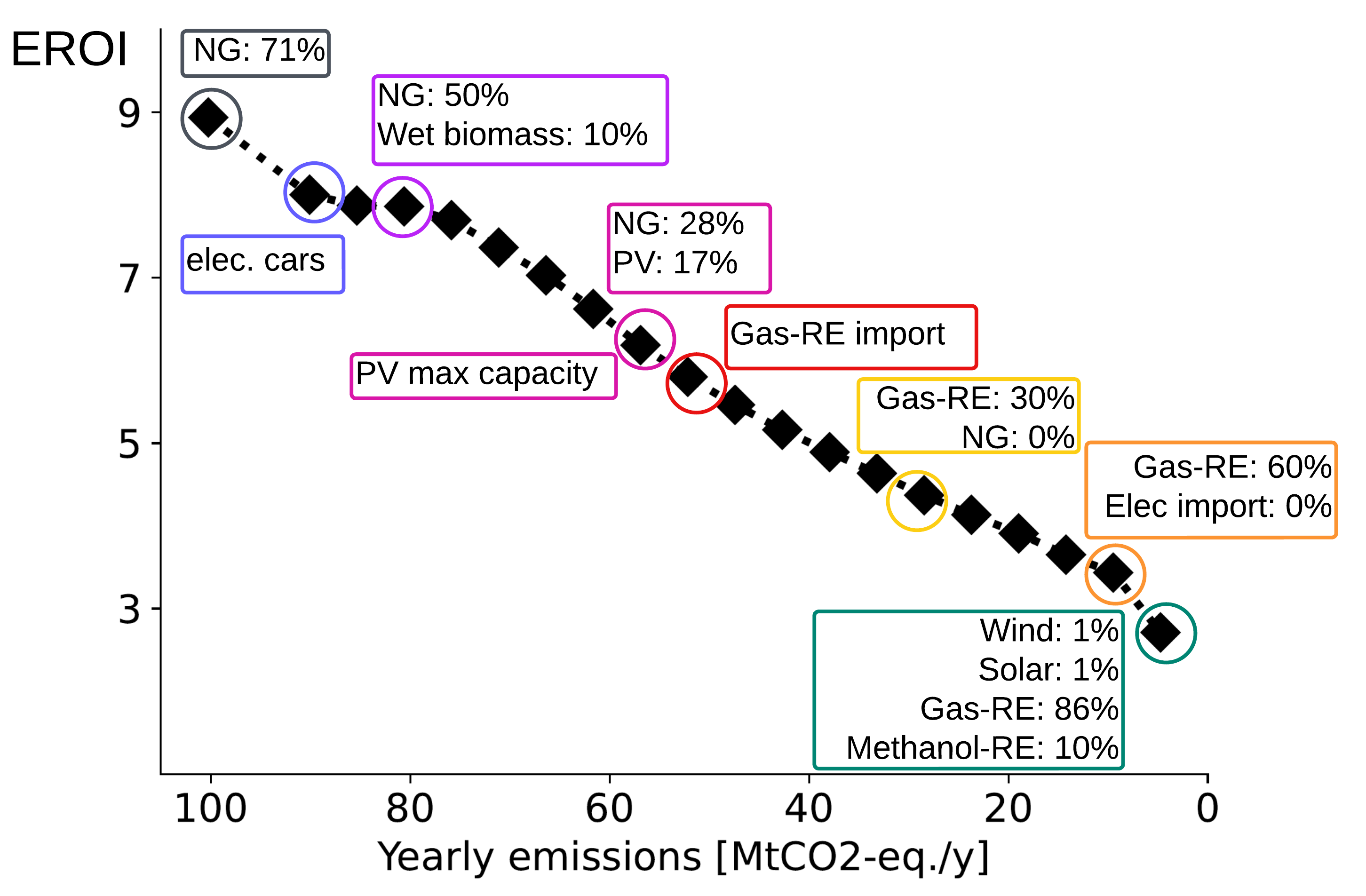}
\caption{EROI - GHG emissions ($\mathbf{GWP_\text{tot}}$) optima with primary energy mix and technologies implementation. The energy transition is composed of seven main steps illustrated with the red circles. Abbreviations: natural gas (NG), electric cars (elec. cars), electricity import (elec import), renewable gas (Gas-RE), renewable methanol (Methanol-RE).
}
\label{fig:eroi-evolution}
\end{figure*}

Non-renewable sources (82.7\%), particularly natural gas (NG) import, dominate the primary energy mix (413 TWh): NG import (292.9 TWh), methanol (38.4 TWh), ammonia (10.2 TWh). Renewables stand only for 17.1\% of the primary energy supply and are split between wind (43.0 TWh), wood (23.4 TWh), solar (4.0 TWh), and hydro (0.5 TWh). The remaining 0.2\% consists of electricity import (0.8 TWh).
Table \ref{table:installed-capacity-reference-case-study} details the major technologies used to supply the demands of Table \ref{table:EUD-2035} in terms of share of production and installed capacity.
The electricity generation relies mainly on gas with CCGT (33.4 TWh) and CHP (65.8 TWh) \textit{vs.} wind (43.0 TWh), PV (4.0 TWh), hydro (0.5 TWh), and imported electricity (0.8 TWh). 
A large part of the electricity production (42.3 TWh) is used to supply heat pumps which supply mainly decentralized and centralized low-temperature heat demands. The gas cogeneration is the most prominent player in supplying the industrial high-temperature heat demand, besides a small share from gas boilers. 
Overall, mobility is also dominated by NG import: (1) passenger mobility is equally divided between private (50\%) and public (50\%) technologies with NG cars (100\% of the private mobility), tramways (30\% of the public mobility), trains (50\% of the public mobility), and NG buses (20\% of the public mobility); (3) NG trucks, trains, and NG boats for the road, train, and boat freight, respectively.
Finally, methanol and ammonia are imported to satisfy the non-energy demand, where a large part of the methanol is used to synthesize high-value chemicals (HVC).
\begin{table}[tb]
\renewcommand{\arraystretch}{1.25}
\centering
\begin{tabular}{llrr} 
 \hline
Demand & Technology & Production & Capacity  \\ \hline
\multirow{3}{*}{\begin{tabular}[c]{@{}l@{}}Electricity\end{tabular}}  & CCGT    & 33.4  &  6.5   \\ 
              & wind    & 43.0  & 16.0    \\ 
              & PV      & 4.0   & 3.9 \\ \hline
\multirow{2}{*}{\begin{tabular}[c]{@{}l@{}}Heat HT\end{tabular}}  & gas CHP  & 59.7  & 8.0 \\
 & gas boiler  & 4.2  & 4.8 \\ \hline
\multirow{2}{*}{\begin{tabular}[c]{@{}l@{}}DHN heat LT\end{tabular}}   & HP  & 45.2  & 12.9 \\
                 & gas CHP  & 7.0  & 4.3 \\ \hline
DEC heat LT & HP  & 92.9  & 31.3 \\ \hline
Private mobility  & gas car  & 97.0  & -\\ \hline
\multirow{3}{*}{Public mobility}  & train  & 48.5  & -\\ 
                                    & tramway  & 29.1  & -\\ 
                                    & gas bus  & 19.4  & -\\ \hline
\multirow{3}{*}{Freight}  & gas truck  & 44.1  & -\\ 
                          & gas boat  & 29.4  & -\\ 
                          & train  & 24.5  & -\\ \hline
\end{tabular}
\caption{Reference scenario-100\%: major technologies used to supply the demands of Table \ref{table:EUD-2035} in terms of production and installed capacity. The private mobility accounts for 50\% of the passengers mobility. 
Units: production of electricity and all types of heat in [TWh], the private and public mobility in [Mpass.-km], the freight mobility in [Mt-km], and the production capacity of electricity and all types of heat in [GW].
Abbreviations: end-use demand (EUD), high temperature (HT), decentralized heat low temperature (DEC heat LT), combined cycle gas turbine (CCGT), combined heat and power (CHP), centralized heat low temperature (DHN heat LT), passenger (pass.),  heat pump (HP), natural gas (NG).}
\label{table:installed-capacity-reference-case-study}
\end{table}

The result of an EROI-optimum for the 2035 Belgian energy system of 8.9 is close to the estimation of the 2018 societal worldwide EROI \citep{dupont2021estimate}. However, this comparison suffers from two limitations. First, in their model, the current global energy system was mainly based on fossil fuels in 2018. Second, the scope of their study differs as it considers the entire world.
EnergyScope TD with the reference scenario indicates that renewable synthetic fuels are too energy-expensive to compete against the fossil equivalent when there is no constraint on GHG emissions.
However, it is essential to remind that such a system results from linear optimization. A slight difference, \textit{e.g.}, efficiency, energy invested in construction or operation, can make the system switch between two completely different solutions but with similar energy invested objective. This is another rationale to account for uncertainties in such a research field, which is investigated in Section \ref{sec:section-5}.

\subsection{Scenarios analysis of the EROI for several GHG emissions targets}\label{sec:eroi-evolution}

This Section conducts an analysis of the Belgian energy system in 2035 by forcing the total annual emissions of the system to decrease by reducing its upper limit, \textit{i.e.}, $\text{gwp}_\text{limit}$ in Eq. (\ref{eq:GWP-tot-target}). In practice, 5\% steps of $\mathbf{GWP_\text{tot}}$ reduction were made from the $\mathbf{GWP_\text{op}}$ (94.9 [Mt\coo-eq./y]) of the ``reference scenario-100\%" presented in Section \ref{sec:ref-case-study}.

First, a summary of the evolution of the EROI of the system is presented. Then, the focus is set on the evolution of GHG emissions in the system. Finally, the evolution of the primary energy uses is depicted.
Appendix \ref{appendix:section-4} provides additional results presented in terms of installed capacities, energy invested, and final energy consumption for the different GHG emissions targets. 

\subsubsection{System evolution summary}

Figure \ref{fig:eroi-evolution} depicts a summary of the EROI optimum values for each GHG emissions scenario with primary energy mix and main technologies changes.
The system shifts from high emissions to low emissions; however, this is a view of the mind because the solutions of each GHG emissions target scenario do not represent a transition path and must be analyzed individually.
In the following, we comment on the energy transition in seven steps, the circles depicted on Figure \ref{fig:eroi-evolution}.

Step 0 - ``reference scenario-100\%" - dark circle depicted in Figure \ref{fig:eroi-evolution}: $\mathbf{GWP_\text{tot}}$ reaches 100.3 [Mt\coo-eq./y], and the system's primary energy mix relies on 71\% of NG and the related technologies to satisfy the electricity, heat, and mobility demand (see Section \ref{sec:ref-case-study}). 

Step 1 - $\text{gwp}_\text{limit} = $ 90.1 [Mt\coo-eq./y]: the system is partially electrified with a shift from NG to electric cars for private mobility. In addition, a part of the electricity production shifts from CCGT to electricity import, reaching 7\% of the primary energy mix. The NG share in the primary energy mix dropped by 10\% and reached 61\%.

Step 2 - $\text{gwp}_\text{limit} = $ 80.6 [Mt\coo-eq./y]: wet biomass is introduced and achieves 10\% of the primary energy mix. Centralized biomass co-generation technology supplies the centralized heat low-temperature demand instead of centralized gas co-generation. There is an additional decrease of 10\% of the NG share in the primary energy mix.

Step 3 - $\text{gwp}_\text{limit} = $ 61.7 [Mt\coo-eq./y]: PV technology and waste resource achieve the maximal available capacity with 59.2 [GWe] and 17.8 [GWh], representing 17\% and 5\% of the primary energy mix, respectively. Waste boilers and direct electricity heating replace the industrial gas boilers. The trucks shift from NG to electricity. Thermal seasonal and daily storage are introduced to cope with the solar and wind seasonal and daily intermittency. 

Step 4 - $\text{gwp}_\text{limit} = $ 56.9 [Mt\coo-eq./y]: synthetic renewable gas (gas-RE) begins to be imported. Then, when $\text{gwp}_\text{limit} = $ 33.2 [Mt\coo-eq./y], the NG disappears from the primary energy mix, and the gas-RE import amounts to 30\% of the primary energy mix.

Step 5 - $\text{gwp}_\text{limit} = $ 19.0 [Mt\coo-eq./y]: the imported methanol and the waste resource ($\text{gwp}_\text{limit} = $ 14.2 [Mt\coo-eq./y]) disappear from the primary energy mix.

Step 6 - $\text{gwp}_\text{limit} = $ 9.5 [Mt\coo-eq./y]: the imported ammonia and electricity vanish from the primary energy mix. The gas-RE import amounts to 60\% of the primary energy mix. CCGT technology replaces the electricity imports to produce electricity by using the gas-RE.

Step 7 - $\text{gwp}_\text{limit} = $ 4.7 [Mt\coo-eq./y]: there is an extreme shift from electric-based to gas-based technologies. Indeed, the wood, wet biomass, wind, and solar energies have almost completely vanished from the primary energy mix. The gas-RE, ammonia-RE, and methanol-RE imports amount to 86\%, 2\%, and 10\% of the primary energy mix, respectively. The cars and trucks shift from electric to gas technologies using the gas-RE. The CCGT produces electricity with gas-RE. Finally, boilers and co-generation using gas-RE satisfy the heat demand.
Note that the value of energy invested in the operation of renewable fuels is difficult to estimate. Thus, when the share of these fuels represents a large amount of the primary energy mix, the value of the EROI of the system becomes more uncertain than when using conventional fossil fuels.

\subsubsection{Breakdown of the system GHG emissions}

\begin{figure}[tb]
\centering
	\begin{subfigure}{0.5\textwidth}
		\centering
		\includegraphics[width=\linewidth]{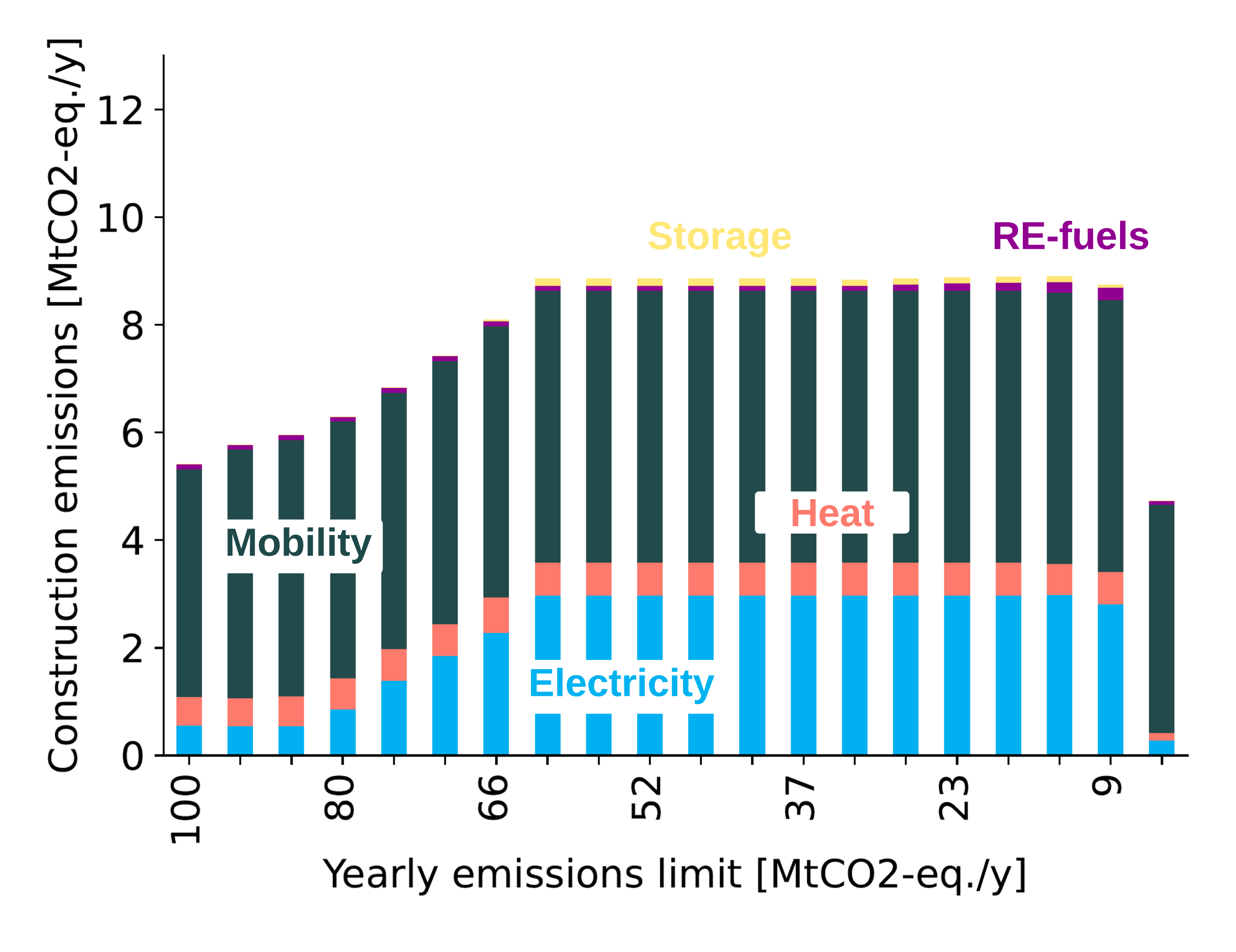}
    \caption{$\mathbf{GWP_\text{constr}}$.}
	\end{subfigure}
	\begin{subfigure}{0.5\textwidth}
		\centering
		\includegraphics[width=\linewidth]{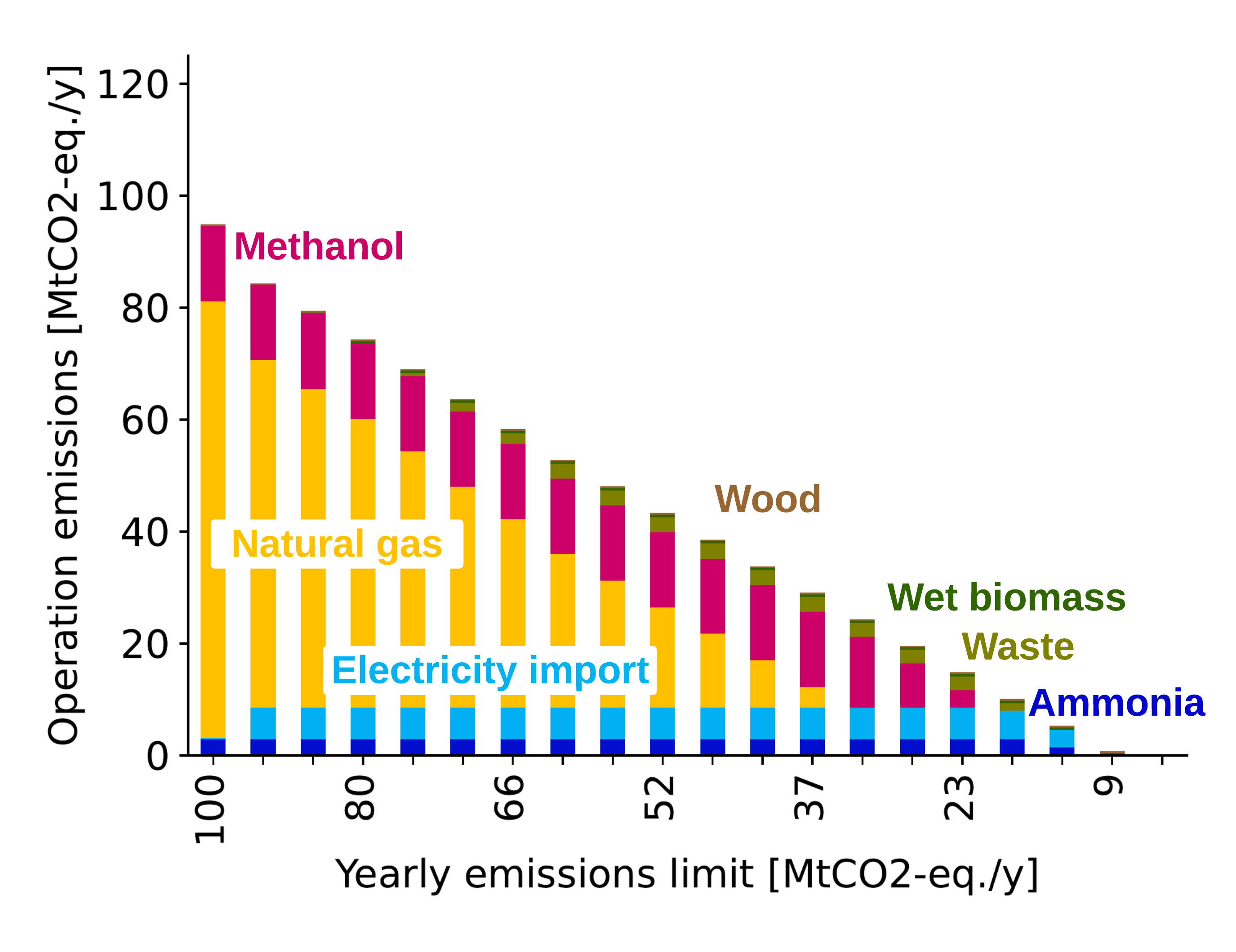}
    \caption{$\mathbf{GWP_\text{op}}$.}
	\end{subfigure}
\caption{$\mathbf{GWP_\text{constr}}$ and $\mathbf{GWP_\text{op}}$ of the system, broken down by technologies and resources for several yearly GHG emissions targets in 2035.}
\label{fig:GWP-breakdown}
\end{figure}

Figure \ref{fig:GWP-breakdown} depicts the $\mathbf{GWP_\text{constr}}$ and $\mathbf{GWP_\text{op}}$ of the system, broken down by technologies and resources for the different yearly GHG emissions targets in 2035.
The GHG construction emissions are mainly driven by electricity and mobility technologies. The PV installation amounts to a significant part of the electricity construction GHG emissions increase between GHG emissions targets of 80.6 and 61.7 [Mt\coo-eq./y]. Private passenger mobility composes the prominent part of the mobility construction GHG emissions with battery-electric cars for GHG emissions targets between 90.1 and 9.5 [Mt\coo-eq./y], and NG cars for the reference case and the GHG emissions target of 4.7 [Mt\coo-eq./y].
The GHG operation emissions mainly comprise non-renewable resources: NG, electricity import, methanol, and ammonia. They decrease with the progressive shift from non-renewable to renewable resources.

\subsubsection{System primary energy evolution}

\begin{figure}[tb]
\centering
\includegraphics[width=\linewidth]{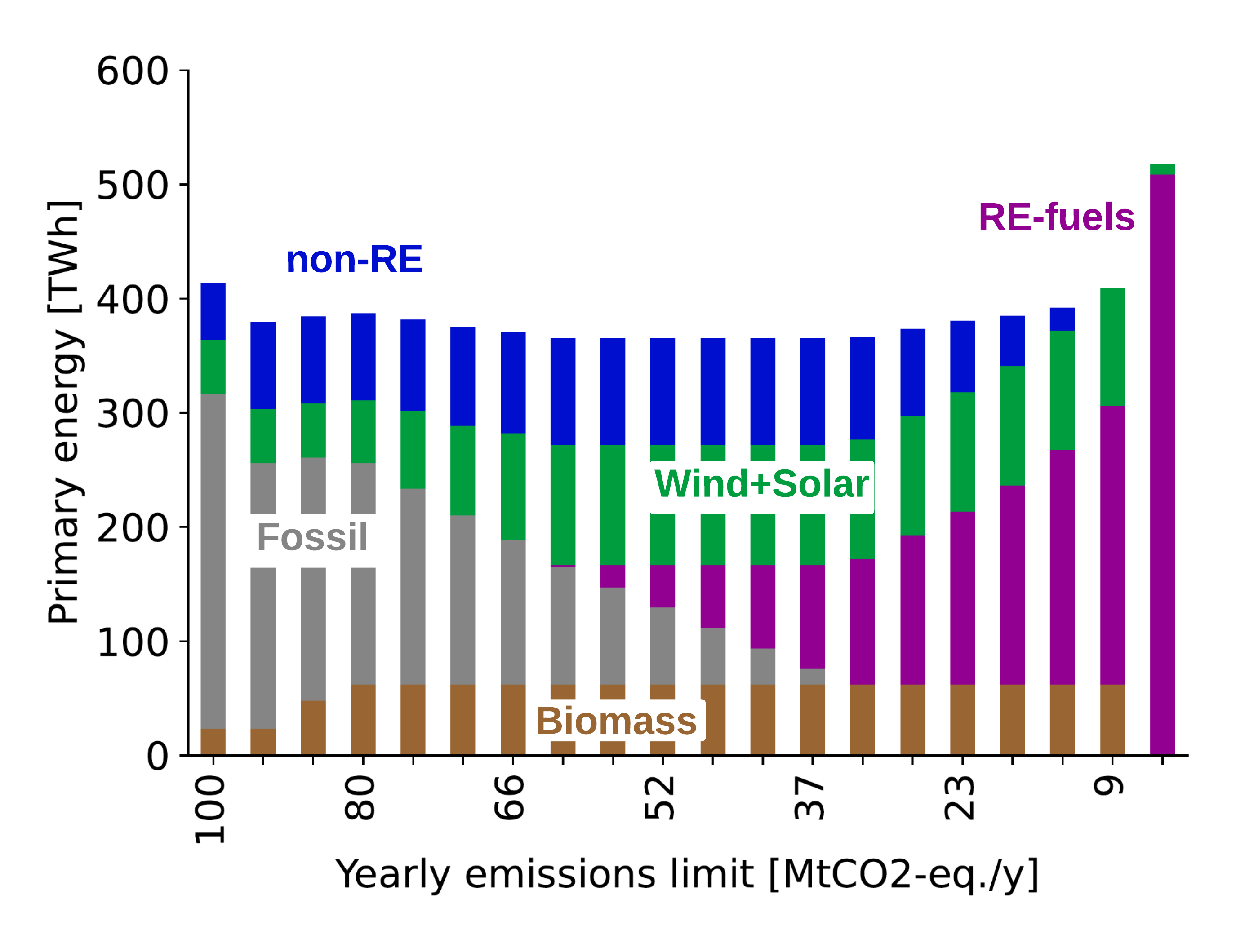}
\caption{Evolution of the system's primary energy mix for several yearly GHG emissions targets in 2035, broken down by resource categories. Abbreviations: non-RE: waste, methanol, and ammonia; Fossil: NG; RE-fuels: gas-RE, methanol-RE, and ammonia-RE; Biomass: wood, and wet biomass.}
\label{fig:GWP-primary-energy-categories}
\end{figure}

Figure \ref{fig:GWP-primary-energy-categories} depicts the evolution of the system's primary energy mix for several yearly GHG emissions targets in 2035, broken down by resource categories. 
Step 0 - ``reference scenario-100\%": the primary energy mix comprises mainly NG with a share of 71\%. Renewable resources are composed of wind, solar, and wood. Onshore and offshore wind resources are used at the maximum available capacity with 10 [GWe] and 6 [GWe]. They amount to 10.4\% of the primary energy mix. The wood represents 5.7\% of the primary energy and is also used at its maximum capacity of 23.4 [TWh/y]. Finally, the PV capacity is 3.9 [GWe] and amounts to only 1\% of the primary energy mix. Indeed, the energy invested in constructing offshore and onshore wind capacities is approximately two times lower than the PV capacity. 

\begin{figure*}[tb]
\centering
	\begin{subfigure}{0.5\textwidth}
		\centering
		\includegraphics[width=\linewidth]{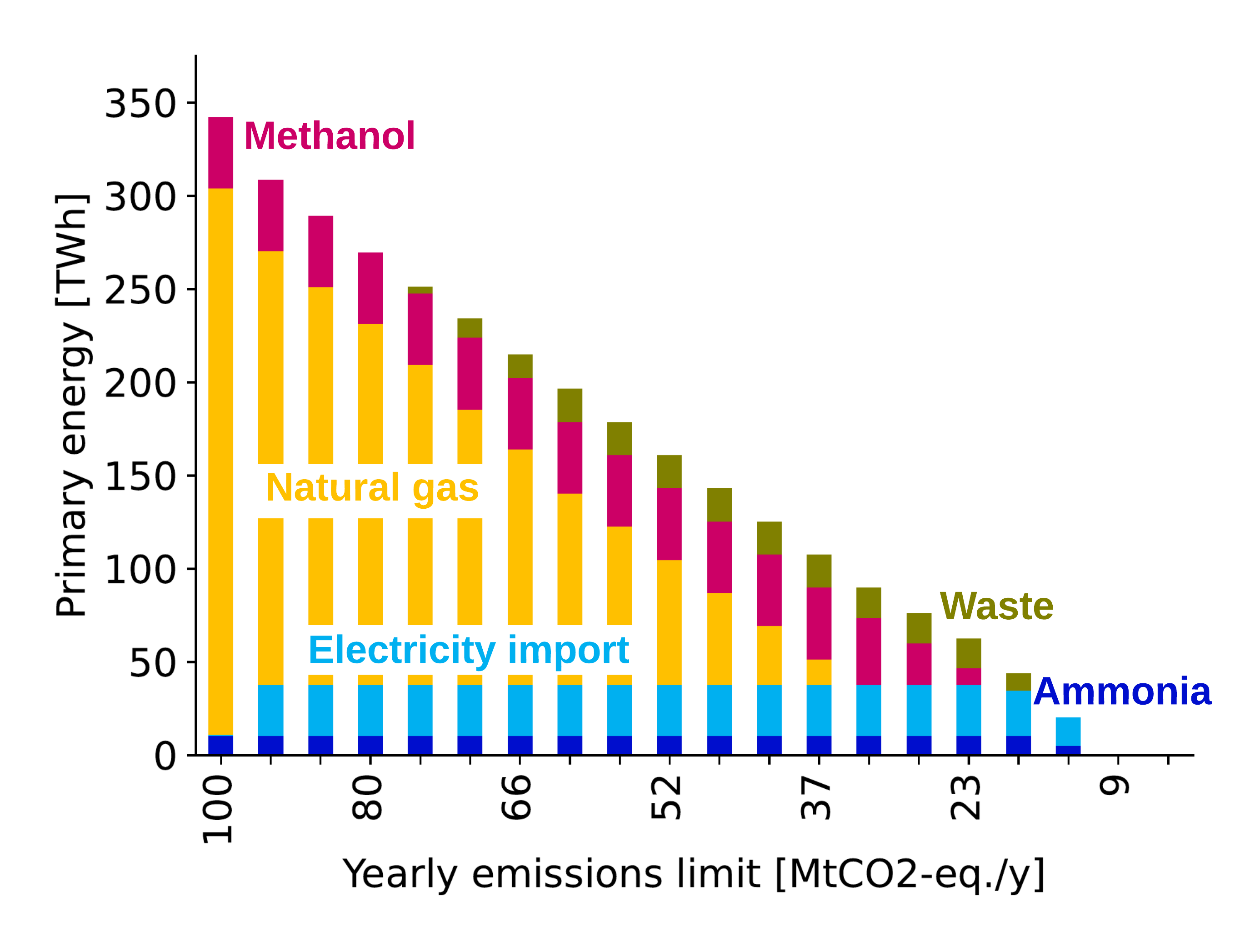}
    \caption{Non-renewable resources.}
	\end{subfigure}%
	\begin{subfigure}{0.5\textwidth}
		\centering
		\includegraphics[width=\linewidth]{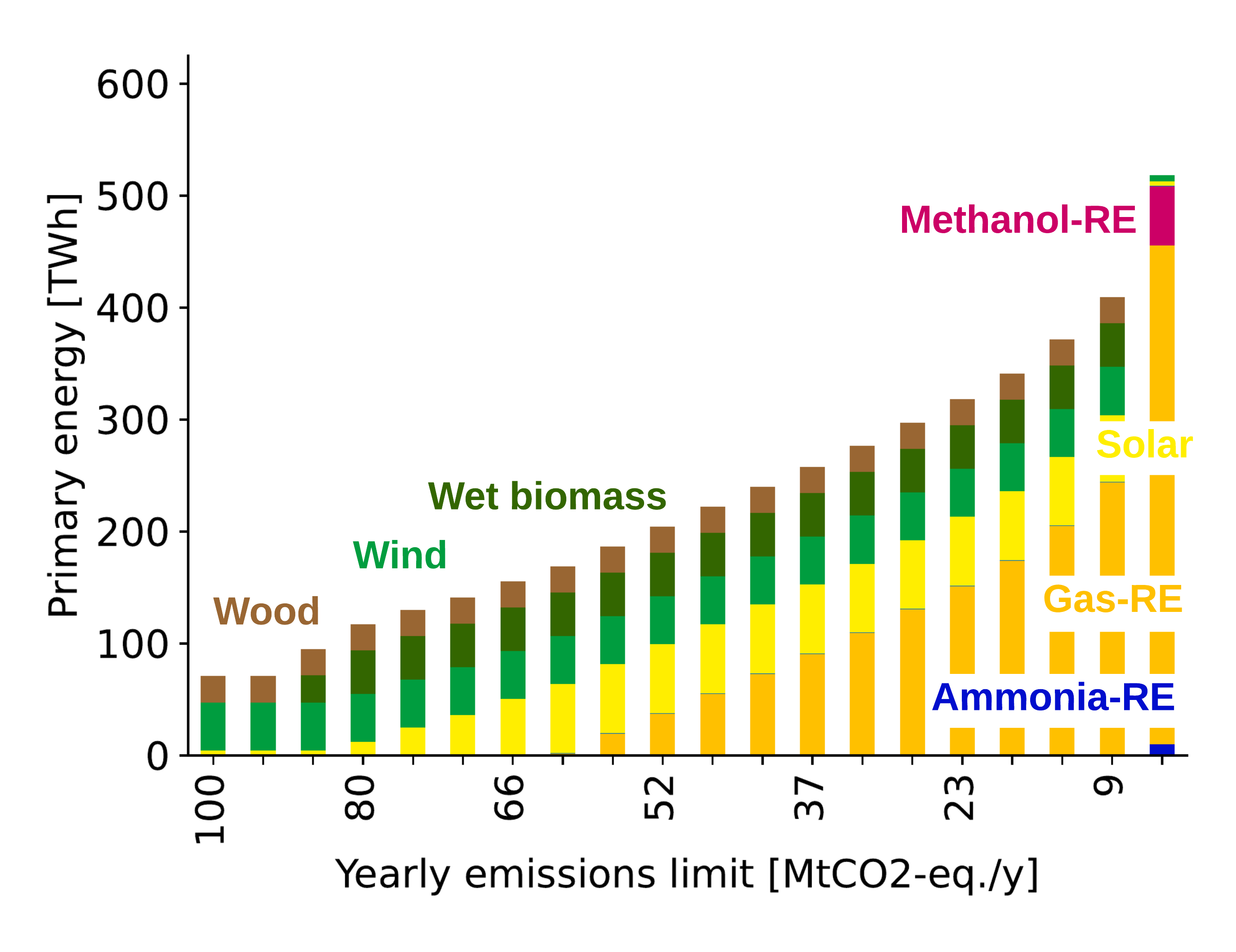}
    \caption{Renewable resources.}
	\end{subfigure}
\caption{Evolution of the system's primary energy mix for several yearly GHG emissions targets in 2035, broken down between non-renewable and renewable resources. Abbreviations: NG: natural gas, Elec. import: electricity import; renewable fuels: gas-RE, methanol-RE, and ammonia-RE; hydro: hydro river.}
\label{fig:GWP-primary-energy-breakdown}
\end{figure*}

Figure \ref{fig:GWP-primary-energy-breakdown} displays the evolution of the system's primary energy mix broken down between non-renewable and renewable resources. The following comments explain the main steps of the decrease in non-renewable resources.
%
Step 1 - $\text{gwp}_\text{limit} = $ 90.1 [Mt\coo-eq./y]: the decrease of NG in the primary energy mix is balanced with electricity import\footnote{The use of imported electricity is highly dependent on its indirect emissions. In this study, electricity imports emit about half of the electricity production of a CCGT. Therefore, imported electricity replaces the production of CCGTs.}, which reaches the maximum importation limit of 27.5 [TWh/y]. The other non-renewable and renewable resources are stable in volume compared to the ``reference scenario-100\%".
Step 2 - $\text{gwp}_\text{limit} = $ 80.6 [Mt\coo-eq./y]: wet biomass and solar renewable resources balance the NG decrease. The wet biomass is used at its maximal capacity of 38.9 [TWh/y]. The PV capacity starts to increase and reaches a capacity of 11.2 [GWe]. Then, from 80.6 to 61.7 [Mt\coo-eq./y], the NG decrease is balanced with increased waste and PV in the primary energy mix. The latter reaches the maximal installed capacity of 59.2 [GWe]. 
Step 3 - $\text{gwp}_\text{limit} = $ 61.7 [Mt\coo-eq./y]: NG amounts to 28.1\%, PV 16.8\%, wind offshore and onshore 11.7\%, wood 6.4\%, biomass 10.6\%, and waste 4.9\% of the primary energy mix.
Steps 4 to 6 - from 61.7 to 9.5 [Mt\coo-eq./y]: the decrease in NG, methanol, waste, and electricity imports is progressively balanced by importing renewable gas (gas-RE). 
Step 7 - $\text{gwp}_\text{limit} = $ 4.7 [Mt\coo-eq./y]: below 9.5 [Mt\coo-eq./y], there are no more non-renewable resources. GHG emissions targets force the system to decrease the construction GHG emissions as the operation GHG emissions are approximately 0 [Mt\coo-eq./y]. Renewable fuels almost completely replace the PV and wind resources, and the system is partially un-electrified to use them. For instance, electric vehicles (mobility private and freight) are replaced by vehicles using synthetic fuels.
That case is purely hypothetical because it assumes that: (1) imported renewable fuels imply lower GHG emissions than renewable resources such as solar and wind. However, the GWP data on renewable fuels are not mature enough to draw such conclusions; (2) importing such large quantities of renewable fuels (approximately 510 [TWh/y]) may be unrealistic.

\subsection{Comparison of the EROI and cost of the system}\label{sec:eroi-vs-cost}

When minimizing the total cost of the system instead of the EROI, without limiting the GHG emissions, EnergyScope TD gives an EROI for the Belgian energy system in 2035 at around 6.3  and a $\mathbf{GWP_\text{tot}}$ of 94.5 [Mt\coo-eq./y]. This can be compared with the EROI of 8.9 and $\mathbf{GWP_\text{tot}}$ of 100.3 [Mt\coo-eq./y] for the ``reference scenario-100\%" where the EROI is maximized.

\begin{figure*}[tb]
\centering
	\begin{subfigure}{0.5\textwidth}
		\centering
		\includegraphics[width=\linewidth]{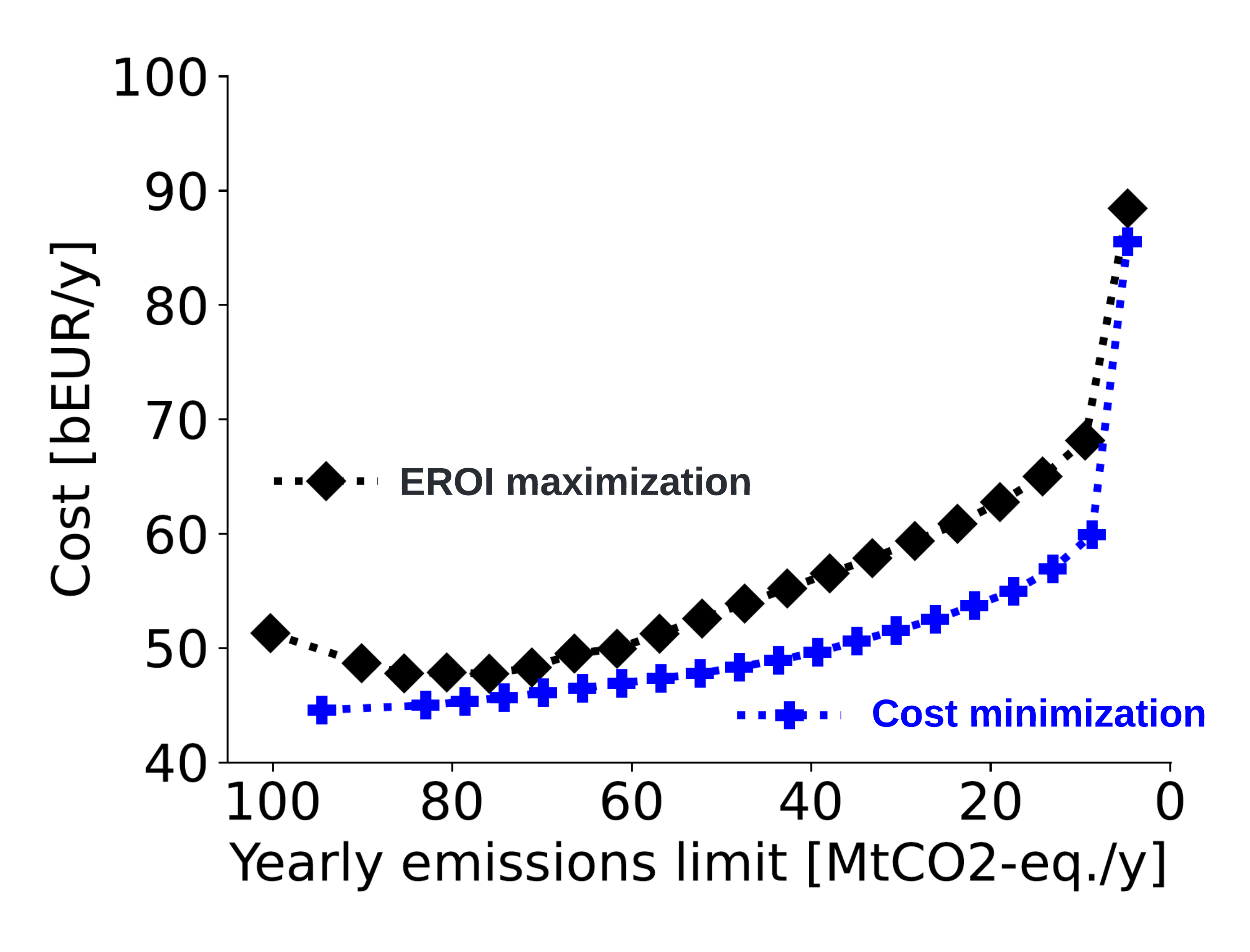}
		\caption{System's cost.}
	\end{subfigure}%
	\begin{subfigure}{0.5\textwidth}
		\centering
		\includegraphics[width=\linewidth]{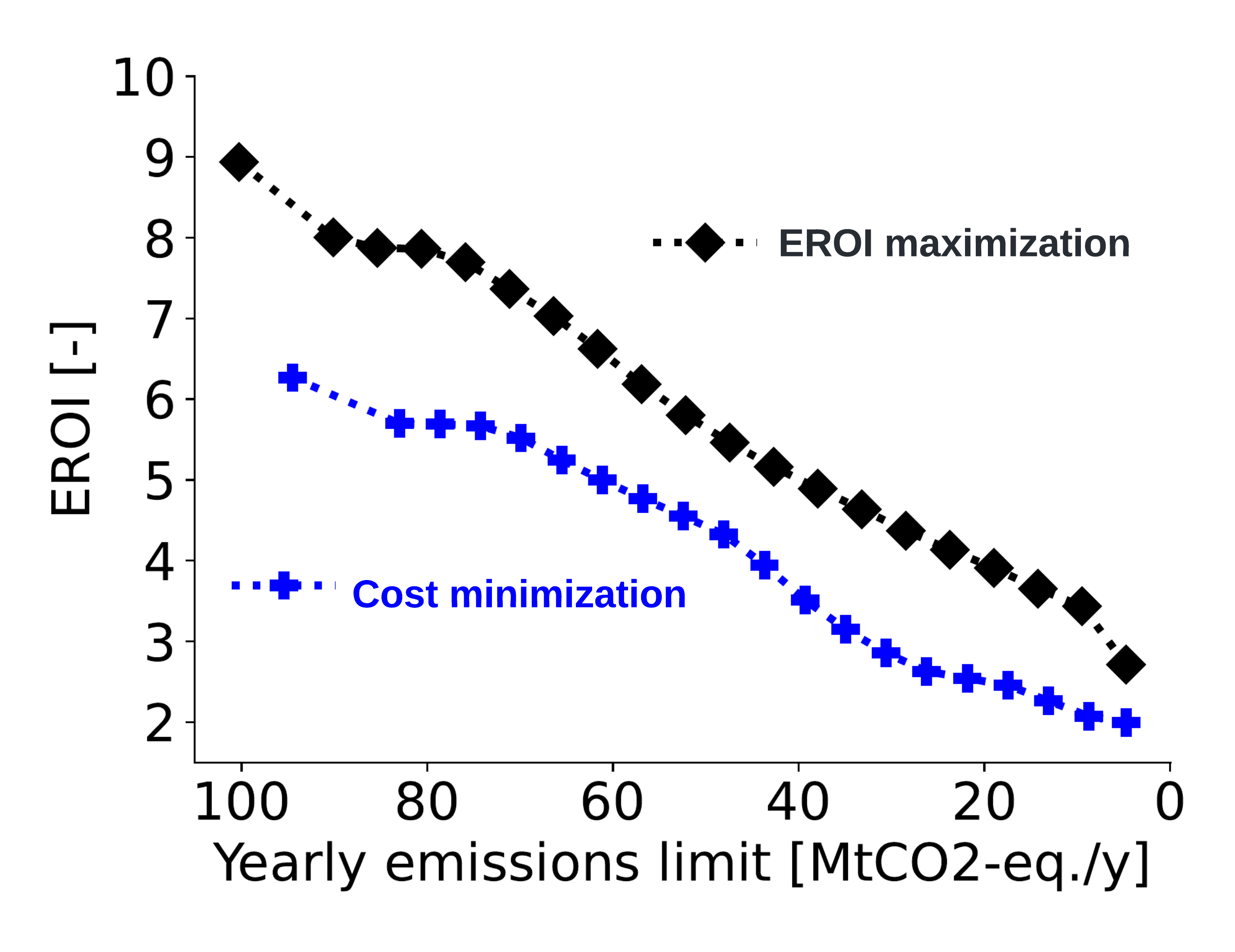}
		\caption{System's EROI.}
	\end{subfigure}
\caption{Comparison of the evolutions of the system's cost and EROI for several scenarios of GHG emissions, when, respectively, minimizing the cost and maximizing the EROI.}
\label{fig:cost-comparison-1}
\end{figure*}

\begin{figure*}[tb]
\centering
	\begin{subfigure}{0.5\textwidth}
		\centering
		\includegraphics[width=\linewidth]{figures/section-4/EI-non-RE-gwp-tot-0.pdf}
		\caption{EROI maximization.}
	\end{subfigure}%
	\begin{subfigure}{0.5\textwidth}
		\centering
		\includegraphics[width=\linewidth]{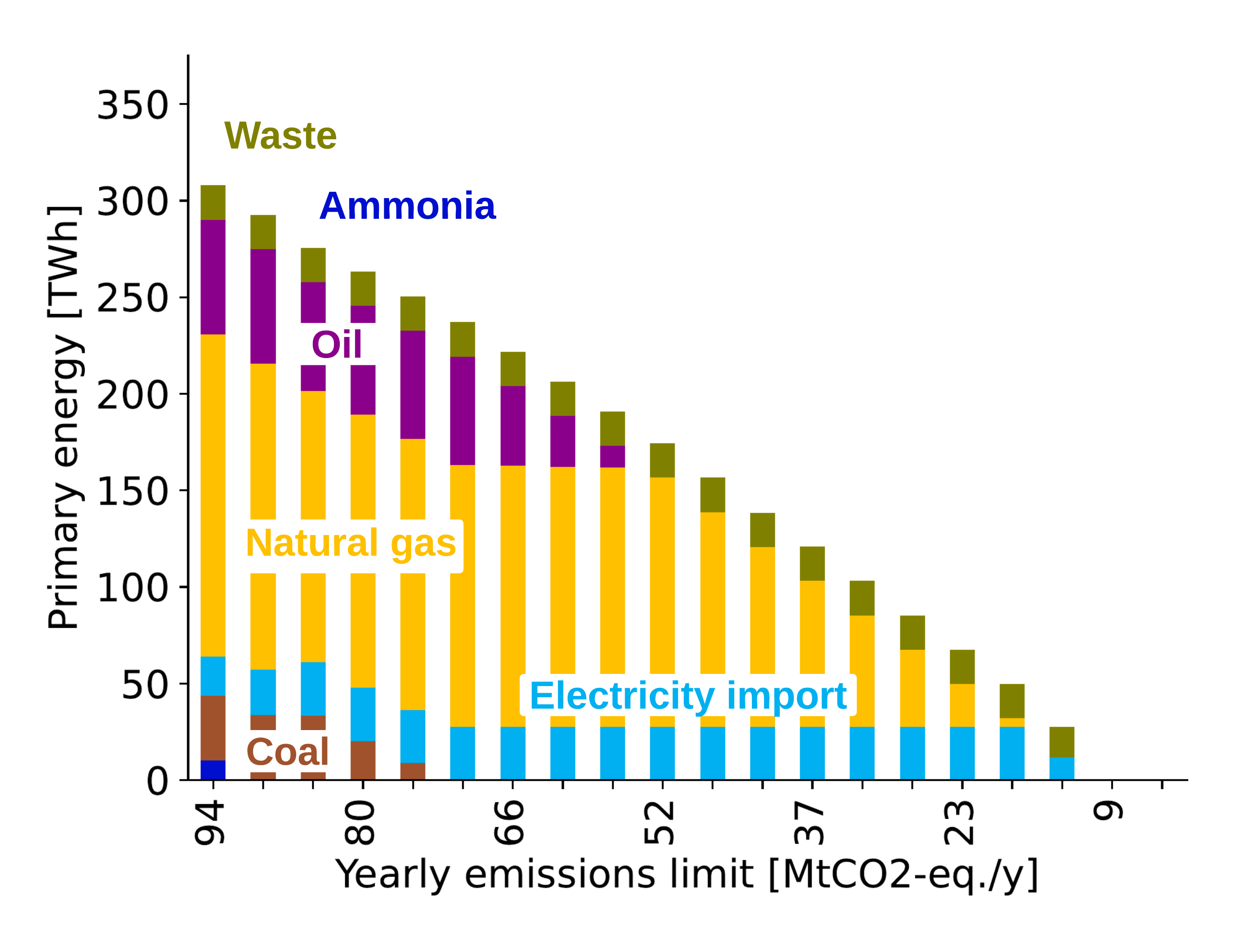}
		\caption{Cost minimization.}
	\end{subfigure}
	\begin{subfigure}{0.5\textwidth}
		\centering
		\includegraphics[width=\linewidth]{figures/section-4/EI-RE-gwp-tot-0.pdf}
		\caption{EROI maximization.}
	\end{subfigure}%
	\begin{subfigure}{0.5\textwidth}
		\centering
		\includegraphics[width=\linewidth]{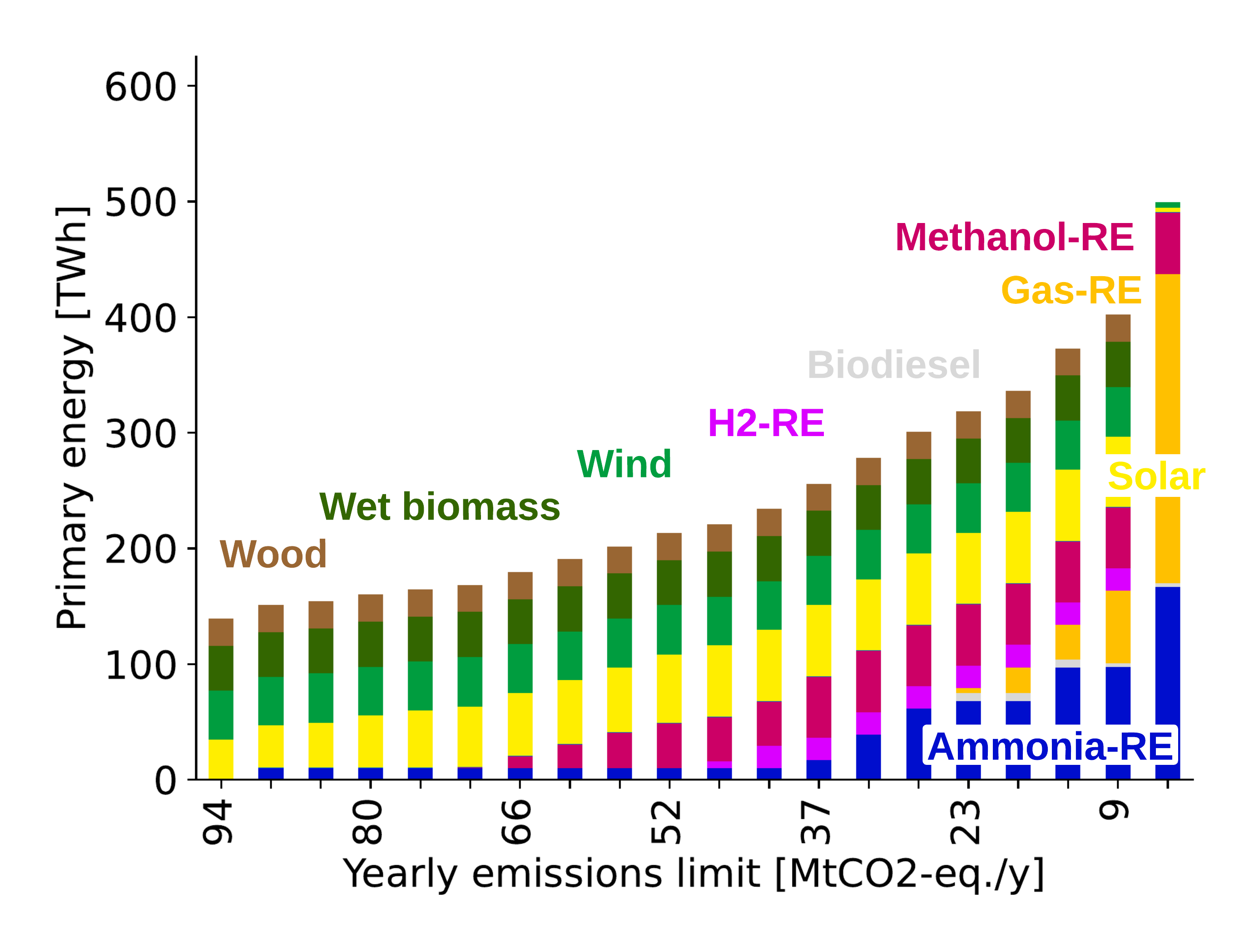}
		\caption{Cost minimization.}
	\end{subfigure}
\caption{Comparison of the system's primary energy mix when minimizing the cost (left) and maximizing the EROI (right), for several scenarios of GHG emissions. The mix is broken down between non-renewable resources (upper) and renewable resources (lower)}
\label{fig:cost-comparison-2}
\end{figure*}

Figure \ref{fig:cost-comparison-1} depicts the evolution of the cost and EROI of the 2035 system for several GHG emissions scenarios when minimizing the cost and maximizing the EROI. The trends are similar. 
%
However, the primary energy mix differs, as illustrated by Figure \ref{fig:cost-comparison-2}. When minimizing the system cost, with no GHG emissions target, the system uses a higher share of domestic renewable resources with 41.9 [TWh/y] of onshore and offshore wind, 33.8 [TWh/y] of PV, 23.4 [TWh/y] of wood, and 38.9 [TWh/y] of wet biomass. In comparison, when maximizing the EROI, the system uses 43 [TWh/y] of wind onshore and offshore, 4 [TWh/y] of PV, 23.4 [TWh/y] of wood, and 0 [TWh/y] of wet biomass. In addition, the system uses a smaller share of NG, which is compensated with more renewable energies, coal, light oil fuel (LFO), waste, and electricity import.
When the GHG emissions decrease, the fossil energies are progressively replaced by imported renewable fuels: ammonia, methanol, H2, and renewable gas. Indeed, domestic renewable energies are almost already used at maximal capacity, except for the PV with 32.2 [GWe] when there is no restriction on GHG emissions. 

Overall, the strategy to decrease GHG emissions is similar when minimizing the system cost and maximizing the EROI by importing an increased share of renewable fuels. 
When GHG emissions are between 56.9 and 4.7 [MtCO2-eq./y], the amounts of domestic renewable energies (PV and wind onshore and offshore) in the system's primary energy are similar.
However, instead of using large quantities of imported renewable gas, the system also employs renewable ammonia, methanol, and H2. The renewable ammonia is employed in CCGT to produce electricity, and the renewable gas in CHP is consumed to satisfy a part of the high-temperature heat demand. The methanol is converted into HVC, and the H2 is used for the freight. Finally, the wet biomass produces renewable gas with the bio-methanation process instead of being directly consumed in CHP, and boilers consume the wood to satisfy the heat high-temperature demand instead of being transformed into methanol.
%
%
The higher the amount of renewable fuel, the lower the EROI of the system. Thus, in both cases, the EROI of the system decreases when constraining the GHG emissions. This result is specific to this case study since Belgium has limited domestic renewable energies, approximately being able to cover at most 30 \% of its primary energy consumption. Thus, first domestic renewable energies are installed and used at their maximal capacities to decrease GHG emissions. However, it is not enough to reach ambitious GHG targets, and the model has no choice but to import renewable fuels to achieve low GHG emissions targets.
In other case studies with a higher potential for domestic renewable energies, such as a vast wind offshore potential (or a high share of nuclear), the trend of EROI could differ. Wind offshore is a technology with relatively low energy invested and cost in construction, such as nuclear. Large shares of these energies could help prevent drastically decreasing the EROI of the system to achieve low carbon targets. However, it is not straightforward to estimate if the trends of the EROI of the system will be similar or not when optimizing the EROI or the cost. It is highly dependent on the data used and the potential of each resource.

\section{Sensitivity analysis of the system's EROI}\label{sec:section-5}

The model relies on numerical input data, which are sometimes highly uncertain, such as the energy invested in the operation of renewable fuels. This uncertainty could influence the key messages of the previous deterministic results. To nuance these messages, two actions are proposed: (i) being transparent on the dataset used (refer to the online documentation), (ii) assessing the impact of uncertainty of the system's EROI for several GHG emissions targets through global sensitivity analysis (GSA), and the Monte Carlo method.
This section is an extension of the works of \citet{limpens2021generating,rixhon2021role} by assessing the uncertainty of the EROI of the system with a GSA using the polynomial chaos expansion (PCE) method \citep{sudret2014polynomial}. 
The PCE approach emphasizes the critical parameters by using Sobol indices and extracting statistical moments, mean and variance, of the EROI of the system.
The implementation of this approach is conducted by using the RHEIA\footnote{\url{https://rheia.readthedocs.io/en/latest/}} \citep{coppitters2020robust} Python library.

This section is organized into two parts. First, the most critical uncertain parameters for the EROI of the system are listed according to their respective Sobol indices based on the GSA results. Then, the EROI of the system is analyzed through its mean, variance, and probability density function (pdf).
Appendix \ref{appendix:section-5} provides the details about the GSA approach and additional results concerning the first and second-order PCE.

\subsection{Critical parameters}

\begin{figure}[tbh!]
\centering
	\begin{subfigure}{0.5\textwidth}
		\centering
\includegraphics[width=1\linewidth]{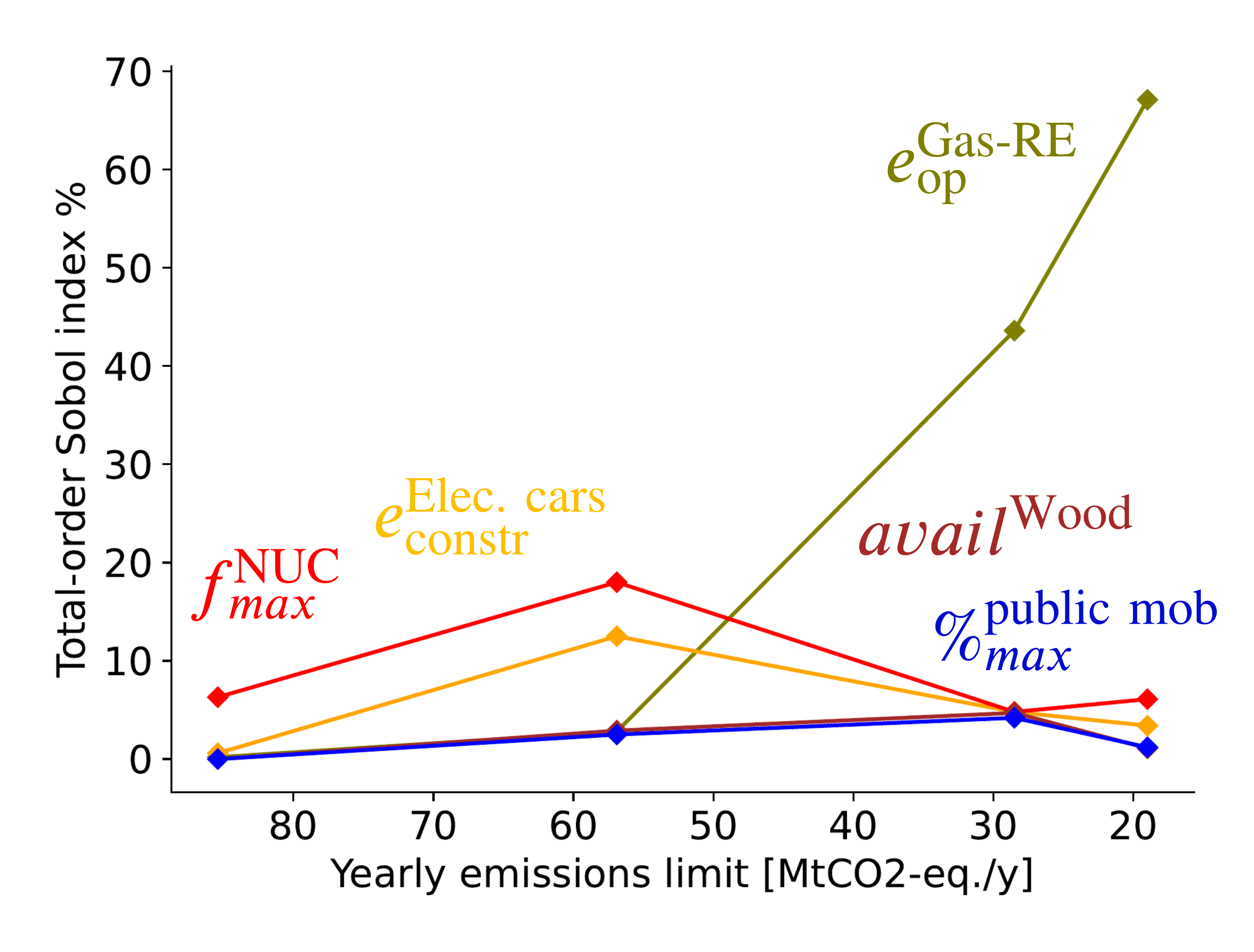}
		\caption{Top-5 parameters for GWP\textsubscript{tot}=28.5 [Mt\coo-eq./y].}
		\label{fig:params_for_28.5}
	\end{subfigure}
	\begin{subfigure}{0.5\textwidth}
		\centering
\includegraphics[width=1\linewidth]{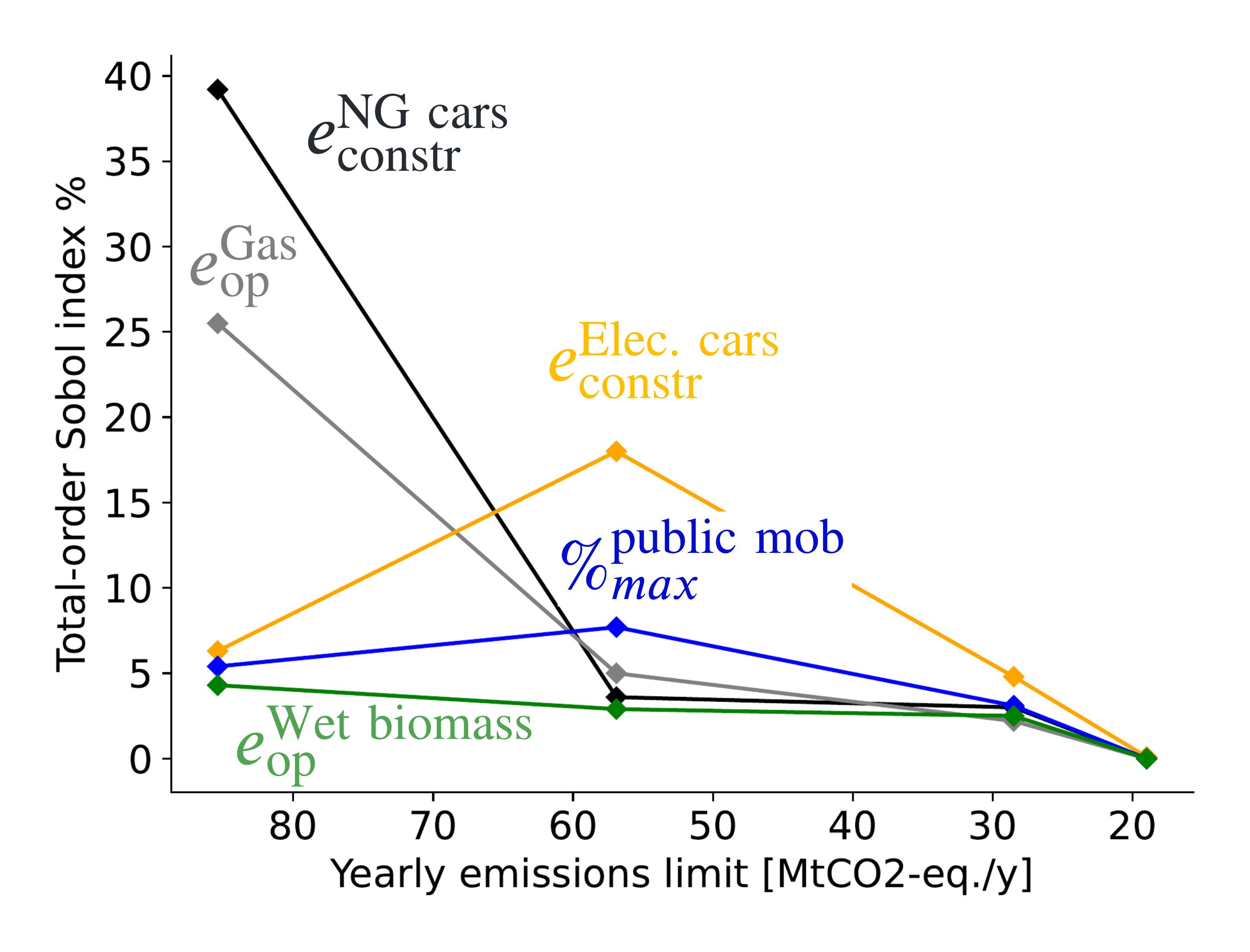}
		\caption{Top-5 parameters for GWP\textsubscript{tot}=85.4 [Mt\coo-eq./y].}
		\label{fig:params_for_85.4}
	\end{subfigure}
\caption{Evolution of the total-order Sobol values of the Top-5 critical parameters for GHG emissions of 28.5 and 85.4 [Mt\coo-eq./y].
Abbreviations: energy invested in the operation of gas-RE cars ($e_\text{op}^\text{Gas-RE}$), energy invested in the construction of electric cars ($e_\text{constr}^\text{Elec. cars}$), nuclear maximal installed capacity ($f_{max}^\text{NUC}$), wood maximal availability ($avail^\text{Wood}$), maximal share of public mobility ($\%_{max}^\text{public mob}$), energy invested in the construction of NG cars ($e_\text{constr}^\text{NG cars}$), energy invested in the operation of NG cars ($e_\text{op}^\text{NG}$),  energy invested in the operation of wet biomass ($e_\text{op}^\text{Wet biomass}$).
}
\label{fig:critical-parameters-evol}
\end{figure}

Table \ref{table:critical-parameters} presents the top-5 critical parameters and their total-order Sobol values [\%] for each GHG emissions target considered. The total-order Sobol value of a parameter indicates its contribution to the variance of the EROI of the system. Table \ref{table:appendix-critical-parameters} in Appendix \ref{appendix:pce-second-order-results} lists all the critical parameters with their total-order Sobol values.
Figure \ref{fig:critical-parameters-evol} illustrates the evolution of the critical parameters as a function of GHG emissions constraints. More precisely, the evolution of the total-order Sobol values of the top-5 parameters for the GHG emissions constraints of 28.5 [Mt\coo-eq./y] (Figure \ref{fig:params_for_28.5}) and 85.4 [Mt\coo-eq./y] (Figure \ref{fig:params_for_28.5}) are represented.
%
%
\begin{table*}[tb]
\renewcommand{\arraystretch}{1.25}
\centering
\begin{tabular}{lrrrr} 
 \hline
  \backslashbox[22mm]{\small{Ranking}}{\small{GWP\textsubscript{tot}}}
& 85.4 & 56.9 & 28.5
& 19.0\\
 \hline
 1& $e_\text{constr}^\text{NG cars}$ 39.2 & $e_\text{constr}^\text{Elec. cars}$ 18.0& $e_\text{op}^\text{Gas-RE}$ 43.6&  $e_\text{op}^\text{Gas-RE}$ 67.1\\
 2& $e_\text{op}^\text{NG}$ 25.5 & $f_{max}^\text{NUC}$ 12.5& $e_\text{constr}^\text{Elec. cars}$ 4.8& $e_\text{constr}^\text{Elec. cars}$ 6.1 \\
 3& $e_\text{constr}^\text{Elec. cars}$ 6.3& $\%_{max}^\text{public mob}$ 7.7& $f_{max}^\text{NUC}$ 4.8 & $f_{max}^\text{NUC}$ 3.4\\
 4& $\%_{max}^\text{public mob}$ 5.4& $e_\text{constr}^\text{PV}$ 5.0& $avail^\text{Wet biomass}$ 4.7& $avail^\text{Wood}$ 2.5 \\
5& $e_\text{op}^\text{Wet biomass}$ 4.3& $e_\text{op}^\text{NG}$ 5.0&$f_{max}^\text{Offshore wind}$ 4.2& $\%_{max}^\text{public mob}$ 2.3 \\ \hline
\end{tabular}
\caption{Top-5 critical parameters and their total-order Sobol values [\%] for several GHG emissions targets [Mt\coo-eq./y].
Abbreviations: operating energy required to use resource $i$ ($e_\text{op}^\text{i}$), energy invested in the construction of a technology $j$ ($e_\text{constr}^\text{j}$), maximal installed capacity of a technology $j$ ($f_{max}^\text{j}$), maximal availability of a resource $i$ ($avail^\text{i}$), maximal share of public mobility ($\%_{max}^\text{public mob}$), electric (elec.) photovoltaic (PV), nuclear (NUC), renewable gas (Gas-RE), natural gas (NG).}
\label{table:critical-parameters}
\end{table*}

%
It is expected that $e_\text{op}^\text{Gas-RE}$ becomes the primary driver of the variation of the EROI of the system when GHG emissions targets decrease, given the increasing share of renewable gas in the primary energy mix. In the model, the value of the energy invested in the operation of renewable gas is 4.4 times greater than its fossil equivalent, making it less competitive; thus, unused with no restrictions on GHG emissions. However, it becomes the most-impacting parameter, up to 67.1\% of the variance of the EROI of the system for the GHG emissions target of 19.0 [Mt\coo-eq./y].
The lower part of Figure \ref{fig:critical-parameters-evol} illustrates the opposite trend for NG. Given its low energy invested in operation, it is a critical resource when the GHG emissions targets are not compelling. The energy invested in the construction of gas cars and NG operation substantially impacts the EROI variance with 39.2\% and 25.5\%, respectively, when the GHG emissions are weakly constrained (85.4 [Mt\coo-eq./y]).

Then, the energy invested in the construction of electric cars $e_\text{constr}^\text{Elec. cars}$ is the second parameter to play a key role in the variance of the EROI of the system with the decrease of GHG emissions. It is the most-impacting parameter on the EROI with 18.0\%, for the target of 56.9 [Mt\coo-eq./y]. Then, it is the second most-impacting parameter with 4.8\% and 6.1\%, for the targets of 28.5 and 19.0 [Mt\coo-eq./y].
Figure \ref{fig:GWP-einv-const} in Appendix \ref{appendix:section-4} depicts the essential impact on the system's EROI of the private mobility for GHG emissions between 61.7 and 9.5 [Mt\coo-eq./y], where the energy invested in construction is mainly composed of mobility and electricity technologies, and more particularly of PV and electric cars. 

The maximum capacity of nuclear power plants $f_{max}^\text{NUC}$ is the third critical parameter. It has lower energy invested in construction than PV, 2600 \textit{vs.} 4400 [GWh/GW], which is similar to wind on/offshore, and a negligible related global warming potential of construction. Thus, the system consistently relies on the maximum capacity of the nuclear power plants. $f_{max}^\text{NUC}$ is the second most-impacting parameter with 12.5\% for GHG emissions target of 56.9 [Mt\coo-eq./y], and the third one with 4.8\% and 3.4\% for targets of 28.5 and 19.0 [Mt\coo-eq./y].

The wet biomass $avail^\text{Wet biomass}$ and wood $avail^\text{Wood}$ availabilities with 4.7\% and 2.5\%, respectively, are the fourth critical parameters for low GHG emissions targets of 28.5 and 19.0 [Mt\coo-eq./y].
The operating energy required to use the wood resource is lower than biomass, 0.049 \textit{vs.} 0.056 [GWh/GWh] for an equivalent global warming potential. The wood is used by the model to produce methanol for satisfying the non-energy demand. Thus, it allows for limiting the methanol importations, which require higher operating energy with 0.08 [GWh/GWh].

Finally, for GHG emissions targets of 28.5 and 19.0 [Mt\coo-eq./y], [Mt\coo-eq./y], the fifth critical parameters are the offshore wind maximal installed capacity $f_{max}^\text{Offshore wind}$ and the maximal share of public mobility $\%_{max}^\text{public mob}$ with 4.2\% and 2.3\%, respectively.
Private car is the most significant partaker in the passenger mobility in Belgium. According to the Federal Planning Bureau \citep{du2015perspectives}, 80\% of the passenger mobility is expected to be supplied by private cars in the future. Therefore, it supports half of the passenger mobility, and the other half is supplied by public transport modes, \textit{i.e.}, buses, trains, and tramways. Thus, the uncertainty on the maximal share of public mobility $\%_{max}^\text{public mob}$ is likely to impact significantly private mobility and the EROI of the system.

\subsection{EROI probability density functions}

The PCE coefficients allow estimating the statistical moments, \textit{e.g.}, mean $\mu$ and variance $\sigma^2$, of the EROI of the system, without additional computational cost. Furthermore, based on the obtained surrogate model and with a few supplementary seconds of computational time, the pdf of the EROI can be estimated by a Monte Carlo approach.

\begin{figure}[tb]
\centering
\includegraphics[width=1\linewidth]{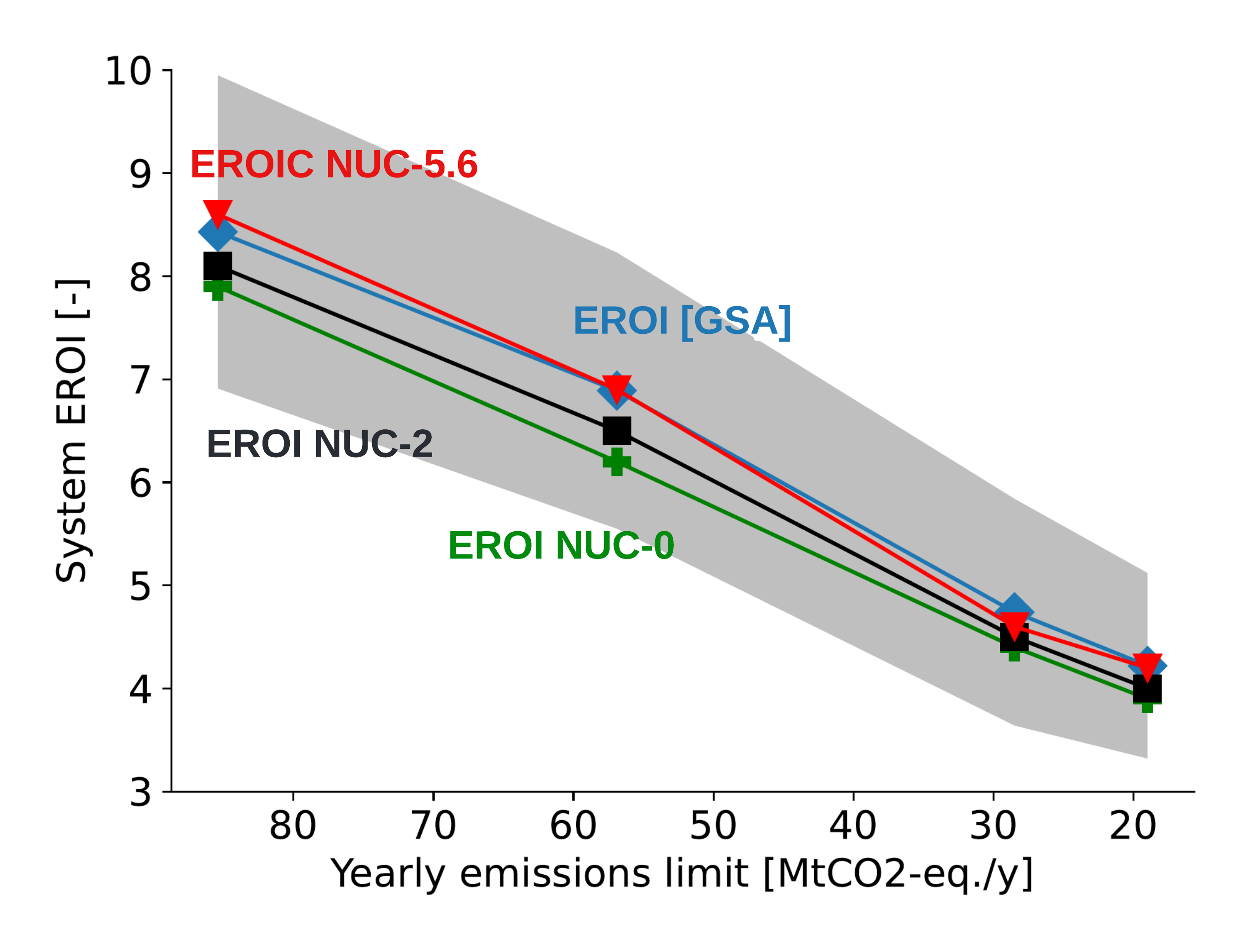}
\caption{The mean values of EROI  $\mu$ of the runs performed during the sensitivity analysis (EROI [GSA], blue curve) with the 95\% confidence interval ($\pm 2 \sigma$ in gray). The EROI values of the three deterministic scenarios: (i) with the nominal value of the parameters corresponding to the complete phase-out of nuclear (EROI [deterministic] NUC-0, green curve); (ii) the two alternative scenarios with the extension of 2 [GWe] (EROI [deterministic] NUC-0, black curve) and 5.6 [GWe] (EROI [deterministic] NUC-5.6, red curve).}
\label{fig:eroi-statistics}
\end{figure}

\begin{figure}[tb]
\centering
\includegraphics[width=1\linewidth]{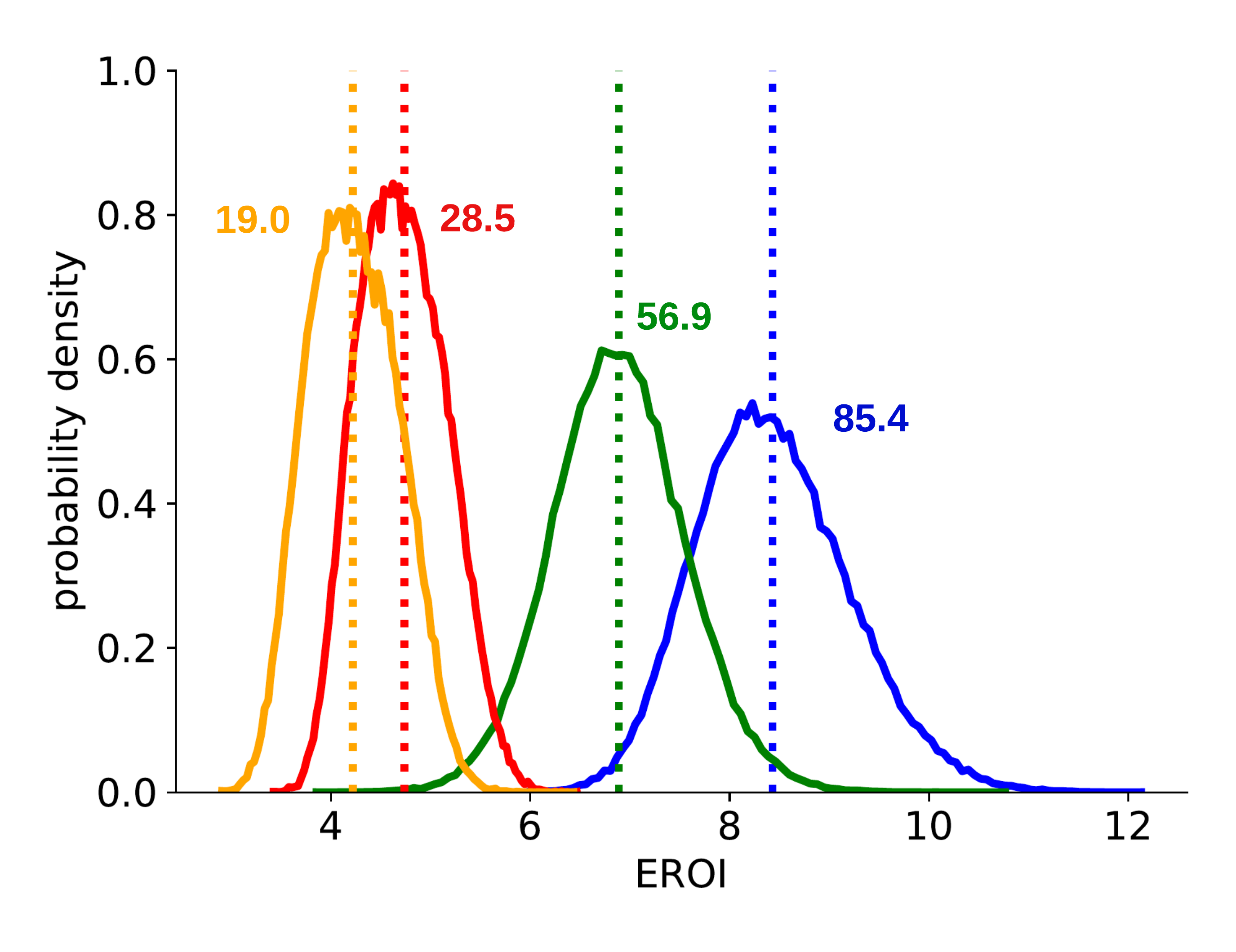}
\caption{Probability density function (plain lines) of the EROI for several GHG emissions targets. The dashed vertical lines provide the EROI mean of the runs performed during the sensitivity analysis: 8.4, 6.9, 4.7, and 4.2 for 85.4, 56.9, 28.5, and 19.0 [Mt\coo-eq./y], respectively.}
\label{fig:compare-pdf-eroi}
\end{figure}

Figure \ref{fig:eroi-statistics} depicts, for the GHG emissions targets considered, the EROI mean $\mu$ (EROI [GSA]) and the evolution of the 95\% ($\pm 2 \sigma$ in gray) confidence interval, along with the EROI values from the deterministic approach EROI [Deterministic] NUC-0, where the maximal installed capacity of nuclear power plants is 0 [GWe]. 
Based on the 2021 policies, Belgium planned to phase out coal and nuclear. However, the 2022 policies\footnote{On Friday, 18 March 2022, the Belgian government decided to extend the two most recent nuclear reactors (Doel 4 and Tihange 3) in operation for another ten years until 2035, corresponding to 2 GWe \citep{afcn2022}.} reconsider the progressive shutdown of the nuclear power plants. Thus, two deterministic scenarios in 2035 with 2 (EROI [Deterministic] NUC-2) and 5.6 [GWe] of maximal installed nuclear capacity are considered. They correspond to the extension of 2 and all out of the seven current nuclear reactors, respectively.
Table \ref{table:eroi-pdf-statistics} presents the mean, the standard deviation of the EROI, and the coefficient of variation (CoV), defined as the ratio between $\sigma$ and $\mu$. It also provides the values of the system's EROI for the three deterministic scenarios: NUC-0, NUC-2, and NUC-5.6.
Finally, Figure \ref{fig:compare-pdf-eroi} presents the EROI pdf for each GHG emissions target using the Monte Carlo approach along with the mean depicted by the dashed vertical line.
\begin{table}[tb]
\renewcommand{\arraystretch}{1.25}
\centering
\begin{tabular}{lrrrr} 
 \hline
$\mathbf{GWP_\text{tot}}$ [Mt\coo-eq./y] & 85.4 & 56.9 & 28.5 & 19.0 \\ \hline
Deterministic NUC-0 & 7.9 & 6.2 & 4.4 & 3.9 \\
Deterministic NUC-2 & 8.1 & 6.5 & 4.5 & 4.0 \\
Deterministic NUC-5.6 & 8.6 & 6.9 & 4.6 & 4.2 \\
Mean $\mu$ & 8.4 & 6.9 & 4.7 & 4.2 \\
Standard deviation  $\sigma$ & 0.76 & 0.67 & 0.55& 0.45 \\
CoV $\sigma / \mu $ [\%] & 9.0 & 9.7 & 11.6 & 10.5 \\ \hline
\end{tabular}
\caption{Deterministic value for the three deterministic scenarios, mean and standard deviation of the EROI resulting from the sensitivity analysis, and the ratio between $\sigma$ and $\mu$, the coefficient of variation (CoV).}
\label{table:eroi-pdf-statistics}
\end{table}

Phasing out of low energy-intensive fossil fuels, particularly NG, by relying on more renewables and importing renewable fuels naturally drives down the EROI of the system. This decrease in EROI raises concerns about a minimal societal EROI value below which a prosperous lifestyle would not be sustainable. The estimation of such value is complex and out of the scope of this study. However, it is not impossible that considering the actual values of energy invested in the construction of renewable technologies and the required operational energy for renewable resources, the EROI of the system could reach this limit before achieving carbon neutrality.
The estimated EROI probability density functions indicate that decreasing the GHG emissions makes the EROI more sensitive to uncertain parameters. Indeed, the 95\% confidence interval ($\pm 2\sigma$) narrows slowlier than the decrease in the mean. This trend is translated in an increase in the CoV with 9.0 \% \textit{vs.} 10.5\% for GHG emissions targets of 85.4 and 19.0 [Mt\coo-eq./y], respectively, as depicted in Table \ref{table:eroi-pdf-statistics}.
These results reinforce the importance of considering the uncertainties of parameters in long-term energy planning. In addition, in this case study, deterministic optimization with the reference scenario (NUC-0) gives an EROI that is 6-7\% lower than the mean of the GSA and cannot provide a confidence interval. The maximal nuclear capacity $f_{max}^\text{NUC}$ is partially responsible for this underestimation, set to 0 [GWe] in the deterministic scenario NUC-0. Indeed, in the two alternative deterministic scenarios with 2 [GWe] and 5.6 [GWe] of maximal installed capacity, the EROI values are closer to the mean values of the stochastic settings.
Indeed, the nuclear maximal installed capacity parameter is considered uncertain in the sensitivity analysis with a uniform distribution between 0 and 5.6 [GWe]. Then, the model systematically uses the nuclear capacity at its maximal value thus increasing the EROI, since the energy invested in constructing nuclear power plants is relatively low and similar to wind on/offshore.

\section{Discussion and limitations}\label{sec:section-6}

This section first discusses the results of Sections \ref{sec:section-4} and \ref{sec:section-5}. Then, it presents the limitations of the model and of the methodology used to perform the sensitivity analysis of the EROI.

\subsection{Discussion}

Overall, this study provides the following primary outcomes when maximizing the EROI of the system: (1) renewable energies (either domestic such as solar and wind, or imported with renewable fuels) are required massively to reach ambitious GHG emissions targets; (2) nuclear energy is not the primary driver of the EROI variance.
\begin{figure}[tb]
\centering
\includegraphics[width=\linewidth]{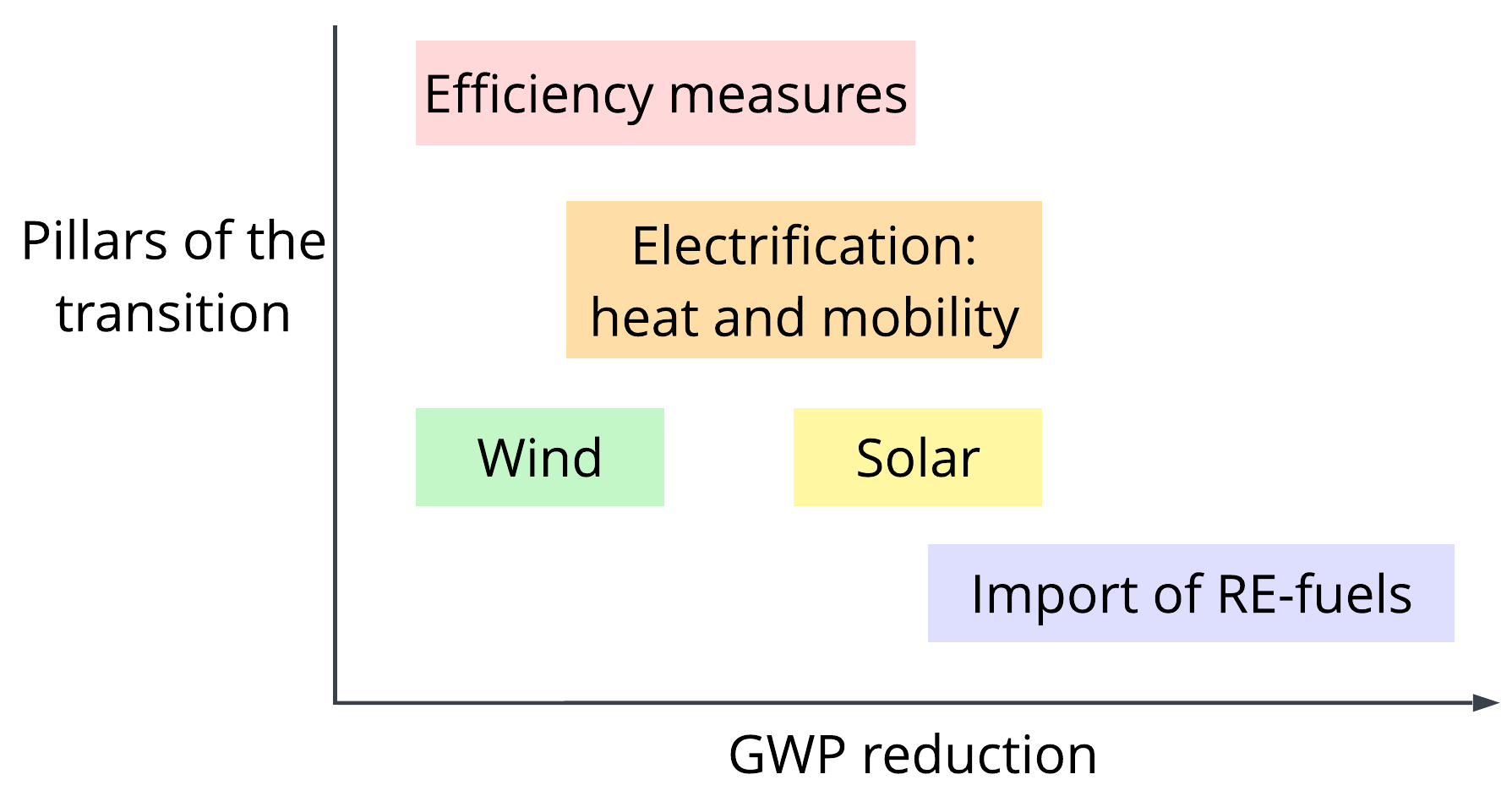}
\caption{The pillars of the energy transition to decrease the GHG emissions when maximizing the EROI of the system.}
\label{fig:pillar-transition}
\end{figure}

Concerning the first point, renewable fuels play an increasingly key role in the Belgian energy transition to satisfy the mobility and heating end-use demands when decreasing GHG emissions. They represent the major part of the system primary energy when $\mathbf{GWP_\text{tot}} \approx 10$ [Mt\coo-eq./y]. However, given the limited domestic renewable potential compared to the end-use demands, Belgium has to import energy-intensive renewable fuels to decrease GHG emissions. The uncertainty in the required operational energy for renewable fuels drives up the variance of the EROI. This result is similar to the simulations with the minimization of the system cost \citep{rixhon2021role,limpens2020belgian}, where the uncertainties of the renewable fuels prices are responsible for the increase of the system cost variance.
Figure \ref{fig:pillar-transition} depicts the key pillars of the energy transition when decreasing the GHG emissions. The system sequentially uses most of the options from the energy Mix scenario presented in \citet{limpens2020belgian}, which is a scenario accounting for an increased amount of renewable resources plus nuclear capacity and geothermal energy, but with a different priority. 
The model begins with low-energy intensive fossil (NG) and domestic renewable resources (wind) when there is no limit on GHG emissions. 
Then, it first improves its energy efficiency in the early stages by reducing the primary energy consumed to meet the demand. The electrification is progressively performed with the electricity imports at their full potential (27.57 [TWh/y]) to improve the electrification of the mobility (private electric vehicles) and heating sectors (heat pumps).
Then, to enhance the electrification while reducing the overall global warming potential, the system uses local PV renewable electricity production up to its full potential of 59.2 [GWe], and wet biomass 38.9 [TWh/y] and waste 17.8 [TWh/y] for heating.
Finally, the model forces the system to import renewable fuels massively to achieve ambitious low GHG emissions targets. When maximizing the EROI, the system uses mainly renewable gas, and when minimizing the cost, it is a mix of H2, renewable gas, and renewable-liquid fuels (ammonia and methanol).
These pillars indicate the main levers to decrease GHG emissions while maximizing the EROI of the system and the research directions to decrease the uncertainties of the parameters of the related technologies and resources. Indeed, efforts should not be distributed equally to decrease the uncertainties of every model parameter. The sensitivity analysis provides the critical uncertain parameters responsible for the main contributions to the EROI variance. A similar conclusion to \citet{rixhon2021role,limpens2020belgian} is drawn: policymakers, industries, and academia should spend time and energy improving the knowledge about renewable fuels by reducing the cost, the energy invested in the operation, and the related uncertainties.

Concerning the second point, the sensitivity analysis reveals the impact of the maximum capacity of the nuclear power plants on the variance of the EROI. The contribution of this parameter reaches a maximum of 12.5\% for the GHG emissions target of 56.9 [Mt\coo-eq./y]. Then, it decreases to 3.4\% at 19.0 [Mt\coo-eq./y] and is never the main driver of the EROI variation for all GHG emissions targets considered. Therefore, nuclear electricity does not compete against local or imported renewables and can be considered complementary to renewables. In addition, it substantially improves the EROI of the system as the model always utilizes nuclear power plants at their maximal capacity.

\subsection{Limitations}

The study limitations are either caused by the model or the uncertainty characterization and quantification conducted in the EROI sensitivity analysis. 

First, we depict three main model limitations:
(1) the snapshot approach \citep{girones2015strategic} limits the concept of a trajectory between several GHG emissions targets. One way to overcome this issue is to consider a pathway \citep[Chapter 7]{limpens2021generating} that could describe the different steps continuously in terms of technologies to implement and resources to exploit;
(2) the unlimited availability of imported renewable fuels regardless of origin. 
For very ambitious GHG emissions targets, $\mathbf{GWP_\text{tot}} \leq 9.5$ [Mt\coo-eq./y], the domestic renewable energies such as solar and wind are replaced by imported renewable fuels and electric technologies by gas-based technologies. Thus, almost all the primary energy is composed of renewable fuels. This case is not realistic as it is improbable that Belgium would be able to import approximately 510 [TWh/y] of such renewable fuels.
Estimating the maximum quantities of renewable fuels that could realistically be imported could be done with different costs and energy invested in operation concerning the origin. For instance, \citet{Colla2022} propose in their study a framework to account for the different origins of biomass imports;
(3) the linear optimization approach makes the results highly sensitive to the input parameters. A slight difference in technology efficiency or energy invested in construction or operation can make the system switch between two solutions which share a similar value for the objective function, but are very different in nature. The sensitivity analysis partially addresses this issue because it relies on the relevant definition of the list of uncertain parameters with their uncertainty range. Another approach could consist in investigating the feasible space near optimality. The study of \citet{dubois2021computing} proposes a generic framework for addressing this issue. It allows looking for solutions that can accommodate different requirements, such as determining necessary conditions on the minimal energy that the system will invest in domestic renewable energies or imported renewable fuels.

Finally, we outline four main limitations concerning the uncertainty characterization and quantification:
(1) the data of the energy invested in the construction of technologies and of the required operational energy for using the resources. The methodology of the data collection relies on \citet{muyldermansmulti}, where the data have been obtained from the \textit{ecoinvent} database \citep{wernet2016ecoinvent}. However, new data and publications could help refine these values, especially for renewable fuels, around which research is booming;
(2) the global warming potential data in the operation of renewable resources, particularly renewable fuels, is assumed to be 0. New data and publications could refine these values, similar to the data for energy invested. Indeed, these investigations are required to study low carbon energy system with $\mathbf{GWP_\text{tot}} \leq 10$ [Mt\coo-eq./y] to avoid having unrealistic results where imported synthetic fuels replace all the domestic renewable;
(3) the choice of uncertain parameters. This study focuses mainly on energy invested related parameters and did not consider other parameters such as the technology efficiencies and lifetimes or the GWP of construction and operation. A complete investigation should be conducted to consolidate the results;
(4) the uncertainty ranges considered are based on \citet{moret2017characterization,rixhon2021role} for the parameters not related to the energy invested in construction and operation. However, these ranges could be updated based on the last data and publications releases. In addition, due to the significant uncertainty of the energy invested in the construction and operation of technologies and resources, we adopted an arbitrary uncertainty range of minus/plus 25\%. Further work should be dedicated to refining this range and adapting it to specific technology and resource.

\section{Conclusion} \label{sec:conclusions}

Most long-term policies to decrease the carbon footprint of our societies consider the cost of the system as the leading indicator in the energy system optimization models. However, the energy transition encompasses economic, technical, environmental, and social aspects. We consider a more comprehensive indicator to address this issue: the EROI of a whole-energy system. 

The primary outcomes of this paper are: (1) the development of a novel and open-source approach by adding the EROI in EnergyScope TD \citep{limpens2019energyscope}, a whole-energy system optimization model, and providing open access to the Python code and the database; (2) the illustration of this approach in a real-world case study: the energy transition of the 2035 Belgian energy system. However, the novel model can be applied to any energy system at the country, regional or world level; (3) the comparison of the results obtained when minimizing the cost and maximizing the EROI; (4) the global sensitivity analysis of the EROI of the system by applying a polynomial chaos expansion method \citep{sudret2014polynomial}. It provides the critical drivers of the variation of the system's EROI; (5) the estimation of the probability density functions of the EROI of the system for several GHG emissions targets.

The main results are five-fold. 
First, the EROI of the Belgian system decreases from 8.9 to 3.9 for GHG emissions targets going from 100 to 19 [MtCO2-eq./y]. These values can be put into perspective with estimated values of: (i) lower bound on the societal EROI, \textit{e.g.}, 5 \citep{hall2009minimum,brandt2017does} or 6 \citep{court2019estimation}; (ii) worldwide EROI, \textit{e.g.}, 8.5 \citep{dupont2021estimate} in 2018 and 12 \citep{CAPELLANPEREZ2019100399} in 2015.
Second, the renewable fuels - mainly imported renewable gas - represent the largest share of the system primary energy mix when GHG emissions decrease due to the lack of endogenous renewable resources such as wind and solar. 
Third, the EROI trend is similar when decreasing GHG emissions when minimizing the cost or maximizing the EROI. In both cases, the EROI values decrease from 8.9 to 3.9 (EROI maximization) \textit{vs.} 6.3 to 2.5 (cost minimization) for GHG emissions targets going from 100 to 19 [MtCO2-eq./y]. In addition, the strategy to decrease GHG emissions is similar when minimizing the cost of the system and maximizing the EROI. It consists of importing an increased share of renewable fuels to reduce GHG emissions. However, there is a difference; instead of using only large quantities of imported renewable gas in the case of EROI maximization, the renewable fuels are more diverse when minimizing the cost with renewable ammonia, methanol, and H2.
Fourth, the sensitivity analysis reveals that the energy invested in the operation of renewable gas is responsible for 67.1\% of the variation of the EROI for the GHG emissions target of 19 [Mt\coo-eq./y]. 
Finally, the estimation of the EROI probability density functions exhibits that decreasing the GHG emissions makes the EROI of the system more variable to uncertain parameters. Indeed, the coefficient of variation, which is the ratio of the standard deviation over the mean, increases with GHG emissions reduction. 

Overall, the decrease in the EROI of the Belgian system with the GHG emissions raises questions about meeting the climate targets without adverse socio-economic impact. Indeed, most countries rely massively on fossil fuels, like Belgium, and they could probably experience such an EROI decline when shifting to carbon neutrality. 
%
%
As pointed out in the introduction, several attempts to determine a lower bound of the societal EROI below which a prosperous lifestyle would not be sustainable have been performed over the past few years. The estimation of this threshold differs from one study to the others. Thus, it is difficult to conclude what does mean an EROI value of 4 for Belgium in terms of lifestyle when reaching GHG emissions targets of around 40 [MtCO2-eq./y]. However, this decreasing value of EROI should encourage the scientific community and the decision-makers to reconsider our current energy and economic models. Indeed, most economic models assume at least a constant demand based on economic growth, if not an increase. Thus, the energy sector is likely to take a more significant share in the economy at the expense of other sectors to sustain such a demand with a decreasing EROI. A decreasing EROI implies a decreasing quantity of services and raises the question: which sectors and services should be favored? A challenging way to answer this question could be to extend the concept of the "Pyramid of Energetic Needs" \citep[Fig. 12]{lambert2014energy}. It was designed to represent the minimum EROI required for conventional oil to be able to perform various tasks required for civilization. Making such a pyramid with the minimum levels of EROI of a carbon-free society to perform these services would greatly help policy-makers.
Thus, we hope this paper will encourage policymakers, industries, and academia to: (i) dedicate more research to assess whole-energy systems with the EROI indicator; (ii) spend more time and energy improving the knowledge about renewable fuels, mainly to decrease the uncertainties related to their cost, availability, and energy invested.

Future works should address the model limitations, \textit{e.g.}, drawing a continuous plan of strategies from today to the carbon neutrality of 2050 instead of the snapshot approaches. They will also focus on refining the data and reducing the uncertainties of the main drivers, \textit{i.e.}, the renewable fuels, of the variation of the EROI of the system.
In addition, this novel model could be extended to: 
(i) assess the EROI of a system composed of several inter-connected countries using EnergyScope Multi-Cell \citep{thiran2020energyscope}, such as Europe, to better take into account the domestic complementarity of renewables. Indeed, as previously stated, the developed EROI-based approach is not limited to a specific country or area; 
(ii) perform multi-criteria optimization in the vein of \citet{muyldermansmulti} with indicators such as the system cost, global warming potential, and EROI; 
(iii) include a macroeconomic model following the approach of \citet{dupont2021estimate} to estimate another indicator, introduced by \citet{FAGNART2016121}, called the net energy ratio (NER) of the economy. The NER is more comprehensive than the EROI and allows assessing the energy embodied in the intermediate and capital consumptions of the entire economy. 

Finally, the comparative study of maximizing the EROI of the system \textit{vs.} minimizing the cost indicates that the EROI of the system remains higher by at least one point when using it as an indicator to plan the system, but with a cost between 5 \% and 15 \% higher. It raises the question: is this additional cost worth it? Indeed, a decline in EROI implies that less energy is available to fuel society due to a decreased efficiency of the energy system. A lower-EROI society requires more sober ways of living and more rational use of energy. Optimizing the EROI instead of the cost makes this sobriety imperative slightly less stringent, to the extent that it still needs to be better characterized and quantified. 
In addition, it is not easy to assess how the economy of society could be when reducing the EROI significantly. The price is a human concept dependent on the economy, and a financial cost depends on the system economy and energy. However, the EROI does not depend on the economic system but only on the energy system. Two societies with the same energy system, thus the same EROI, could have two different economic systems. With the energy available in excess, society makes choices to use this energy to provide some services and goods. With a decreasing EROI, the field of economic possibilities may also decrease. 
Thus, the EROI should be taken into account by policymakers when planning the transition as a metric for energy sobriety. 
The trade-off between EROI decrease, \textit{i.e}, more sobriety, and economic cost would require further investigation by researchers from several research fields, including social sciences. 

\backmatter

\bmhead{Acknowledgments}

The authors would like to acknowledge the authors and contributors of the EnergyScope TD model and the RHEIA Python library. In addition, we acknowledge professor Herv\'e Jeanmart, President of the Institute of Mechanics, Materials, and Civil engineering (IMMC) at the Catholic University of Louvain. He has facilitated the collaboration of the teams and provided helpful insights and discussions about the present work.
In addition, the authors would like to thank the editor and the reviewers for the comments that helped improve the paper.

\section*{Declarations}

\bmhead{Conflict of interest} There is no conflict of interest to be reported.

\bmhead{Availability of code and data}

Code repository and data, and the latest documentation are available on Github: \url{https://github.com/energyscope/EnergyScope}, \url{https://energyscope-td.readthedocs.io/en/master/}.

\bmhead{Authors' contributions}

The first draft of the manuscript was written by Jonathan Dumas and all authors commented on previous versions of the manuscript. All authors helped improve the paper quality based on the reviewer's comments and approved the final manuscript.
Jonathan Dumas did the validation of the results and conducted the formal analysis.
Jonathan Dumas, Antoine Dubois, Paolo Thiran, Pierre Jacques, Francesco Contino, and Gauthier Limpens did the conceptualization of the approach and the methodology.
Jonathan Dumas, Antoine Dubois, and Gauthier Limpens did the data curation and the investigation.
Jonathan Dumas, Antoine Dubois, Paolo Thiran, and Gauthier Limpens worked on the software.
Finally, Jonathan Dumas and Gauthier Limpens did the supervision and project administration.




\begin{appendices}

\section*{Acronyms}
\begin{supertabular}{l p{0.65\columnwidth}}
	Name & Description \\
	\hline
	Ammonia-RE   & Synthetic renewable ammonia. \\
	CCGT  & Combined cycle gas turbine. \\
	CHP   & Combined heat and power. \\
	DEC heat LT   & Decentralized heat low temperature. \\ 
	DHN   & District heating networks. \\ 
	DHN heat LT  & Centralized heat low temperature, through the DHN network. \\ 
	EROI  & Energy return on investment. \\
	ES-TD & EnergyScope Typical Days. \\
	EUD   & End-use demand. \\
	FEC   & Final energy consumption. \\
	FC    & Fuel cell. \\
	Gas-RE & Synthetic renewable gas. \\
	GEO   & Geothermal. \\
	GHG   & Greenhouse gas. \\
	GWP   & Global warming potential. \\
	H2-RE   & Synthetic renewable H2. \\
	HP    & Heat pump. \\
	HVC   & High value chemicals. \\
	IPCC  & Intergovernmental Panel on Climate Change. \\
	LCA   & Life cycle assessment. \\
	LFO   & Light oil fuel. \\
	LP    & Linear programming. \\
	Methanol-RE   & Synthetic renewable methanol. \\
	NG    & Natural gas. \\
	PCE   & Polynomial chaos expansion. \\
	pdf   & Probability density function. \\
	PHS   & Pumped hydro storage. \\
	RE    & Renewable. \\
	RES   & Resources. \\
	TECH  & Technologies. \\
	PV    & Photovoltaic. \\

\end{supertabular}

\subsection*{Variables ($\mathbf{\text{bold}}$) and parameters}
\noindent The snapshot approach implicitly considers variables and parameters over a year. For instance, $\mathbf{E_\text{in,tot}}$ is the annual system energy invested expressed in [GWh/y]. However, for the sake of clarity, we omit the year unit. The variables are in $\textbf{bold}$ with the [unit] specified in brackets $[\quad]$. \\

\begin{supertabular}{l p{0.43\columnwidth}}
	Name [unit] & Description \\
	\hline
$\mathbf{E_\text{in,tot}}$ [GWh]   & System energy invested. \\
	$\mathbf{E_\text{constr}}(j)$ [GWh]  & System energy invested in construction of technology $j$. \\
	$\mathbf{E_\text{op}}(j)$ [GWh]  & System energy invested in operation of resource $i$. \\
	$\mathbf{F}(j)$ [GW or GWh]   & Installed capacity of technology $j$. \\
	$\mathbf{F_t}(i,h,td)$ [GWh]   & Quantity of resource $i$ used at hour $h$ of typical day $td$. \\
	$\mathbf{GWP_\text{tot}}$ [Mt\coo-eq.]   & System GHG emissions. \\
	$\mathbf{GWP_\text{constr}}(j)$ [Mt\coo-eq.]   & GHG emissions of construction of technology $j$. \\
	$\mathbf{GWP_\text{op}}(i)$ [Mt\coo-eq.]   & GHG emissions of resource $i$. \\
	$e_\text{constr}(j)$ [GWh/GW]   & Energy invested in construction of technology $j$. \\
	$e_\text{op}(i)$ [GWh/$\text{GWh}_\text{fuel}$]   & Energy invested in operation of resource $i$. \\
	${gwp}_\text{constr}(j)$ [kt\coo-eq.]   & GHG emissions for the construction of technology $j$. \\
	${gwp}_\text{op}(i)$ [kt\coo-eq.]   & GHG emissions for the operation of resource $i$. \\
	$\text{gwp}_\text{limit}$ [Mt\coo-eq.]   & GHG emissions target. \\
	$t_\text{op}(h,td)$ [hour]   & Time period duration of hour $h$ of typical day $td$. \\
	$\text{lifetime}(j)$ [years]   & Lifetime of technology $j$. \\
\end{supertabular}

\subsection*{Sets and indices}
\begin{supertabular}{l p{0.7\columnwidth}}
	Name & Description \\
	\hline
	$j$ & Technology index. \\
	$i$ & Resource index. \\
	$h$ & Hour index. \\
	$td$ & Typical day index. \\
	$eud$ & End-use demand index. \\
    $T\_H\_TD(t)$ & Hour $h$ and typical day $td$ associated to the time period $t$.\\
	$TECH$ & Set of technologies. \\
	$RES$ & Set of resources. \\
	$\mathcal{T}$ & Set of all periods of the year. \\
\end{supertabular}

\section{Table \ref{tab:contributions} justifications}\label{appendix:intro}

The EU reference scenario 2020 \citep{EU2020} uses the Price-Induced Market Equilibrium System (PRIMES) \citep{primeseu} model, which is not in open-access, and the data used as input are not available. It is an update of the previous version published in 2016 \citep{eu2016}. The model is multi-sectors as the ``Reference Scenario" projects the impact of macro-economic, fuel price, and technology trends and policies on the evolution of the EU energy system, transport, and GHG emissions. However, this study considers only one scenario, \textit{i.e.}, the ``Reference scenario". In addition, this study does not consider the EROI, and there is no sensitivity analysis.

In the study \citep{devogelaer2021bon}, the Federal Planning Bureau discusses what role offshore wind can play in helping Belgium achieve climate neutrality by the middle of the century. The analysis is multi-sectors by considering the electricity, H2, and gas sectors. The model used is Artelys Crystal Super Grid \citep{csgartelys}, which is not in open-access, and the data used for the study are not available. In particular, this report examines the development of joint hybrid offshore wind projects that both provide renewable energy capacity and can serve as interconnectors linking different countries. Two different scenarios are defined and studied; thus, this study is considered partially multi-scenarios. However, this study does not consider the EROI, and there is no sensitivity analysis.

The study \citep{meinke2017energy} uses the TIMES/MARKAL model, a reference in scenario analysis. This model is open access \citep{fishbone1981markal}, but the different versions for each country are not open. This study has adapted the model to the Belgian case and is unavailable. The main assumptions are detailed in the report with some input parameters. However, there is no proper access to all the input data. The TIMES Belgium model includes different technology portfolios for different supply and demand sectors of the energy system and is consequently multi-sectors. The model generates a set of five scenarios where assumptions on three parameters, namely the import capacity for electricity, the fossil fuel prices, and the phase-out of nuclear energy, are being altered. Finally, the scenario analysis with the TIMES Belgium model is based on a system cost optimization approach; thus, it does not consider the EROI.

The study \citet{elia2017} analyzes both short-term and long-term policy options on the future energy mix for Belgium on the path towards 2050. It proposes the ``base case scenario", ``decentral scenario", and ``large-scale RES scenario". On top of these scenarios, different sensitivities are assessed at the 2030 and 2040 time horizons, resulting in additional scenarios. The assumptions of each scenario are detailed, but the input data are not available, and there is no sensitivity analysis. The report focuses on the electricity sector with renewables (PV, onshore and offshore wind, biomass, hydro, and geothermal) and thermal (CCGT, nuclear, and CHP) generation plants, electric demand (heat pumps, electric vehicles), and considers interconnections with neighboring countries. However, it is not multi-sectors as it does not model the non-electric demand of the transportation, heating, and non-energy sectors. The electricity market simulator developed by RTE, Antares \citep{doquet2008new,antares} is used to perform the electricity market and adequacy simulations. Antares is open-source and calculates the most-economic unit commitment and generation dispatch. Finally, the scenario analysis with the Antares model is based on a system cost optimization approach; thus, it does not consider the EROI.

The objective of the Energy Pathways to 2050 report \citep{rte2021} is to construct and evaluate several possible options for the evolution of the French power system (generation, network, and consumption) to achieve carbon neutrality. To this end, several scenarios are proposed based on different assumptions, from 100\% renewable generation technologies to a mix of renewable and nuclear capacities. Each scenario is detailed with the assumptions in the report, and the dataset used to conduct the study is open-access \citep{rte2021_data}. The open-source power system model, Antares \citep{doquet2008new,antares}, describes the production capacities, the network, and the sources of consumption in all European countries, to simulate the production, consumption, and exchanges per country at hourly intervals in all the countries of the European Union. The study does not conduct a global sensitivity analysis of the input parameters. However, it performs several variations of some key parameters to assess the variation in the cost of the system. Finally, the Antares model uses a cost optimization approach.

The studies \citep{limpens2020belgian,rixhon2021role} analyze the Belgian energy system in 2035 and 2050, respectively, for different GHG emissions targets using the multi-sectors open-source model EnergyScope TD \citep{limpens2019energyscope}. They optimize the design of the overall system to minimize its costs and emissions. The input data of the model are open-access on the Github repository. A sensitivity analysis is conducted in \citet{rixhon2021role} by implementing the PCE approach to emphasize the influence of the parameters on the total cost of the system. They point out Belgium's lack of endogenous renewable resources to achieve ambitious GHG emissions targets. Thus, additional potentials shall be obtained by importing renewable fuels, electricity or deploying geothermal energy.

\section{FEC calculation}\label{appendix:section-3}

This appendix provides the details to derive the FEC from the simulation results to be used to calculate the EROI of the system following Eq. (\ref{eq:eroi-final}).

The set of end-use demands EUD comprises: (1) electricity; (2) heat: heat high temperature, decentralized heat low-temperature, and centralized heat low-temperature; (3) non-energy: ammonia, HVC, and methanol; (4) mobility: freight boat, freight rail, freight road, passenger private, and public.
The end-use demand is expressed in [Mpkm/y] (millions of passenger-km) for passenger mobility, [Mtkm/y] (millions of ton-km) for freight mobility, [GWh/y] for heating, [GWh/y] for non-energy, and [GWhe/y] for electricity end-uses. 

The system FEC is the sum of the final energy consumption related to each end-use demand (eud):
\begin{align}
    \text{FEC} &= \sum_{eud \in \text{EUD}} \text{FEC}(eud).
\end{align}
Then, for a given end-use demand (eud) the energy balance is
\begin{align}
    eud + \sum_{i \in I} c_i(eud) &= \sum_{j \in J} p_j(eud),
\label{eq:fec-balance}
\end{align}
with $c_i(eud)$ the consumption of this end-use demand by technology $i$, and $p_j(eud)$ the production of this end-use demand by technology $j$. For instance, several technologies can produce heat at high temperatures, such as gas boilers or CHP. Furthermore, some technologies use heat at high temperatures as input material, such as technology to produce HVC.

First, let us consider the case where no technology uses this end-use demand as input material: $I = \emptyset$. Then, from Eq. (\ref{eq:fec-balance}) the FEC related to this end-use demand is
\begin{align}
    \text{FEC}(eud) &= \sum_{j \in J}\text{FEC}_j(eud).
\end{align}
If $j$ is a technology, it produces $p_j(eud)$ and possibly other outputs, such as electricity or hydrogen, by consuming gas, electricity, or biomass. Then, $\text{FEC}_j(eud)$ is defined as follows 
\begin{align}
    \text{FEC}_j(eud) &= \frac{p_j(eud)}{ p_j(eud) + \sum \text{outputs}^j }  \sum \text{inputs}^j.
\label{eq:fec-calculation-technology}
\end{align}
For instance, in the model, when considering the heat high-temperature end-use demand, the technology gas CHP industry consumes 2.1739 GWh of gas to produce 1 GWh of heat high-temperature and 0.9565 GWh of electricity. In this case, the FEC of gas of this technology to produce 1 GWh of heat at high-temperature is
\begin{align}
    \text{FEC} &= \frac{1}{ 1 + 0.9565 } 2.1739 \approx 1.111.
\end{align}
If $j$ is a resource such as methanol or ammonia, then
\begin{align}
    \text{FEC}_j(eud) &= p_j(eud).
\label{eq:fec-calculation-resource}
\end{align}
For instance, the methanol end-use demand can be partially satisfied with imports.

Let us consider the case where at least one technology uses this end-use demand as input material: $I \ne \emptyset$. Then, the consumptions $c_i(eud)$ are taken into account as follows to estimate the FEC correctly
\begin{align}
    \tilde{p}_j(eud) &= p_j(eud) - \sum_{i \in I} c_i(eud) \frac{p_j(eud)}{\sum_{j \in J} p_j(eud)}.
\label{eq:fec-calculation-correction-factor}
\end{align}
Finally, the different $\text{FEC}_j(eud)$ are estimated as previously described by replacing $p_j(eud)$ in Eq. (\ref{eq:fec-calculation-technology}) and Eq. (\ref{eq:fec-calculation-resource}) by $\tilde{p}_j(eud)$ defined in Eq. (\ref{eq:fec-calculation-correction-factor}).

\section{Reference scenario additional results}\label{appendix:section-4}

This appendix presents the results of the simulation with the reference scenario (see Section \ref{sec:eroi-evolution}) in terms of installed capacities for several GHG emissions targets, energy invested, and FEC. 

\subsection{Assets installed capacity evolution}

\begin{figure}[tb]
\centering
	\begin{subfigure}{0.25\textwidth}
		\centering
		\includegraphics[width=\linewidth]{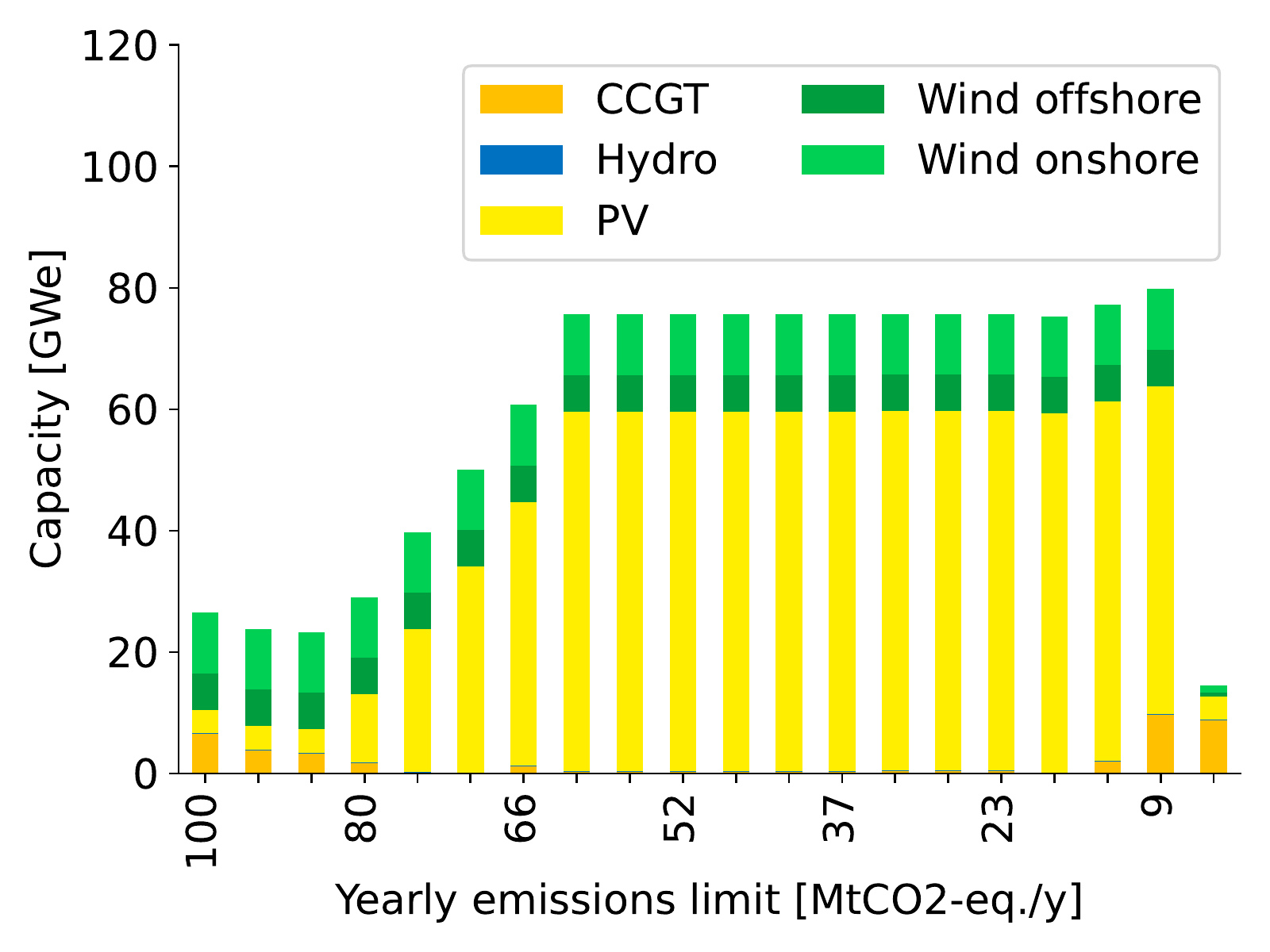}
    \caption{Electricity production.}
	\end{subfigure}%
	\begin{subfigure}{0.25\textwidth}
		\centering
		\includegraphics[width=\linewidth]{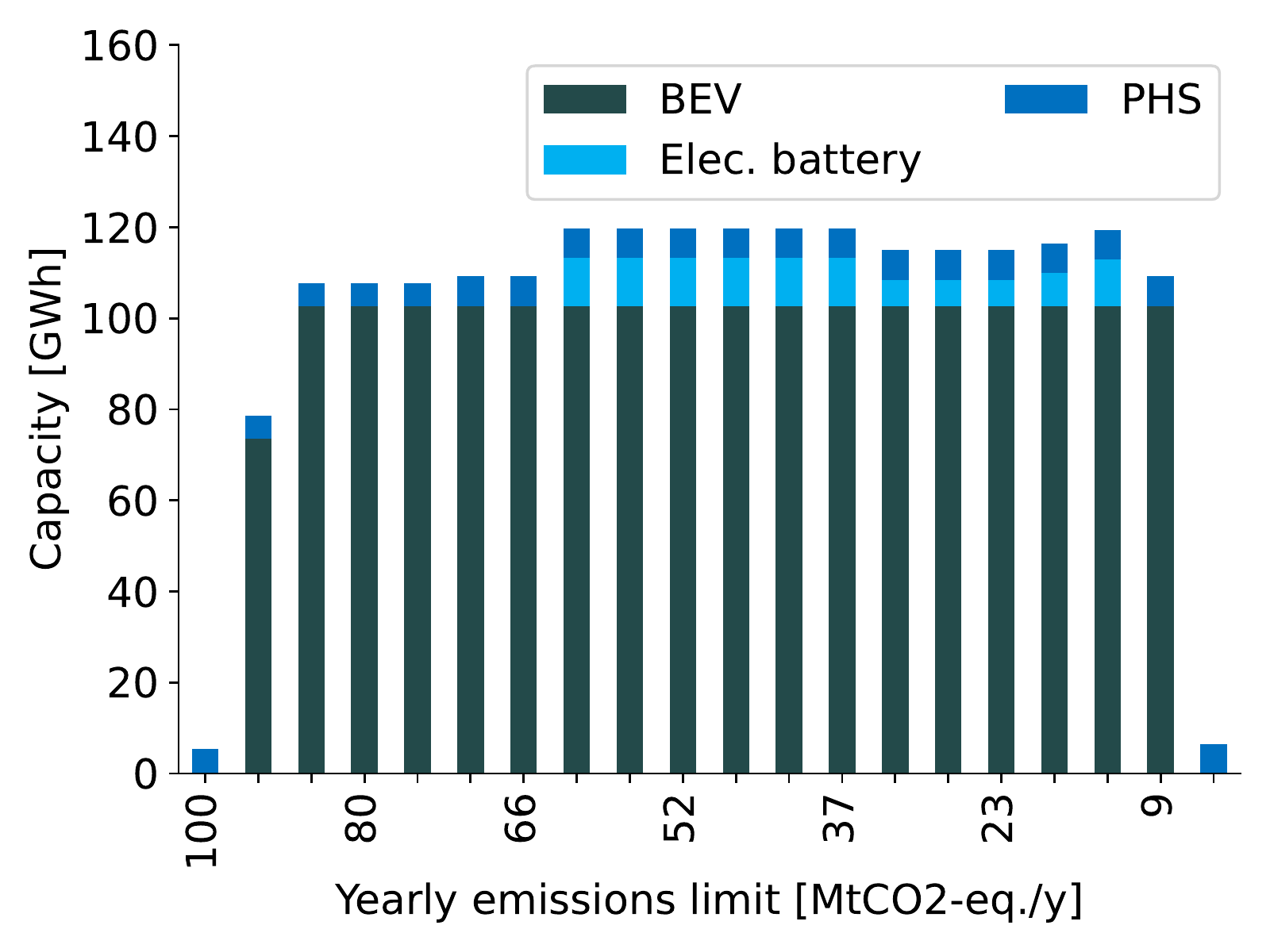}
    \caption{Electric storage.}
	\end{subfigure}
	\begin{subfigure}{0.25\textwidth}
		\centering
		\includegraphics[width=\linewidth]{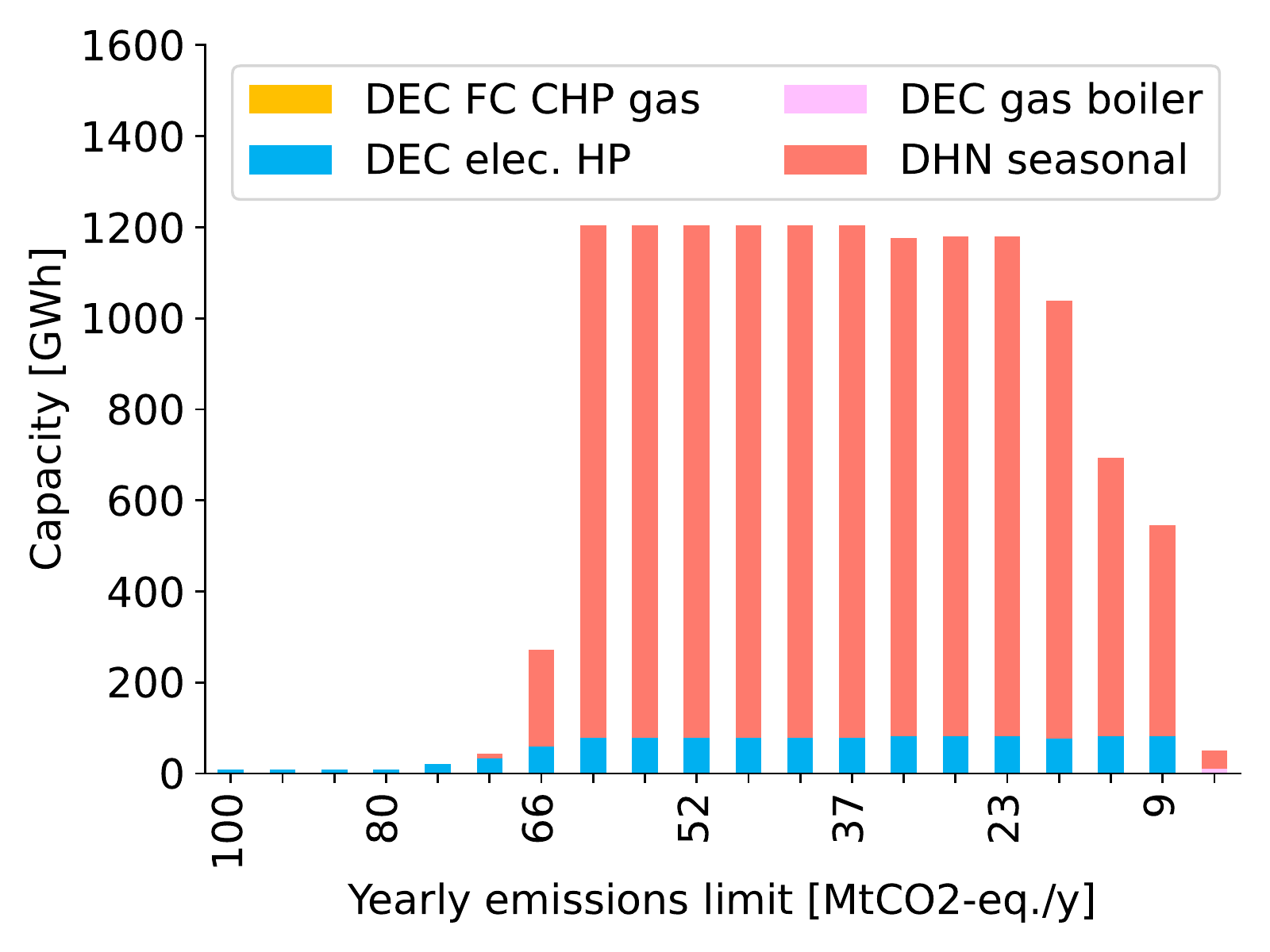}
    \caption{Thermal storage.}
	\end{subfigure}%
		\begin{subfigure}{0.25\textwidth}
		\centering
		\includegraphics[width=\linewidth]{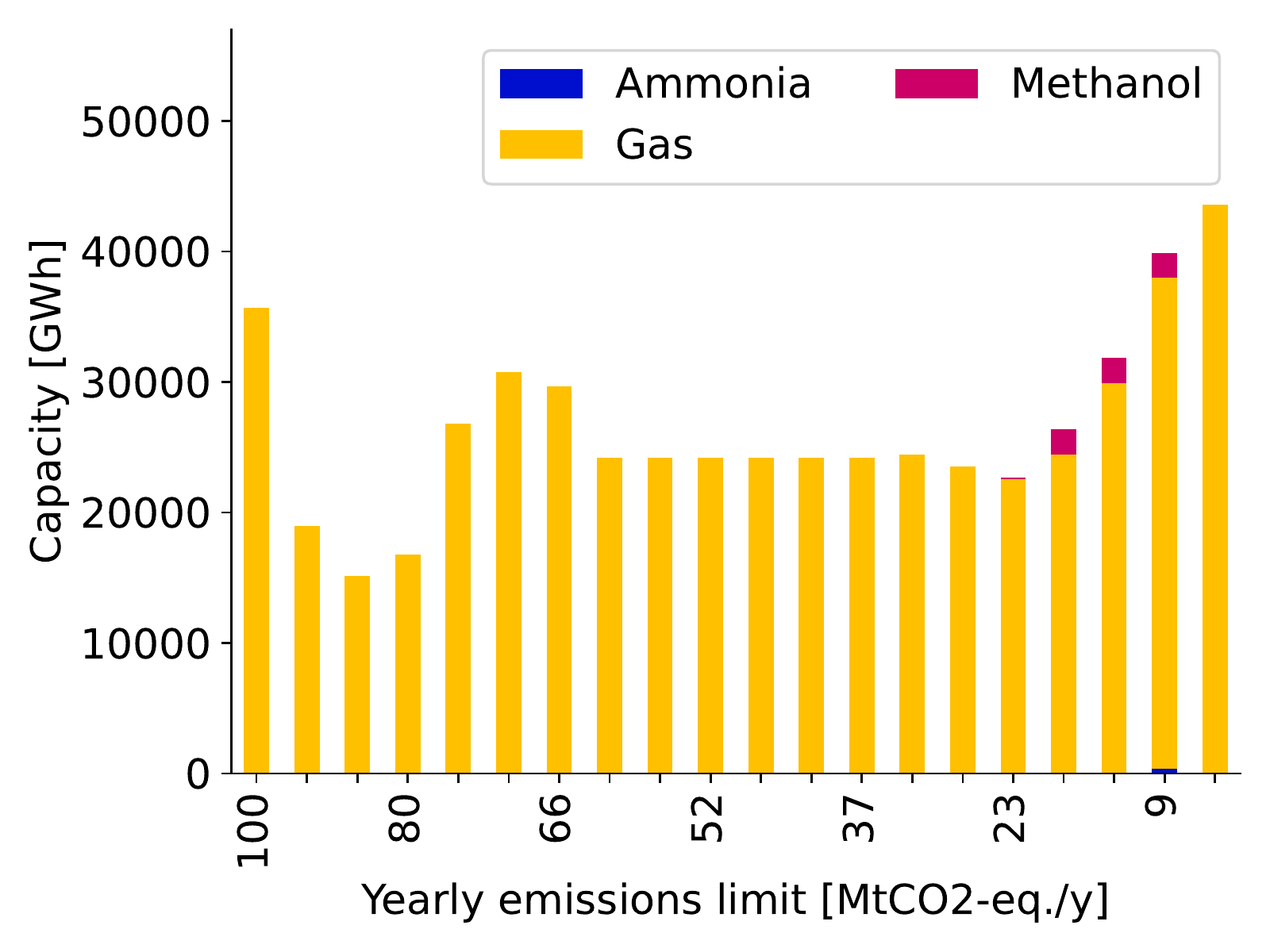}
    \caption{Other storage.}
	\end{subfigure}	
	\begin{subfigure}{0.25\textwidth}
		\centering
		\includegraphics[width=\linewidth]{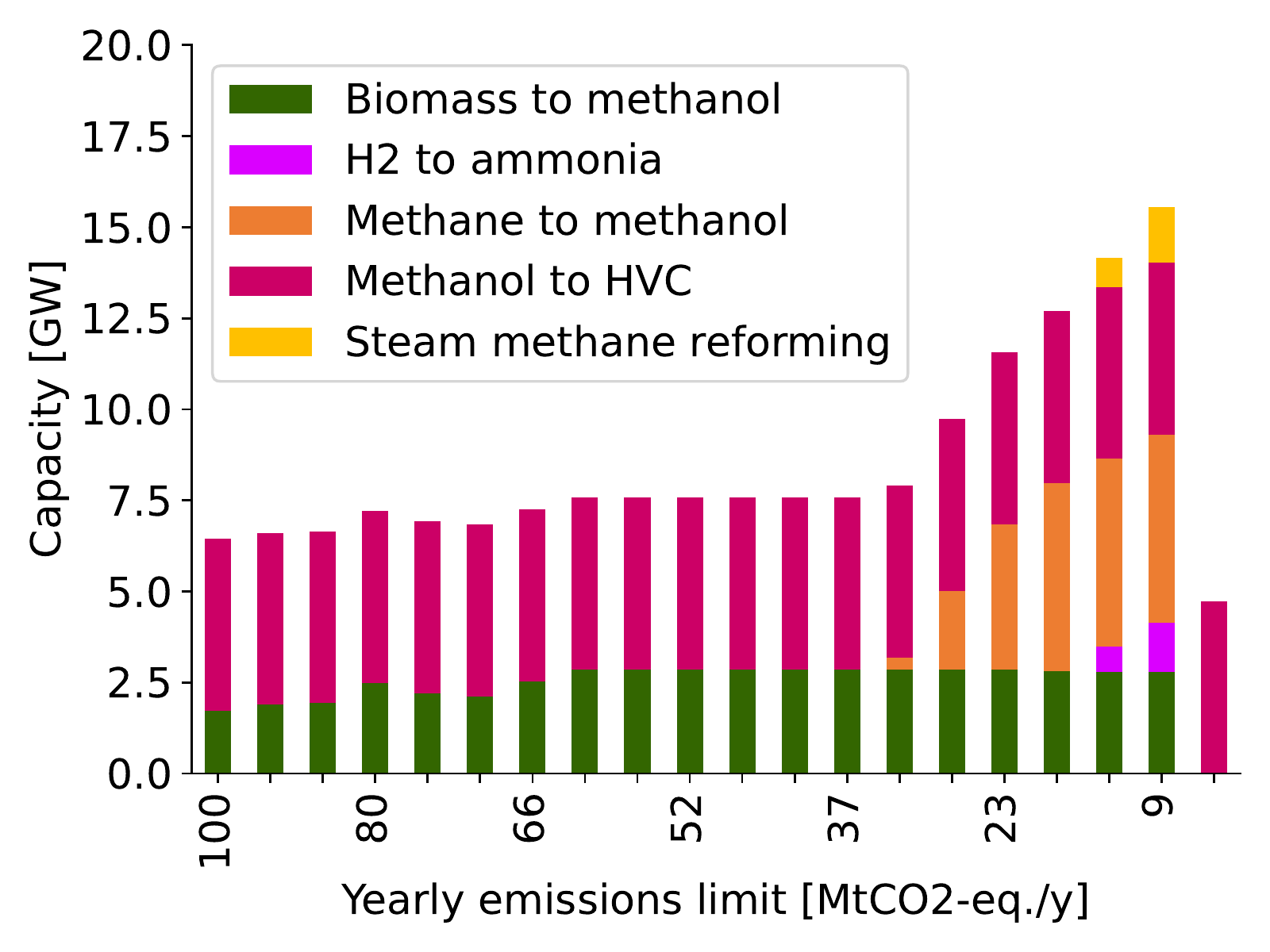}
    \caption{Renewable fuels.}
	\end{subfigure}
\caption{Evolution of the electricity production, storage (electric, thermal, gas, ammonia, and methanol), and renewable fuels asset installed capacities breakdown by technology for several scenarios of GHG emissions in 2035.
Abbreviations: combined cycle gas turbine (CCGT), photovoltaic (PV), electric (elec.), battery of electric vehicle (BEV), pumped hydro storage (PHS), decentralized electrical heat pump (DEC elec. HP), Decentralised fuel cell cogeneration gas (DEC FC CHP gas), Decentralised boiler gas (DEC gas boiler), centralized seasonal (DHN seasonal), high value chemicals (HVC).
}
\label{fig:GWP-comparison-asset-1}
\end{figure}

Figure \ref{fig:GWP-comparison-asset-1} depicts the installed capacities of electricity production, storage (electric, thermal, gas, ammonia, and methanol), and renewable fuels technologies.
The PV technology drives the evolution of electricity production assets by replacing the CCGTs and reaching the maximal available capacity of 59.2 [GWe]. Notice that the onshore and offshore wind technologies are already at their maximal capacities in the ``reference scenario-100\%". When the GHG emissions are below 9.5 [Mt\coo-eq./y], the PV and wind capacities are replaced mainly by CCGT, which uses renewable gas (gas-RE). 
The electric storage is composed of daily storage: pumped hydro storage\footnote{In Belgium, it is mainly the Coo-Trois-Ponts hydroelectric power station.} (PHS) and batteries of electric vehicle (BEV). The PHS is already at its maximal available capacity in the ``reference scenario-100\%", and the batteries of electric vehicles are used when GHG emissions are below 66.4 [Mt\coo-eq./y] to cope with uncertainty related to the increasing share of PV and wind power in the primary energy mix.
With the shift from thermal to electric cars, batteries (BEV) can interact with the electricity layer (vehicle-to-grid) when GHG emissions decrease. They provide additional flexibility to cope with a primary energy mix that relies on an increasing share of intermittent renewable energy as the GHG emissions decrease. Then, when the GHG emissions are below 9.5 [Mt\coo-eq./y], electric cars are replaced by thermal cars, which use renewable fuels (gas-RE). 
When the GHG emissions target is below 66.4 [Mt\coo-eq./y], the PV installed capacity is maximal (59.2 [GWe]), and the system relies on a high share of intermittent renewable energy: solar and wind to produce electricity. The centralized and decentralized low heat temperature demands are mainly satisfied by heat pumps. Therefore, the system uses an increased capacity of seasonal centralized thermal and daily decentralized thermal storage technologies to cope with seasonal and intraday intermittent electricity production. When the GHG emissions target is below 9.5 [Mt\coo-eq./y], the primary energy mix comprises renewable fuels, including renewable gas. Thus, the gas boiler technology mainly satisfies the centralized and decentralized low heat temperature demands, and the capacities of centralized and daily decentralized thermal storage technologies are close to 0 [GWh].
The seasonal gas storage is filled with NG in the ``reference scenario-100\%". Its capacity decreases when GHG emissions are $\approx$ 85.4 [Mt\coo-eq./y] with the electrification of the private mobility. Then, its capacity increases when GHG emissions are $\approx$ 71.1 [Mt\coo-eq./y] and reaches a constant value with GHG emissions from 61.7 to 33.2 [Mt\coo-eq./y]. Finally, its capacity increases progressively when GHG emissions decrease below 33.2 [Mt\coo-eq./y] to satisfy the seasonality of the heating, mobility, and electricity demands that rely heavily on renewable gas.
The HVC end-use demand, which amounts to most non-energy demand, is satisfied with technology that converts methanol into HVC. The methanol is imported and synthesized from biomass when the GHG emissions are between 100.3 - 33.2 [Mt\coo-eq./y]. When the GHG emissions are between 33.2 and 9.5 [Mt\coo-eq./y], the methanol imports are replaced by technologies to synthesize methanol from imported renewable gas. Finally, when the GHG emissions are below 9.5 [Mt\coo-eq./y], there are only renewable methanol imports. 

\begin{figure}[tb]
\centering
	\begin{subfigure}{0.25\textwidth}
		\centering
		\includegraphics[width=\linewidth]{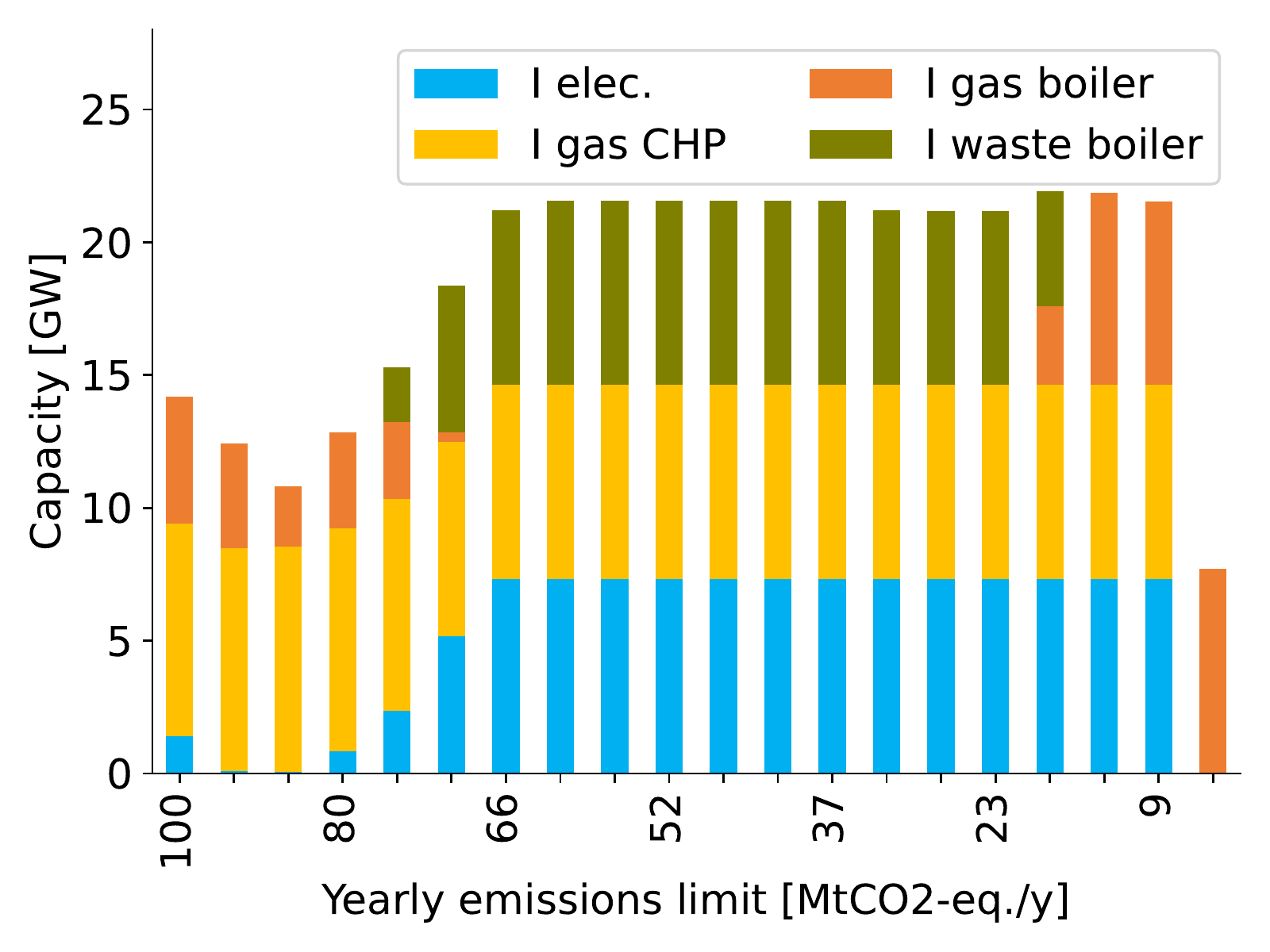}
    \caption{Heat high T.}
	\end{subfigure}%
		\begin{subfigure}{0.25\textwidth}
		\centering
		\includegraphics[width=\linewidth]{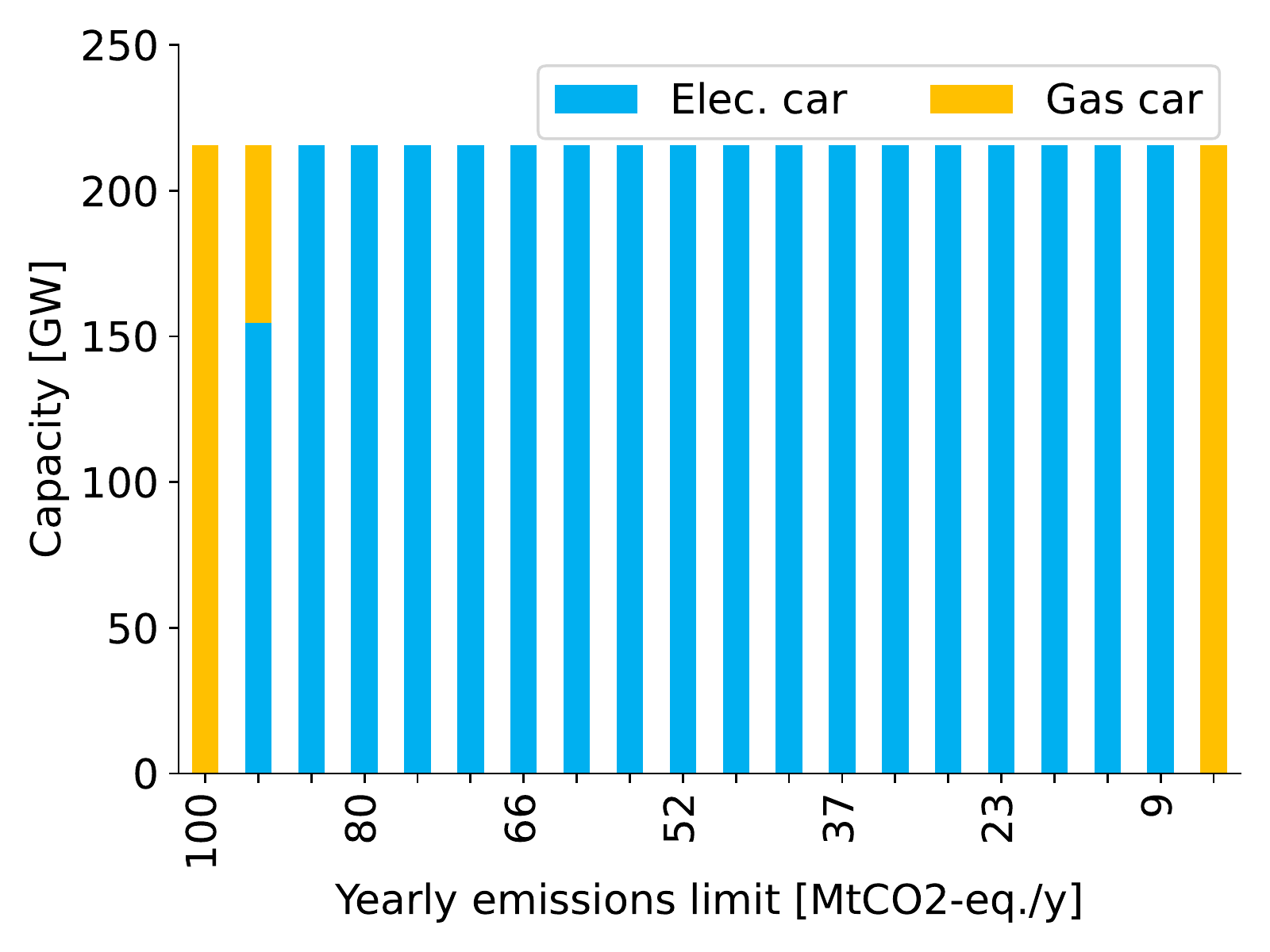}
    \caption{Private mobility.}
	\end{subfigure}
	\begin{subfigure}{0.25\textwidth}
		\centering
		\includegraphics[width=\linewidth]{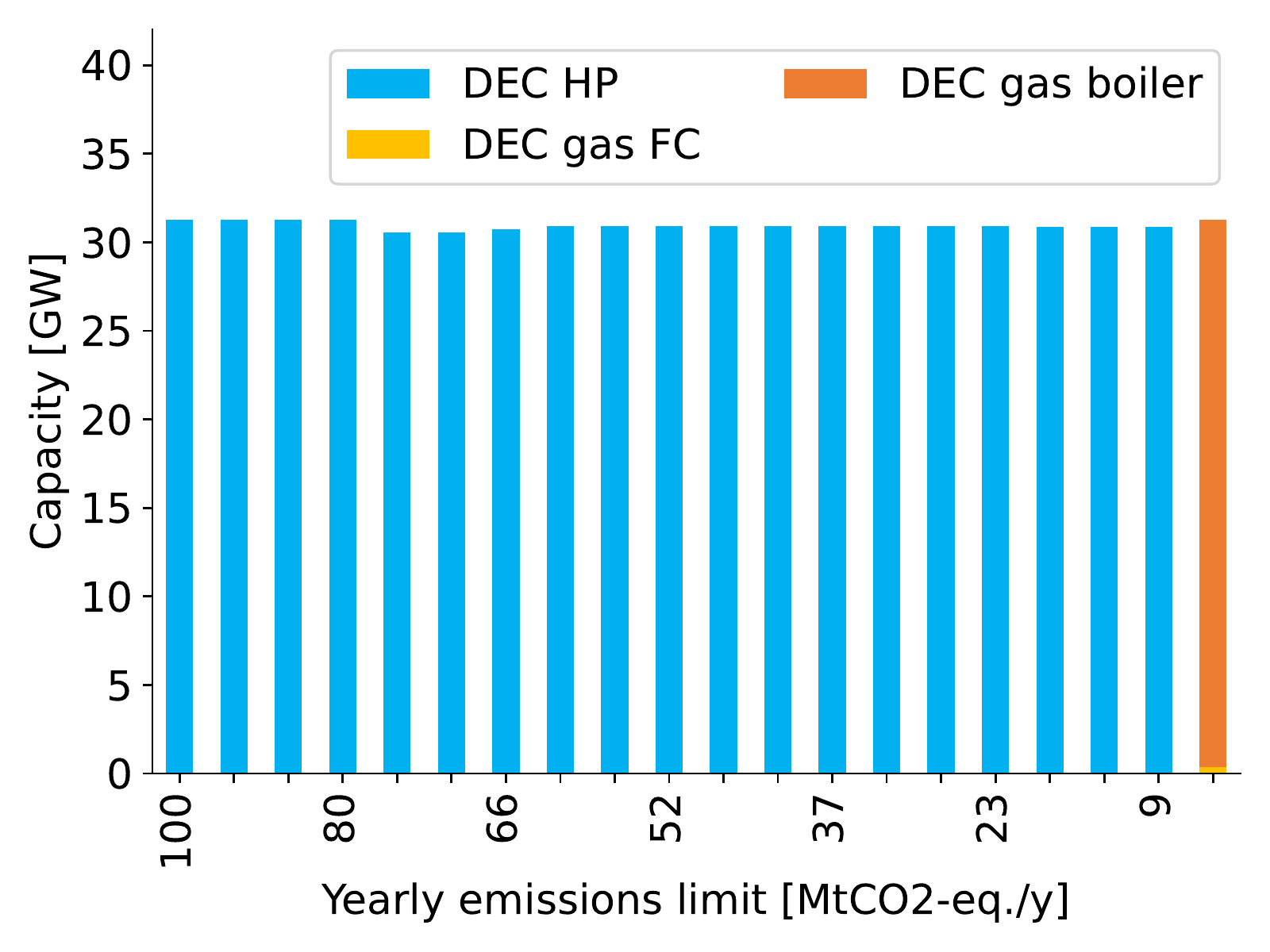}
    \caption{Heat low T DEC.}
	\end{subfigure}%
		\begin{subfigure}{0.25\textwidth}
		\centering
		\includegraphics[width=\linewidth]{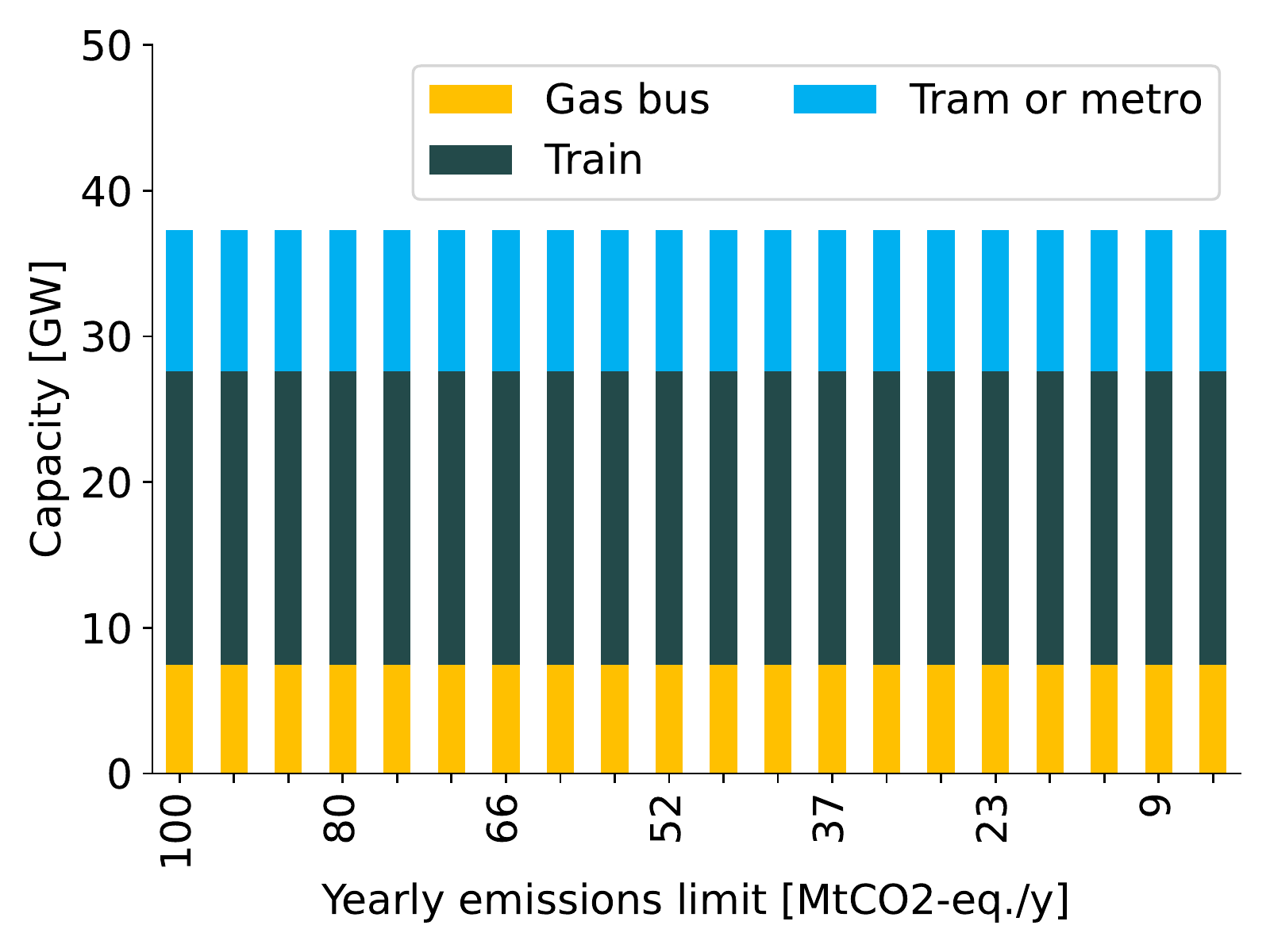}
    \caption{Public mobility.}
	\end{subfigure}
	\begin{subfigure}{0.25\textwidth}
		\centering
		\includegraphics[width=\linewidth]{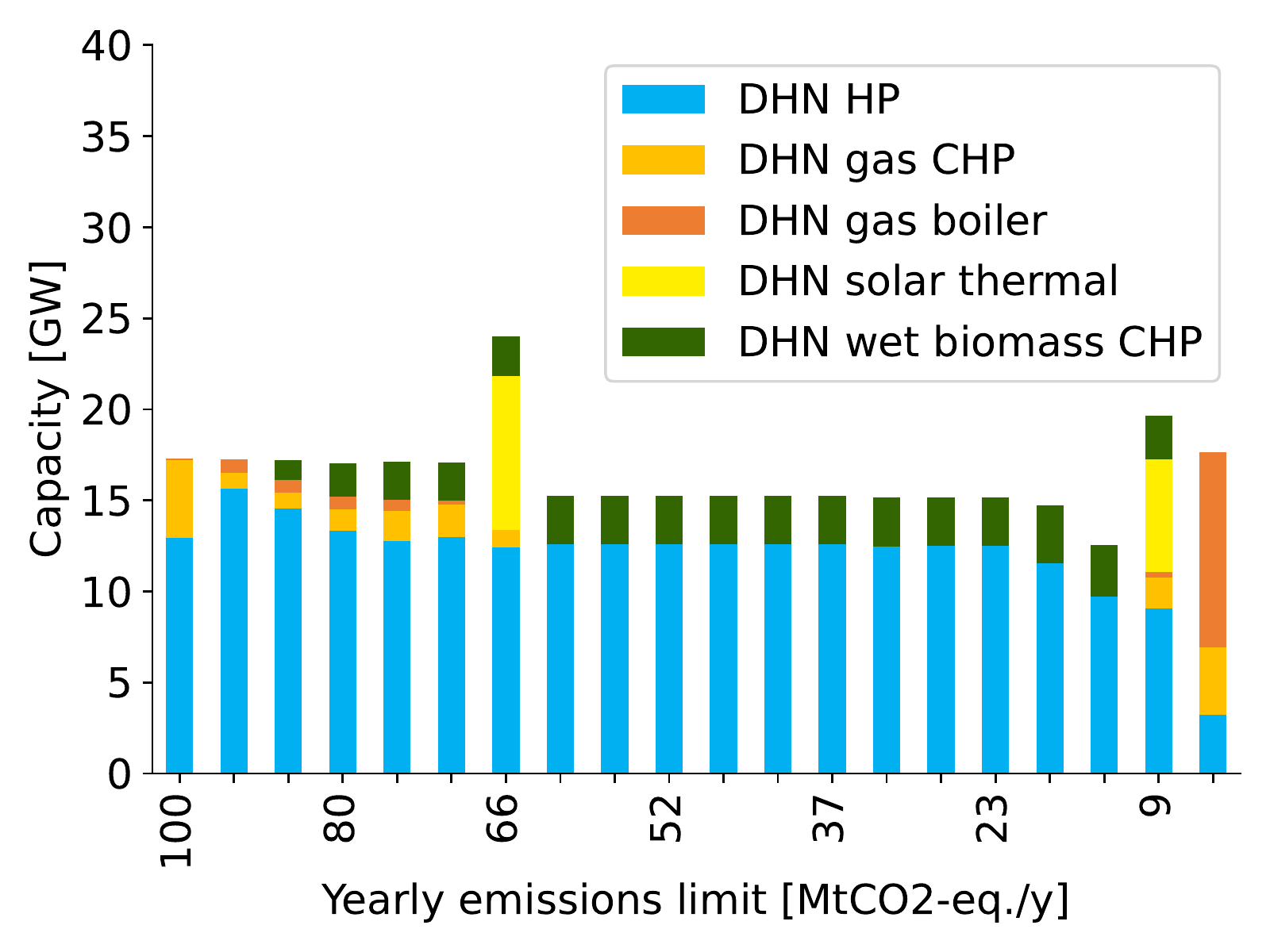}
    \caption{Heat low T DHN.}
	\end{subfigure}%
	\begin{subfigure}{0.25\textwidth}
		\centering
		\includegraphics[width=\linewidth]{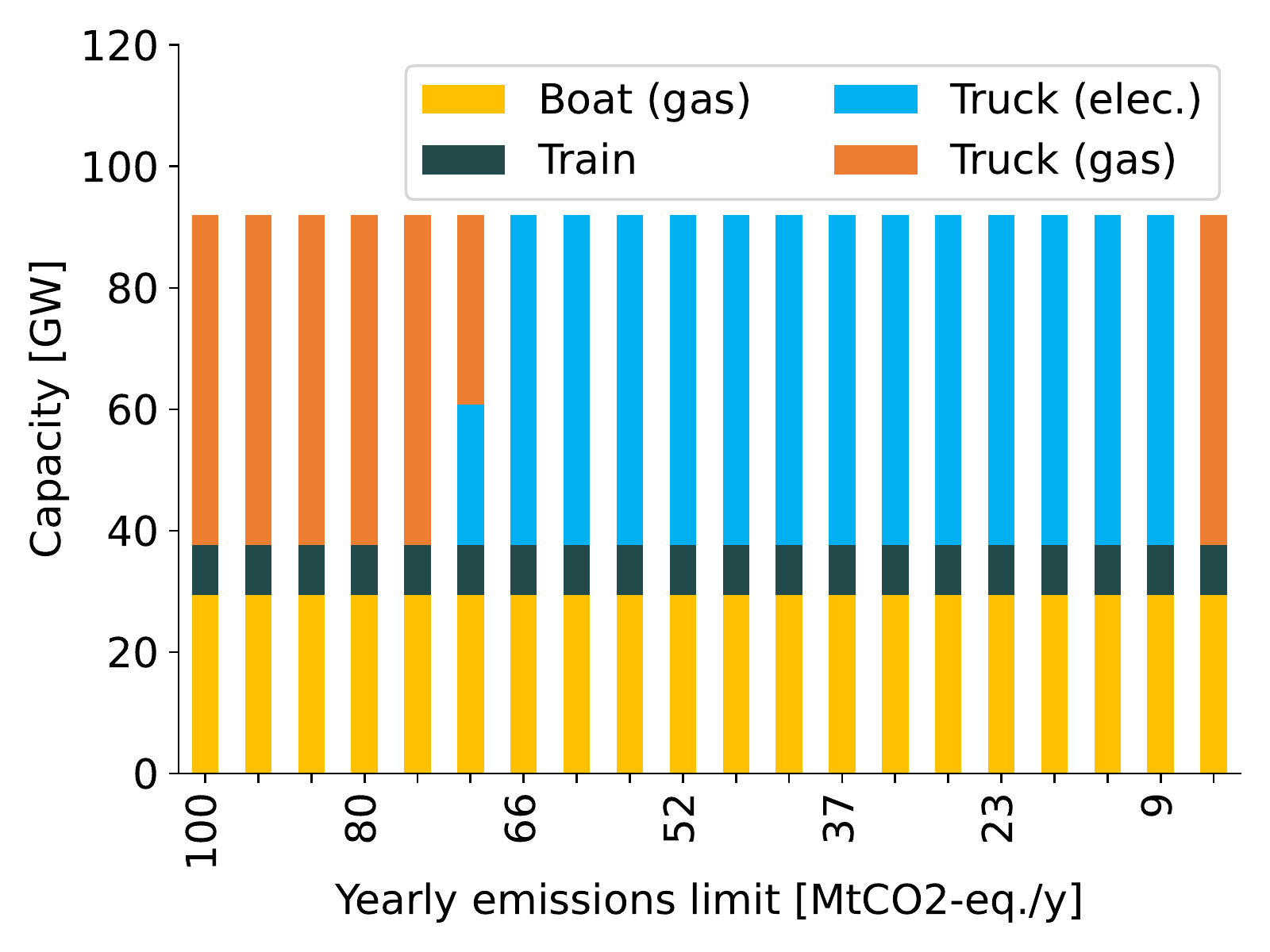}
    \caption{Freight mobility.}
	\end{subfigure}
\caption{Evolution of the heat and mobility asset installed capacities breakdown by technology for several scenarios of GHG emissions in 2035.
Abbreviations: industry (I), combined heat and power (CHP), centralized heat low temperature (Heat low T DHN), decentralized heat low temperature (Heat low T DEC), electric (elec.), heat pump (HP), fuel cell (FC).
}
\label{fig:GWP-comparison-asset-2}
\end{figure}

Figure \ref{fig:GWP-comparison-asset-2} depicts the installed capacities of heating and mobility technologies.
The industrial gas boilers are replaced by waste boilers and electrical resistors (I elec.) when GHG emissions reach 66.4 [Mt\coo-eq./y]. It corresponds to the shift from a primary energy mix composed mainly of NG to less intensive carbon energies, such as solar, wind, and waste, to satisfy the heat high-temperature end-use demand. When the GHG emissions are below 14.2 [Mt\coo-eq./y], the waste boilers are replaced by gas boilers that use renewable fuels, including imported renewable gas. Finally, when the GHG emissions are below 9.5 [Mt\coo-eq./y], the gas boiler technology is exclusively used with imported renewable gas.
The decentralized low heat temperature end-use demand is always satisfied with heat pumps, except when GHG emissions decrease below 9.5 [Mt\coo-eq./y]. In this case, decentralized gas boilers are used with imported renewable gas.
Overall, the centralized heat low-temperature demand is mainly satisfied with centralized electricity heat pumps when GHG emissions are $>$ 14.2 [Mt\coo-eq./y]. The decrease of centralized gas CHP is first balanced by centralized electricity heat pumps and then by centralized biomass CHP. When GHG emissions target is below 9.5 [Mt\coo-eq./y], centralized gas CHP, and gas boiler technologies use imported renewable gas to satisfy the demand.
Private mobility relies on electric cars from GHG emissions of 85.4 to 9.5 [Mt\coo-eq./y] and gas cars using imported renewable gas when GHG emissions decrease below 9.5 [Mt\coo-eq./y]. The trend is similar for freight trucks. The freight boat first uses NG and then renewable gas when GHG emissions decrease. The freight trains rely only on electricity in EnergyScope TD; thus, there is no technology change for this mobility type.

\subsection{Energy invested evolution}

\begin{figure}[tb]
\centering
	\begin{subfigure}{0.25\textwidth}
		\centering
		\includegraphics[width=\linewidth]{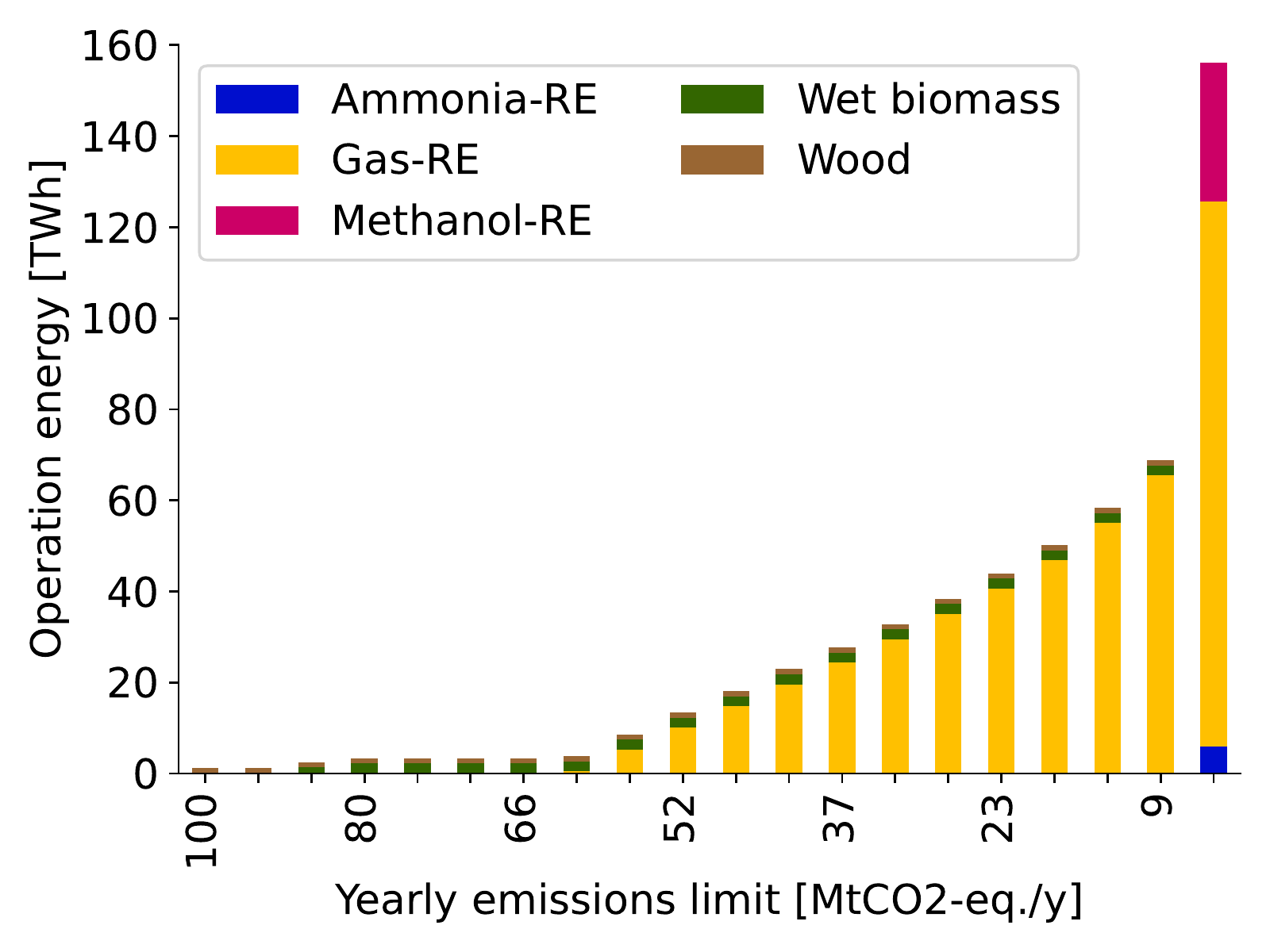}
    \caption{RE resources.}
	\end{subfigure}%
	\begin{subfigure}{0.25\textwidth}
		\centering
		\includegraphics[width=\linewidth]{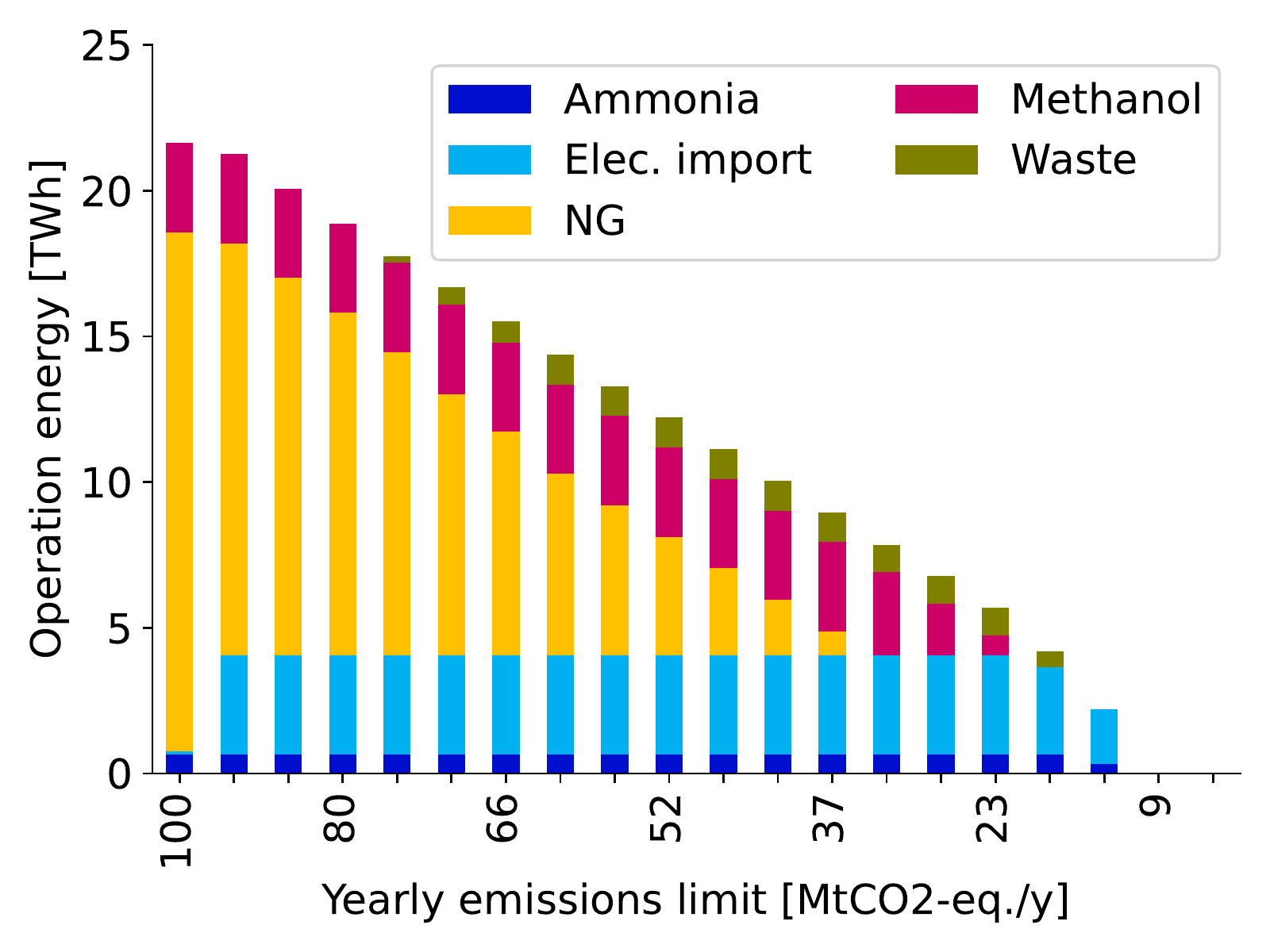}
    \caption{Non-RE resources.}
	\end{subfigure}
\caption{System energy invested in operation ($\mathbf{E_\text{op}}$) evolution in 2035 for several GHG emissions targets breakdown between renewable (RE) (left) and non-renewable resources (right). Abbreviations: RE-fuels: gas-RE, methanol-RE, and ammonia-RE; electricity (Elec.).}
\label{fig:GWP-einv-op}
\end{figure}

\begin{figure}[tb]
\centering
	\begin{subfigure}{0.25\textwidth}
		\centering
		\includegraphics[width=\linewidth]{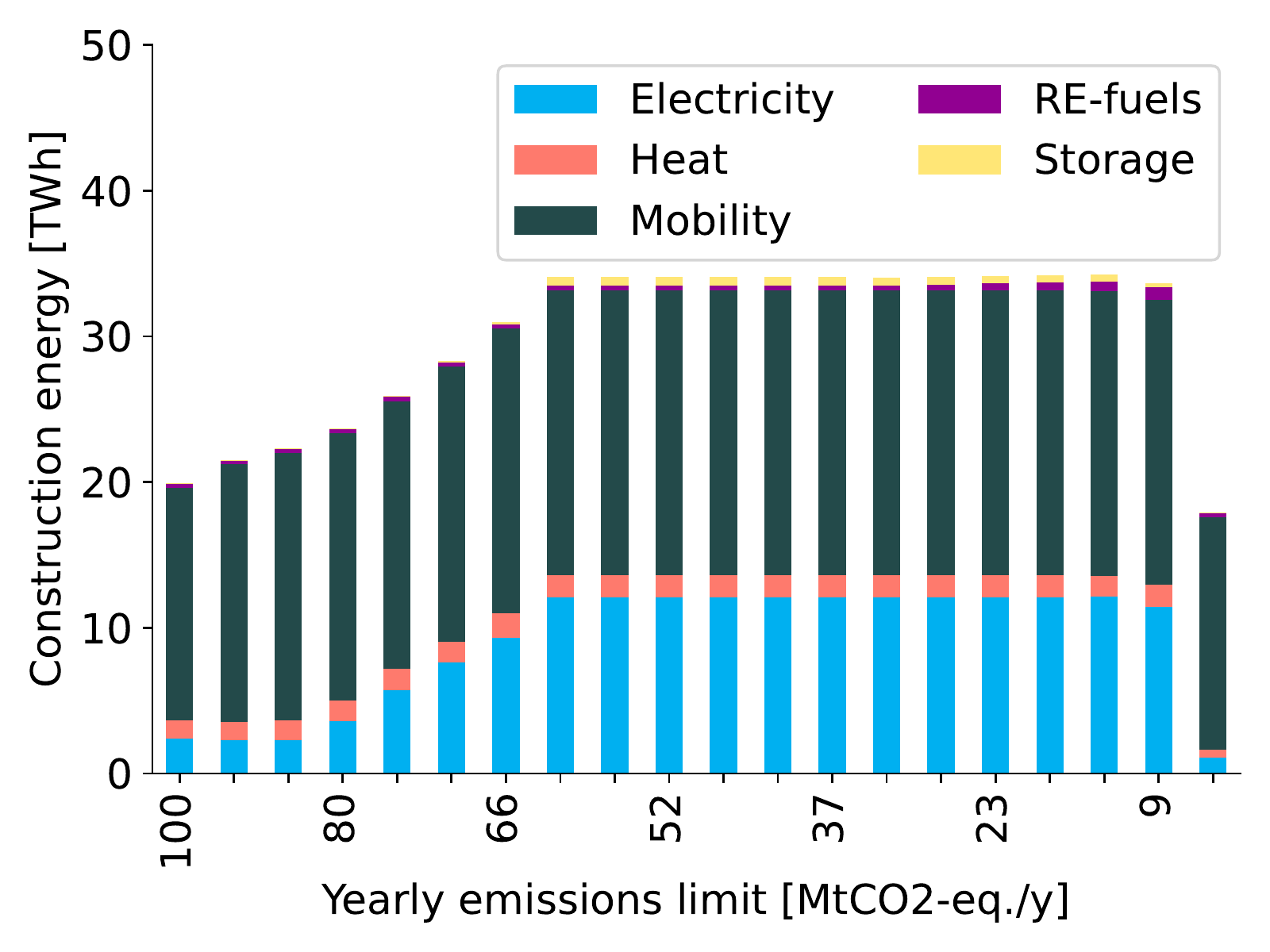}
    \caption{All technologies.}
	\end{subfigure}
	   \begin{subfigure}{0.25\textwidth}
		\includegraphics[width=\linewidth]{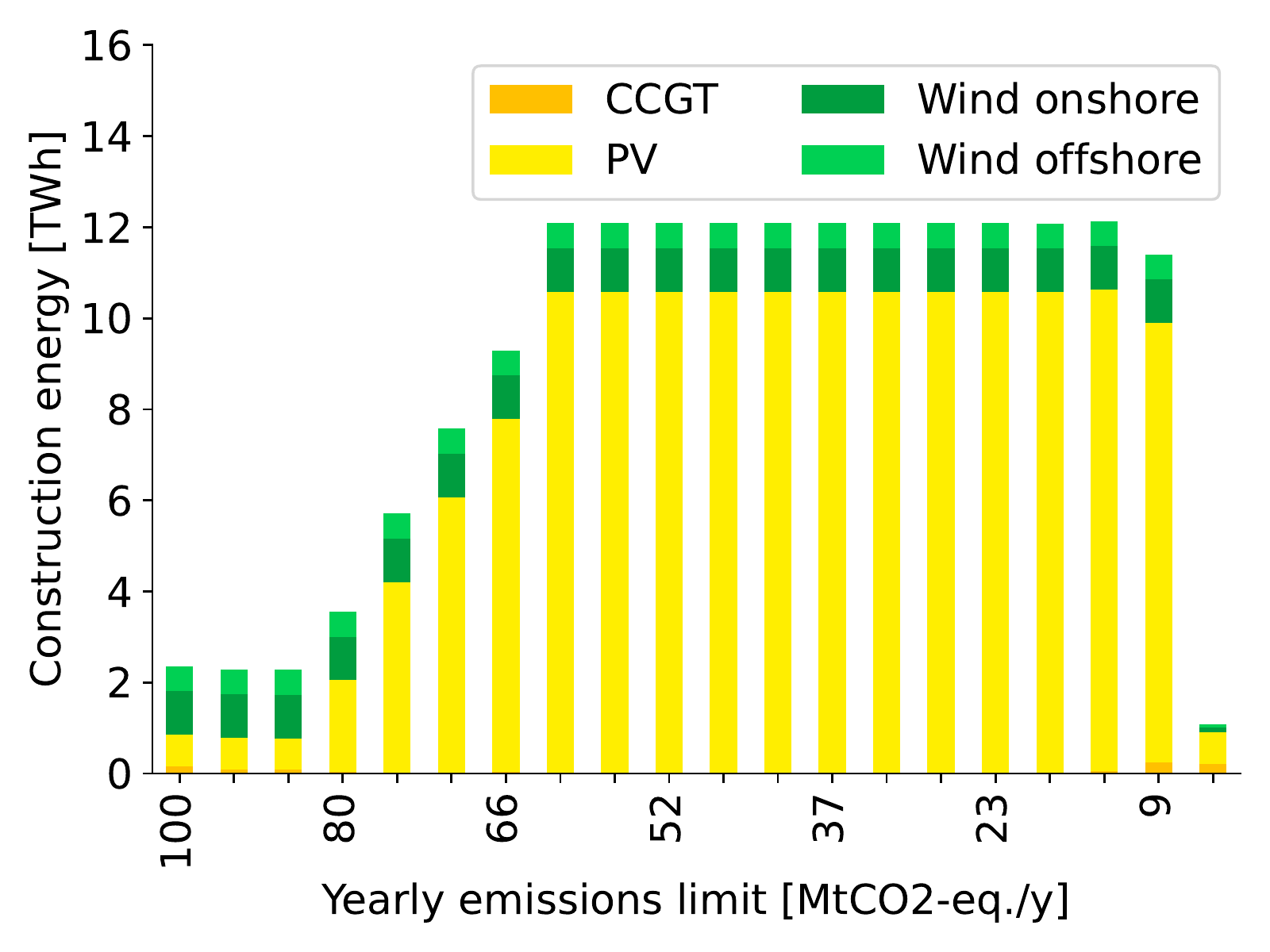}
		\caption{Electricity.}
	\end{subfigure}%
	\begin{subfigure}{0.25\textwidth}
		\centering
		\includegraphics[width=\linewidth]{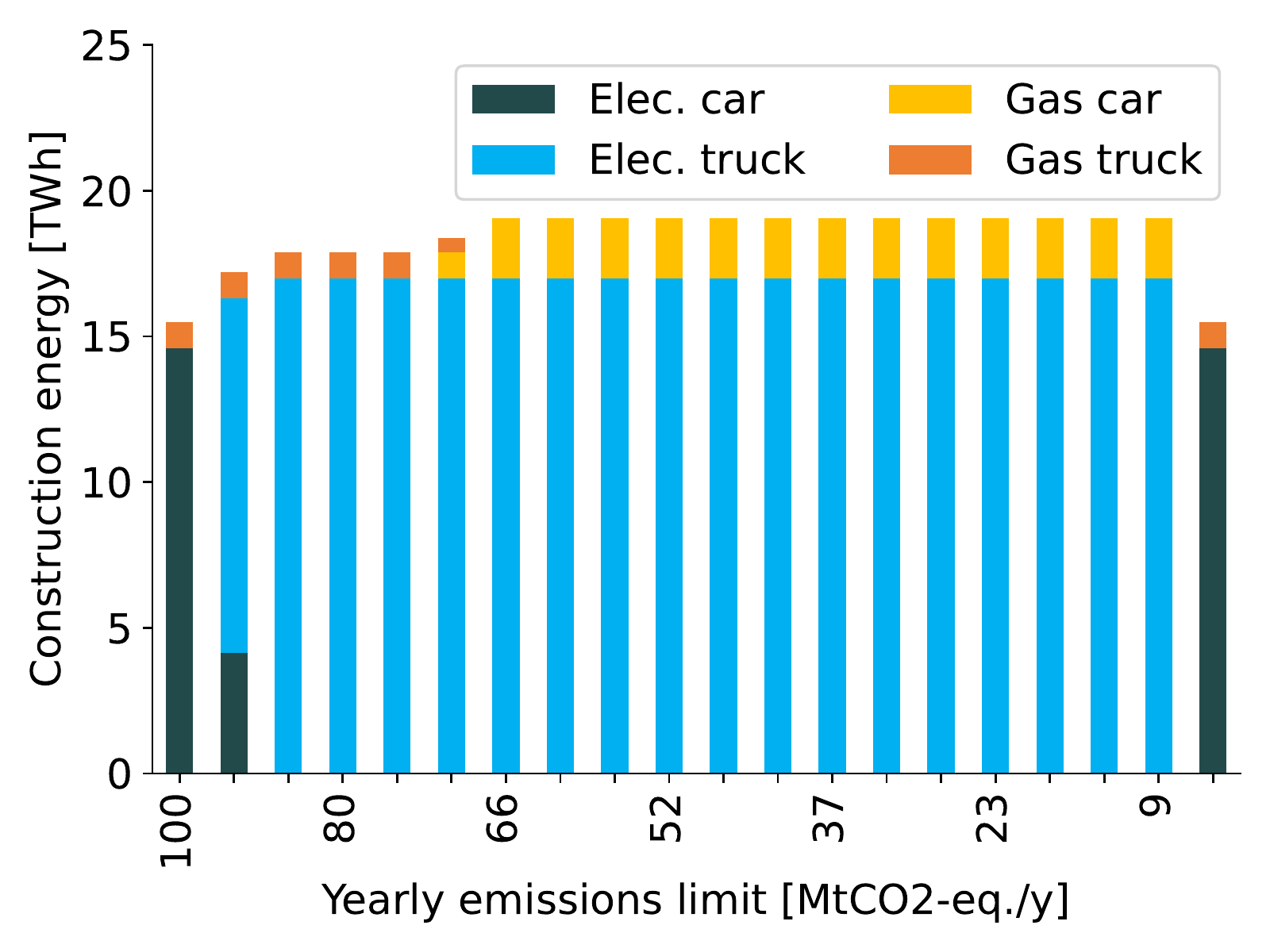}
		\caption{Mobility.}
	\end{subfigure}
\caption{System energy invested in construction ($\mathbf{E_\text{constr}}$) evolution in 2035 for several GHG emissions targets (upper), and breakdown between electricity (lower left) and mobility (lower right) technologies.
Abbreviations: RE-fuels: gas-RE, methanol-RE, and ammonia-RE; natural gas (NG); electric (Elec.).
}
\label{fig:GWP-einv-const}
\end{figure}

Figures \ref{fig:GWP-einv-op} and \ref{fig:GWP-einv-const} depict the operation ($\mathbf{E_\text{op}}$) and construction ($\mathbf{E_\text{constr}}$) system energy invested breakdown by resources, between renewable and non-renewable, and technologies (all technologies, and between electricity and mobility technologies). 

The total system energy invested ($\mathbf{E_\text{in,tot}}$) increases with the limitation of GHG emissions due to the increase of $\mathbf{E_\text{constr}}$ and $\mathbf{E_\text{op}}$, and is driven by the increase of renewable fuels.
The $\mathbf{E_\text{op}}$ increase is mainly due to the shift from NG to renewable gas, and it exceeds the construction system energy invested when GHG emissions are below 47.4 [Mt\coo-eq./y].
The main drivers of the $\mathbf{E_\text{constr}}$ increase are the PV technology and the shift from NG cars and trucks to electric cars and trucks. 

\subsection{FEC evolution}

\begin{figure}[tb]
\centering
	\begin{subfigure}{0.25\textwidth}
		\centering
		\includegraphics[width=\linewidth]{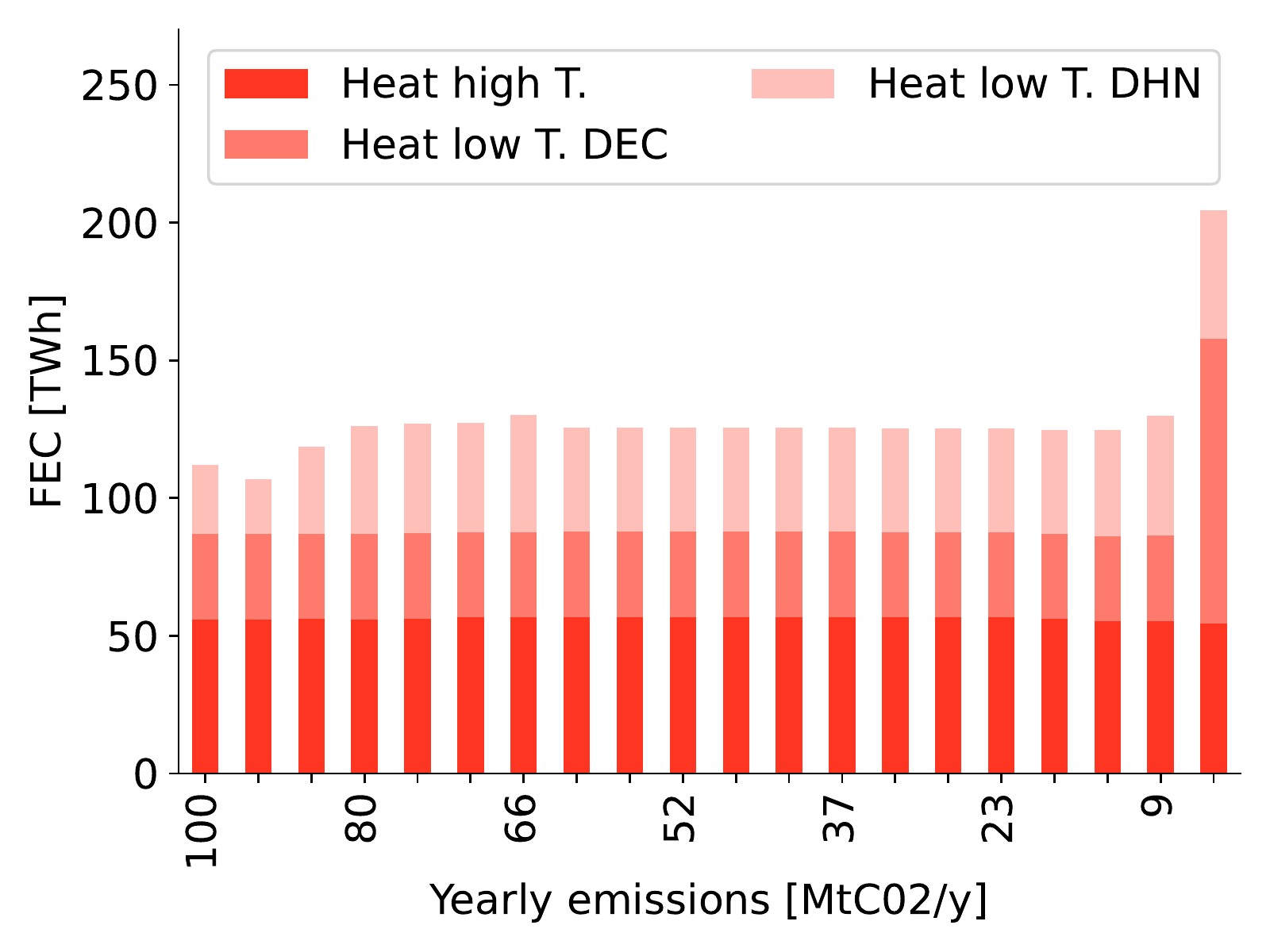}
		\caption{FEC heat.}
	\end{subfigure}%
	\begin{subfigure}{0.25\textwidth}
		\centering
		\includegraphics[width=\linewidth]{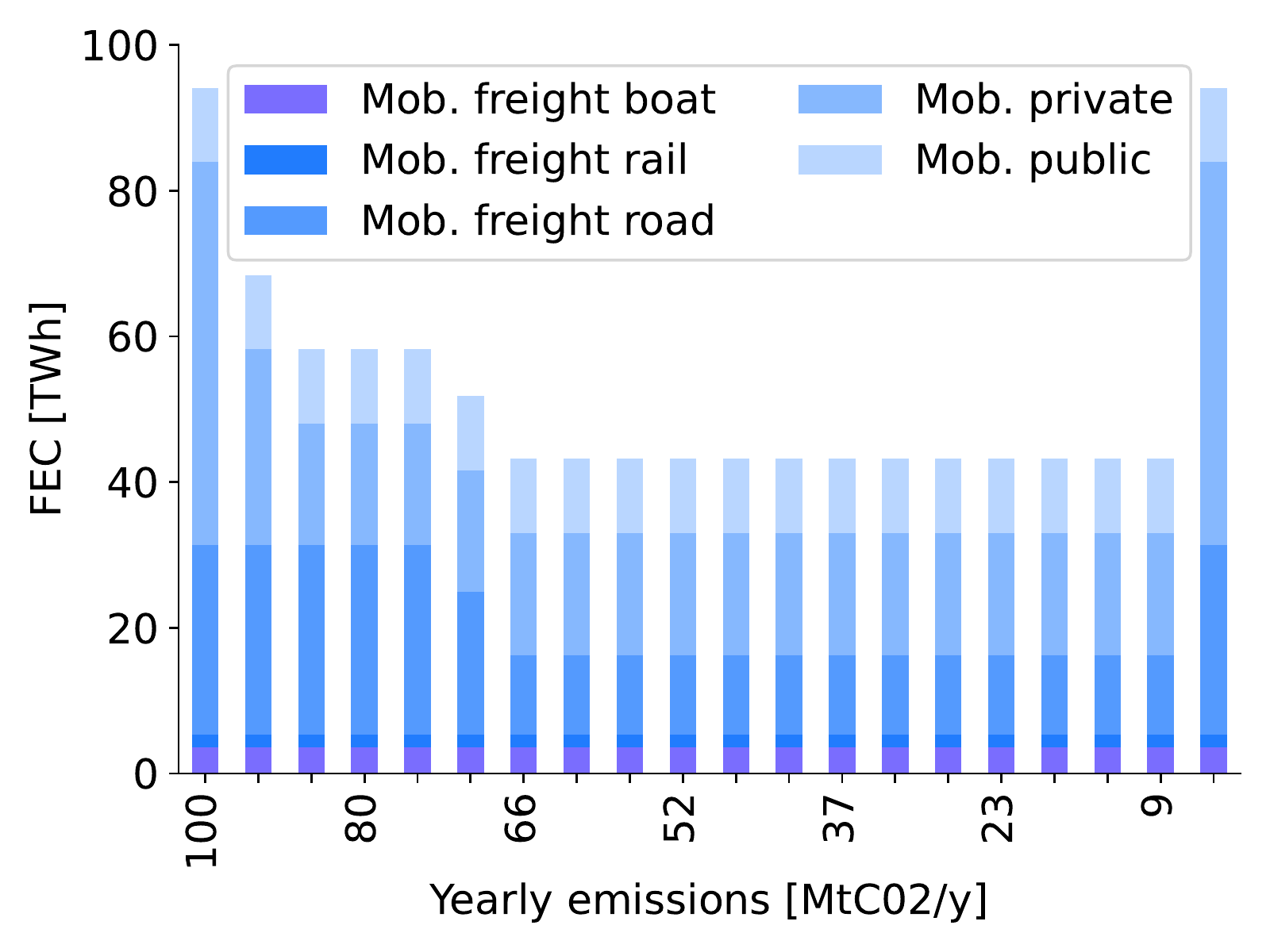}
		\caption{FEC mobility.}
	\end{subfigure}
	\begin{subfigure}{0.25\textwidth}
		\centering
		\includegraphics[width=\linewidth]{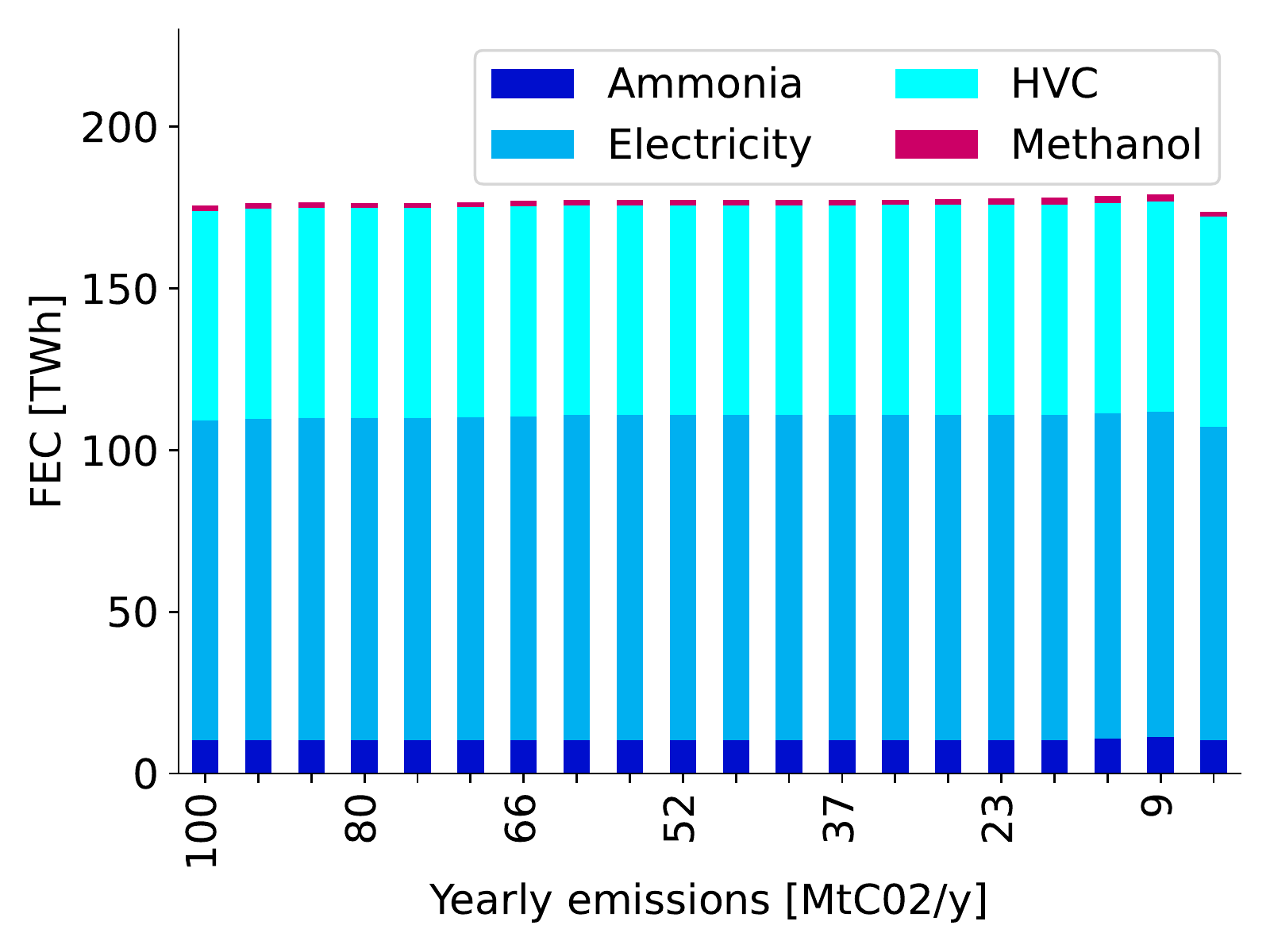}
		\caption{FEC electricity and non-energy.}
	\end{subfigure}
\caption{Final energy consumption (FEC) evolution for several GHG emissions targets in 2035 breakdown by end use demand: heat (upper left), mobility (upper right), non-energy and electricity (lower). Abbreviations: heat high temperature (Heat high T.); decentralized heat low temperature (Heat low T. DEC); centralized heat low temperature (Heat low T. DHN); mobility (Mob.); high value chemicals (HVC).}
\label{fig:FEC-comparison-1}
\end{figure}

Figure \ref{fig:FEC-comparison-1} depicts the evolution of the FEC breakdown by end-use demand: heat, mobility, non-energy, and electricity. First, the FEC decreases with the shift from NG cars to better efficient electric cars. Then, it increases slightly due to the shift from centralized gas co-generation to centralized bio hydrolysis CHP technology that uses more primary energy to produce the same amount of heat low-temperature end-use demand. Finally, it decreases with the shift from NG trucks to more efficient electric trucks.
When GHG emissions achieve 4.7 [Mt\coo-eq./y], the heat low-temperature decentralized FEC increases with the shift from electric technologies such as heat pumps to gas boilers that use renewable gas.
Then, the FEC of mobility, public and freight road, increases due to the shift from electric vehicles to renewable fuel vehicles.

\section{Sensitivity analysis}\label{appendix:section-5}

\begin{figure}[tb]
\centering
\includegraphics[width=1\linewidth]{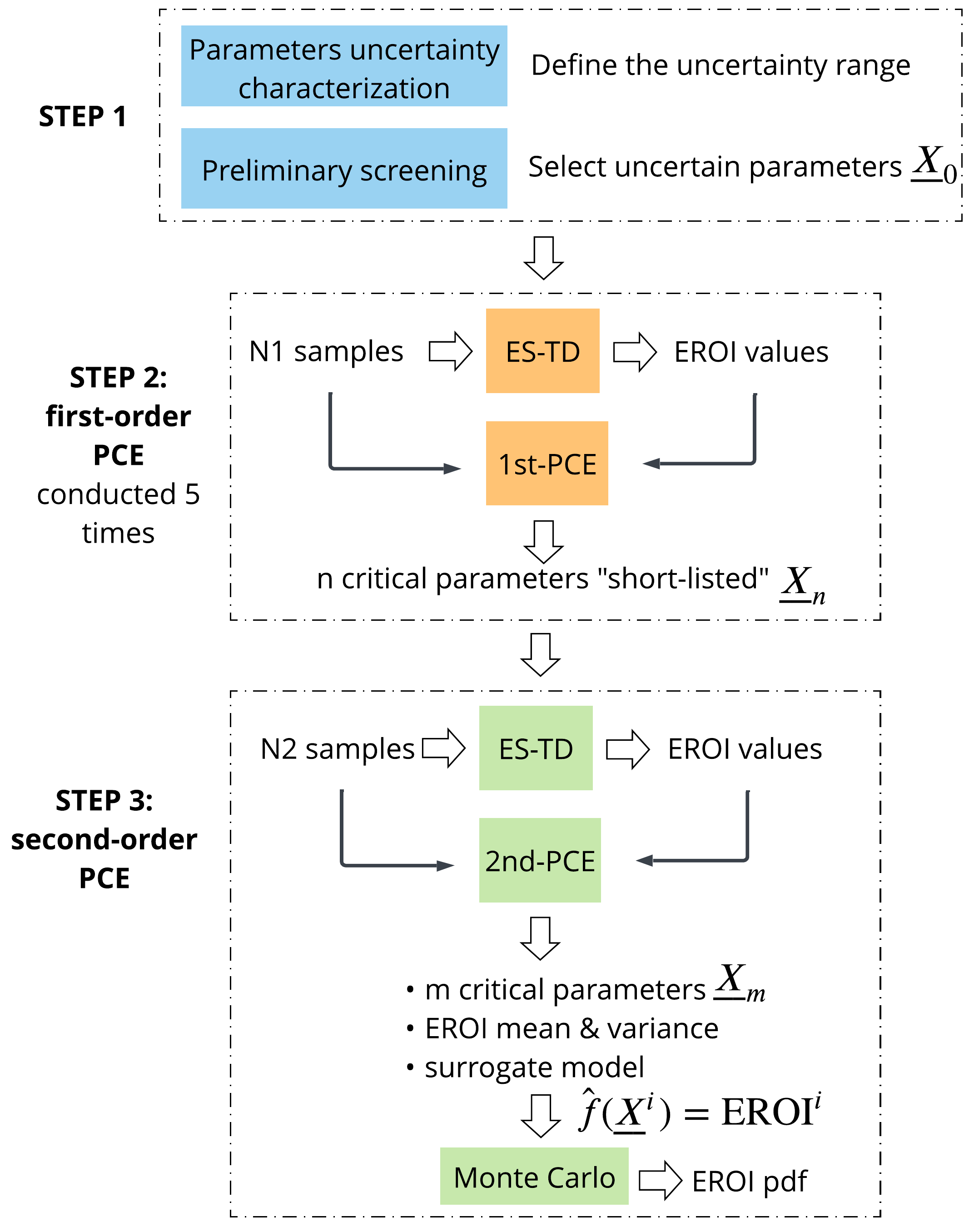}
\caption{Framework of the EROI system sensitivity analysis. The steps 2 and 3 are conducted for each GHG emissions target considered.}
\label{fig:uq-methodo}
\end{figure}

This appendix details the sensitivity analysis methodology and provides additional results. Figure \ref{fig:uq-methodo} depicts the framework of the EROI system sensitivity analysis and how the PCE method is implemented. First, uncertain parameters are defined with their respective uncertainty ranges. Then, the PCE framework is applied to generate a surrogate model and retrieve the critical uncertain parameters. Finally, the mean and variance of the system's EROI are estimated, and the surrogate model is used to perform Monte Carlo sampling to estimate the EROI pdf for several GHG emissions targets.

This appendix is organized as follows. First, \ref{appendix:gsa-framework} details the framework used to perform the sensitivity analysis of the system's EROI. Then, \ref{appendix:uq-uncertainty-characterization} presents the set of uncertain parameters considered with their respective uncertainty ranges. Finally, \ref{appendix:pce-first-order-results} illustrates the selection process using the first-order PCE to build a shortlist of uncertain parameters for the second-order PCE. 

\subsection{Sensitivity analysis methodology}\label{appendix:gsa-framework}

The PCE approach provides a computationally efficient alternative to the Monte Carlo simulation for uncertainty quantification to address the ``curse of dimensionality" pointed out by \citet{rixhon2021role}. Indeed, given limited information about the uncertainty of the parameters for long-term energy planning models, the PCE method constructs a series of multivariate orthonormal polynomials used as a surrogate model $\hat{f}$.
It is a closed-form function that takes as input a vector composed of the values of the realization $i$ of the $N$ uncertain parameters considered $\underline{X}^i = [X_1^i, \cdots, X_n^i]^\intercal$ and outputs the EROI of the system
\begin{align}
    \hat{f}(\underline{X}^i) &= \text{EROI}^i.
\label{eq:surrogate-model}
\end{align}
Depending on the number of uncertain parameters, the polynomial order can be increased and is, therefore, more accurate. A few hundred evaluations are required for tens of uncertain parameters to have a first-order polynomial. However, thousands of evaluations are required to obtain a third-order polynomial. Then, the surrogate model Eq. (\ref{eq:surrogate-model}) allows: (1) to extract statistical moments such as the mean and variance using the coefficients and the total-order Sobol indices. They illustrate the contribution of each uncertain input parameter to the variance of the quantity of interest, in our case, the EROI of the system, including the mutual interactions; (2) to conduct a Monte-Carlo evaluation where millions of samples are generated, and the associated results are calculated instantaneously. It provides the accurate estimation of the EROI pdf for several GHG emissions targets.

The first step, depicted in Figure \ref{fig:uq-methodo}, is the uncertainty characterization which consists of defining a set of uncertain parameters and their uncertainty ranges. In this study, we use the extension to the Belgian energy system uncertainty characterization performed by \citet{limpens2021generating} based on the work of \citet{moret2017characterization}. 
A preliminary screening was performed to determine which parameters had no impact and resulted in a set of 138 uncertain parameters $\underline{X}_0 = [X_1, \cdots, X_{138}]^\intercal$.
%
The second step consists of conducting a first-order PCE on these 138 uncertain parameters to build a shortlist of $n$ critical parameters $\underline{X}_n$ used for a second-order PCE. This step is performed five times to increase the confidence of the result. A parameter is considered negligible if its total-order Sobol index is close to 0 for all five cases. 
%
The last step consists of conducting a second-order PCE to identify the $m$ critical uncertain parameters $\underline{X}_m$ based on $\underline{X}_n$ and estimating an accurate total-order Sobol index for each of them. Then, a Monte-Carlo evaluation is performed using the surrogate model $\hat{f}$ to estimate the pdf of the EROI for several GHG emissions targets. Steps 2 and 3 are conducted for each several GHG emissions targets.

\subsection{Uncertainty characterization}\label{appendix:uq-uncertainty-characterization}

Accounting for uncertainties in energy system long-term planning is crucial \citep{mavromatidis2018uncertainty} to obtain robust designs against uncertainty. However, the insufficient quantity and quality of available data is frequently a limitation. This challenge is addressed in \citet{moret2017characterization} by developing an application-driven method for uncertainty characterization, allowing the definition of ranges of variation for the uncertain parameters. These ranges were initially defined for the Swiss energy system and have been adapted for Belgium \citep{limpens2021generating,rixhon2021role}. Similarly, this study assumes that all the uncertain parameters are independent and uniformly distributed between their lower and upper bounds. 

Table \ref{table:sensitivity-analysis-parameters} summarizes the uncertainty ranges for the different groups of technologies and resources considered in the sensitivity analysis. Following the approach developed by \citet{moret2017characterization} and capitalizing on the works of \citet{limpens2021generating,rixhon2021role} the uncertainty intervals are defined. A preliminary screening, including all the parameters of the model, allowed to obtain an initial list of 138 parameters to be used for the first-order PCE. 

There are four categories of uncertain parameters: end-use demands, technologies, resources, and others. 
The uncertainty in the yearly end-use demands is split by energy sectors. The electricity demand, space heating demand, and industrial demand are related to the yearly industrial demand uncertainty $\text{endUses}_{year}^I$, which has the most extensive range. The freight and passenger mobility are related to the uncertainty of transport $\text{endUses}_{year}^{TR}$. 
Technologies are defined through different parameters: the energy conversion efficiency $\eta$, the investment cost $c_{inv}$, the construction energy investment $e_\text{constr}$, the yearly $c_p$ and hourly $c_{p,t}$ load factors, the potential $f_{max}$, the maintenance cost $c_{maint}$, and the lifetime. This study does not consider the energy conversion efficiency, investment and maintenance costs, yearly capacity factors, and lifetime as uncertain parameters. In addition, the energy invested in the construction of storage technologies is not taken into account.
Intermittent renewable energy is limited in its potential ($f_{max}$ of PV, onshore and offshore wind) and the hourly capacity factor ($c_{p,t}$ of PV, solar, onshore and offshore wind, and hydroelectricity). 
The electricity $\%_{loss}^{elec}$ and heat $\%_{loss}^{heat}$ network losses are considered uncertain parameters.
Resources are characterized by an operating cost $c_{op}$, not considered uncertain in this study, energy invested in operation $e_\text{op}$, and availability $avail$. Most resources have unlimited availability except biomass, waste, and electricity imported. The availability of local resources (wood, waste, and biomass) are uncertain parameters.
Finally, there is a limited installed capacity $f_{max}$ imposed arbitrarily for nuclear $f_{max}^\text{NUC}$ [GWe], electricity $f_{max}^\text{GEO \ elec}$ [GWe] and heat $f_{max}^\text{GEO \ DHN}$ [GWth] production from geothermal. This work also accounts for uncertainties on the upper bounds of mobility $\%_{max}^\text{public mob}$ [\%], $\%_{max}^\text{train freight}$ [\%] and $\%_{max}^\text{boat freight}$ [\%], the upper bound of heat that can be covered by district heating network $\%_{max}^\text{DHN heat}$, and the capacity of electrical interconnection with neighbours $elec_{max}^\text{import}$ [GWe].
\begin{table}[tb]
\renewcommand{\arraystretch}{1.25}
\centering
\begin{tabular}{llrr} 
 \hline
Category & rep. param. & min & max   \\   \hline
$\text{endUses}_{year}^I$ & $\text{endUses}_{year}^I$ & -10.5\% & +5.9\%  \\ 
$\text{endUses}_{year}^{TR}$ & $\text{endUses}_{year}^{TR}$ & -3.4\% & +3.4\%  \\  
$avail$ & $avail^\text{Waste}$ & -32.1\% & +32.1\%  \\  
$f_{max}$ & $f_{max}^\text{PV}$ & -24.1\% & +24.1\%  \\  
$c_{p,t}$ & $c_{p,t}^\text{PV}$ & -11.1\% & +11.1\%  \\  
$\%_{loss}$ & $\%_{loss}^{elec}$ & -2\% & +2\%  \\ 
$e_\text{constr}$ & $e_\text{constr}^{cars}$ & -25\% & +25\%  \\  
$e_\text{op}$ & $e_\text{op}^{RE-fuels}$ & -25\% & +25\%  \\  \hline
 \multirow{8}{*}{\begin{tabular}[c]{@{}l@{}}$Others$ \end{tabular}}   & $f_{max}^\text{NUC}$ & 0.0 & 5.6  \\ 
         & $f_{max}^\text{GEO \ elec}$  & 0.0 & 2.0  \\ 
         & $f_{max}^\text{GEO \ DHN}$  & 0.0 & 2.0  \\ 
         & $\%_{max}^\text{public mob}$ & 45.0\% & 55.0\%  \\ 
         & $\%_{max}^\text{train freight}$ & 22.5\% & 27.5\%  \\ 
         & $\%_{max}^\text{boat freight}$& 27.0\% & 33.0\%  \\ 
         & $\%_{max}^\text{DHN heat}$ & 33.3\% & 40.7\%  \\ 
         & $elec_{max}^\text{import}$  & 8,749 & 10,692  \\ 
\hline
\end{tabular}
\caption{Application of the uncertainty characterization method \citep{moret2017characterization} to EnergyScope TD when maximizing the EROI of the system. Uncertainty is characterized for one representative parameter (rep. param.) per category. 
Due to the lack of data in the literature for 2035, the uncertainty intervals of $e_\text{op}$ and $e_\text{constr}$ are by default absolute uniform interval U$[-25 \%, +25\%]$.
Abbreviations: photovoltaic (PV), district heating network (DHN), industry (I), transport (TR), nuclear (NUC), geothermal (GEO), electricity (elec).}
\label{table:sensitivity-analysis-parameters}
\end{table}

\subsection{First-order PCE results}\label{appendix:pce-first-order-results}

The second step, depicted in Figure \ref{fig:uq-methodo}, consists of using the first-order PCE to build shorter lists of uncertain parameters for the second-order PCE. 
This selection is performed for each GHG emissions target considered in the sensitivity analysis. It relies on good practice \citep{turati2017simulation}, by selecting the parameters which have at least, over the five runs, \textit{i.e.}, to ensure redundancy, one total-order Sobol index above the threshold $=1/d$, where $d$ = 138 is the number of uncertain parameters at the pre-selection phase. These parameters short-listed are named critical parameters and considered for the rest of the study in the second-order PCE.

\begin{figure}[tb]
\centering
\includegraphics[width=\linewidth]{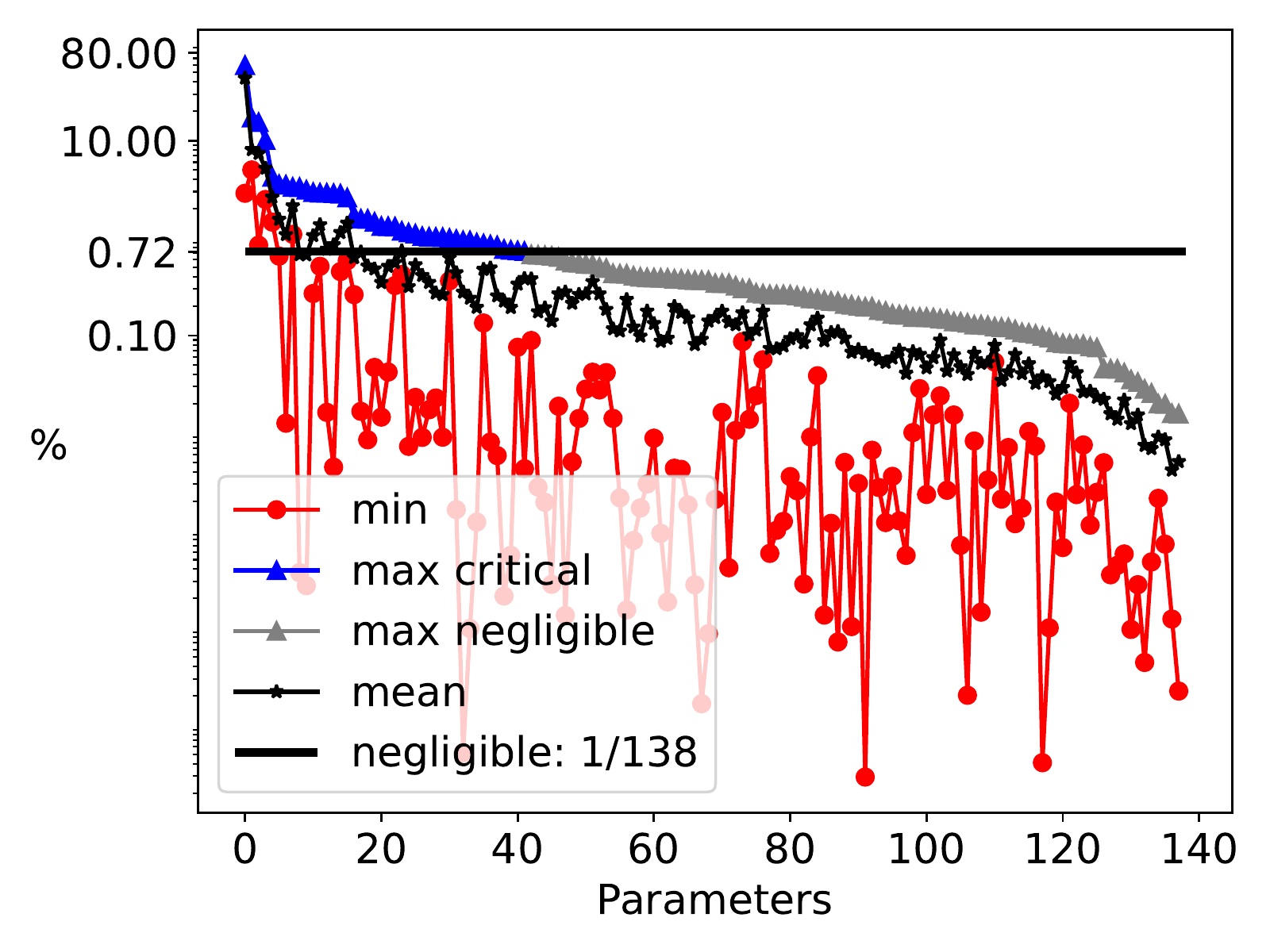}
\caption{Illustration of the selection process using first-order PCE for $\mathbf{GWP_\text{tot}} \leq 28.5$ [Mt\coo-eq./y]. The y-axis is logarithmic. For each parameter, the blue mark indicates the parameter is critical, and the grey mark that it is negligible. It corresponds to the maximal value of the Sobol index over the five runs. A parameter is critical if its maximal value of the Sobol index over the five runs is above threshold $=1/d$.
The red (black) mark is the minimum (mean) value of the Sobol index over the five runs.}
\label{fig:sensitivity-analysis-pce-order-1}
\end{figure}

Figure \ref{fig:sensitivity-analysis-pce-order-1} illustrates this selection process using the first-order PCE for the GHG emissions target of 28.5 [Mt\coo-eq./y]. In this scenario, 42 parameters are identified (blue marks) as critical to be used in the second-order PCE. The red marks are the minimum values of the Sobol index for each parameter over the five runs, and the black marks are the mean of their five Sobol index values.
%
%

\subsection{Second-order PCE results}\label{appendix:pce-second-order-results}

The final step, depicted in Figure \ref{fig:uq-methodo}, consists of using the second-order PCE on the parameters short-listed to limit the error below 1\% \citep{coppitters2020robust} on the EROI statistical moments: mean $\mu$ and variance $\sigma$.
\begin{table}[tb]
\renewcommand{\arraystretch}{1.25}
\centering
\begin{tabular}{lrrrr} 
 \hline
GHG target [Mt\coo-eq./y] & 85.4 & 56.9 & 28.5 & 19.0 \\ \hline
\# parameters short-listed & 64 & 55 & 42 & 45 \\
\# critical parameters & 9 & 27 & 17 & 5 \\\hline
\end{tabular}
\caption{Number (\#) of short-listed and critical parameters using the first-order and second-order PCE.}
\label{table:pce-order-2-results}
\end{table}

\begin{figure*}[tb]
\centering
\begin{subfigure}{0.5\textwidth}
\centering
\includegraphics[width=\linewidth]{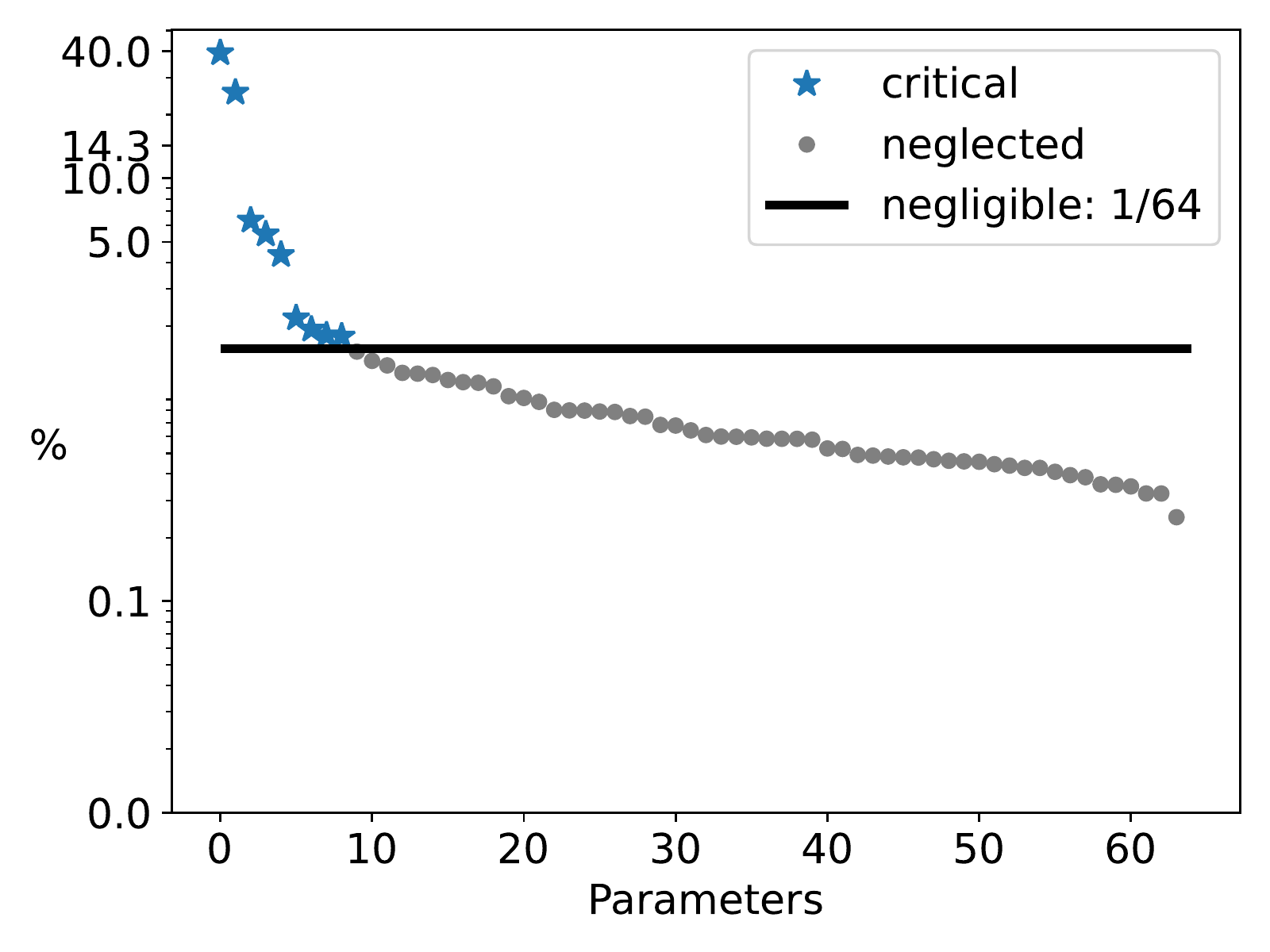}
\caption{$\mathbf{GWP_\text{tot}} \leq 85.4$ [Mt\coo-eq./y].}
\end{subfigure}%
\begin{subfigure}{0.5\textwidth}
\centering
\includegraphics[width=\linewidth]{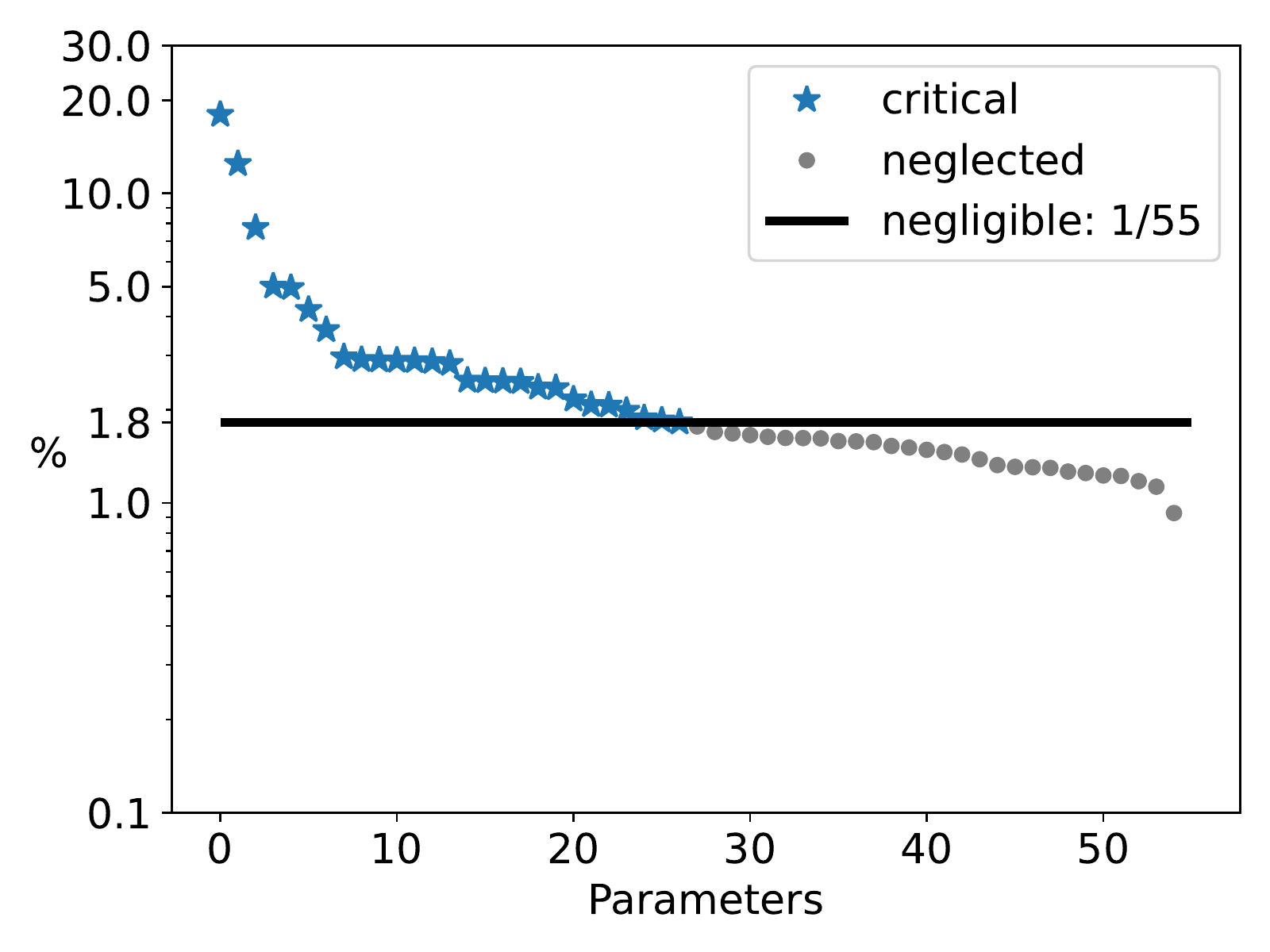}
\caption{$\mathbf{GWP_\text{tot}} \leq 56.9$ [Mt\coo-eq./y].}
\end{subfigure}
\begin{subfigure}{0.5\textwidth}
\centering
\includegraphics[width=\linewidth]{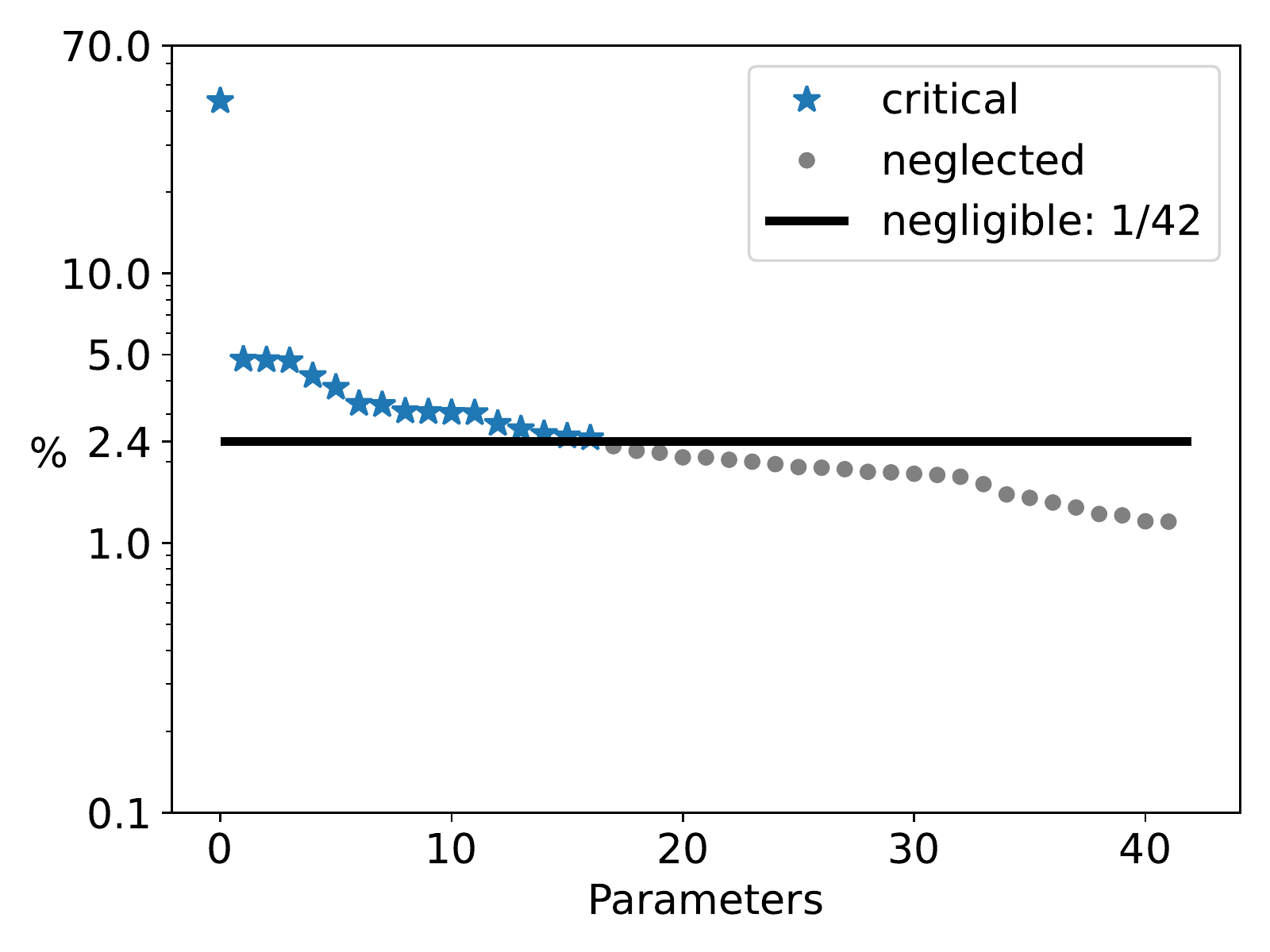}
\caption{$\mathbf{GWP_\text{tot}} \leq 28.5$ [Mt\coo-eq./y].}
\end{subfigure}%
\begin{subfigure}{0.5\textwidth}
\centering
\includegraphics[width=\linewidth]{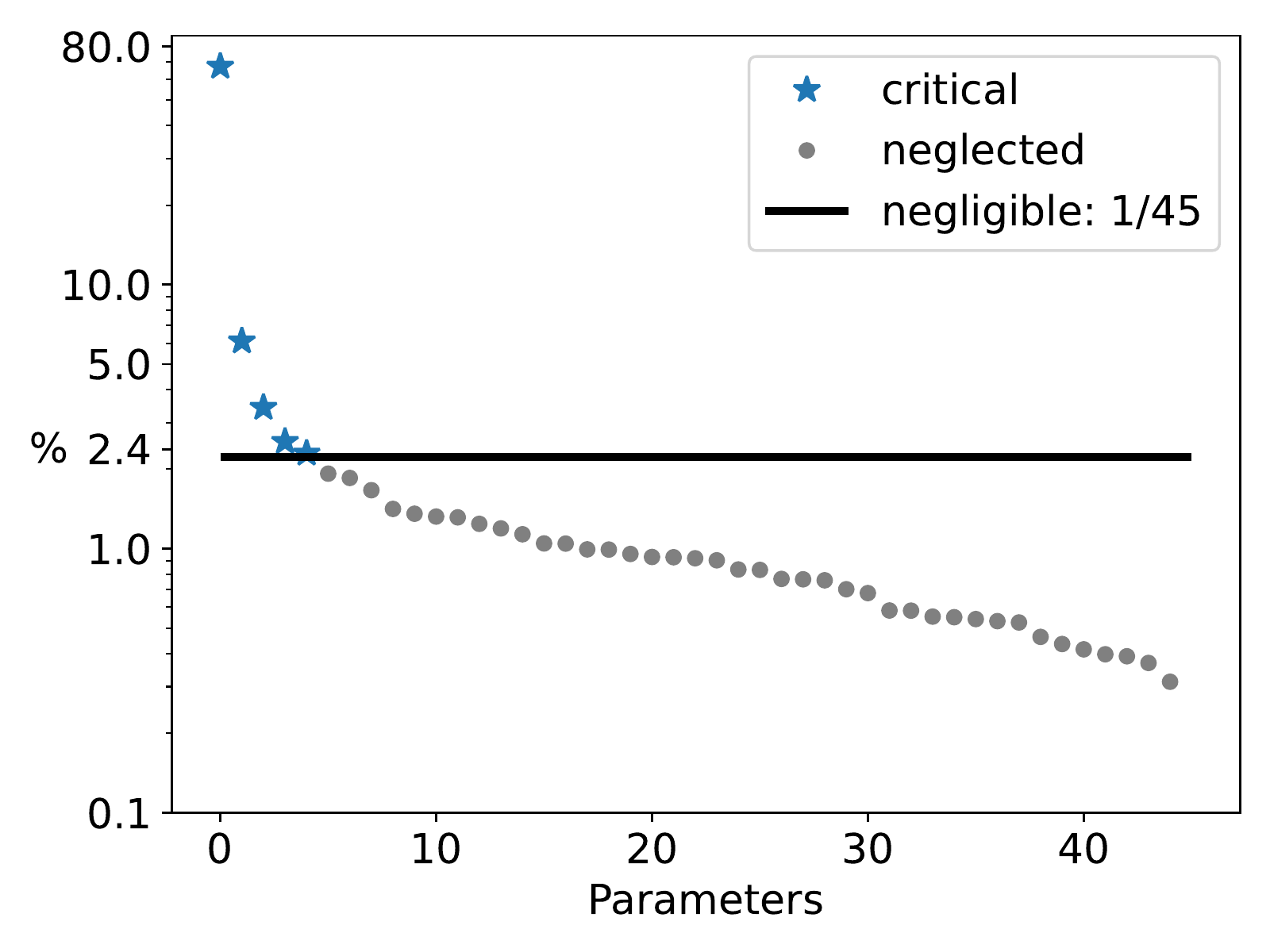}
\caption{$\mathbf{GWP_\text{tot}} \leq 19.0$ [Mt\coo-eq./y].}
\end{subfigure}
\caption{Sobol indices of the parameters using the second-order PCE. Parameters are sorted and critical ones (in blue) have an index above the threshold $1/d$, with $d$ the number of uncertain parameters considered in the second-order PCE. The y-axis is logarithmic.}
\label{fig:sensitivity-analysis-pce-order-2}
\end{figure*}

Figure \ref{fig:sensitivity-analysis-pce-order-2} depicts the selection of the critical parameters using the second-order PCE for the GHG emissions targets considered. Table \ref{table:pce-order-2-results} presents the number (\#) of short-listed and critical parameters using the first-order and second-order PCE. Finally, Table \ref{table:appendix-critical-parameters} lists the critical parameters and their Sobol index for the GHG emissions targets considered.
\begin{table*}[tb]
\renewcommand{\arraystretch}{1.25}
\centering
\begin{tabular}{lrrrr} 
 \hline
Ranking & 85.4 & 56.9 & 28.5 & 19.0 \\ 
 \hline
1 & $e_\text{constr}^\text{NG cars}$ 39.2& $e_\text{constr}^\text{Elec. cars}$ 18.0& $e_\text{op}^\text{Gas-RE}$ 43.6&  $e_\text{op}^\text{Gas-RE}$ 67.1\\
2& $e_\text{op}^\text{NG}$ 25.5& $f_{max}^\text{NUC}$ 12.5& $e_\text{constr}^\text{Elec. cars}$ 4.8& $e_\text{constr}^\text{Elec. cars}$ 6.1 \\
 3&$e_\text{constr}^\text{Elec. cars}$ 6.3& $\%_{max}^\text{public mob}$ 7.7& $f_{max}^\text{NUC}$ 4.8 & $f_{max}^\text{NUC}$ 3.4\\
4& $\%_{max}^\text{public mob}$ 5.4& $e_\text{constr}^\text{PV}$ 5.0& $avail^\text{Wet biomass}$ 4.7& $avail^\text{Wood}$ 2.5 \\
5& $e_\text{op}^\text{Wet biomass}$ 4.3& $e_\text{op}^\text{NG}$  5.0& $f_{max}^\text{Offshore wind}$ 4.2& $e_\text{constr}^\text{NG cars}$ 2.3 \\
6& $e_\text{constr}^\text{DHN wet biomass CHP}$ 2.2& $avail^\text{Wood}$ 4.2& $\text{endUses}_{year}^I$ 3.8 & - \\
7& $e_\text{constr}^\text{Elec. trucks}$ 1.9& $e_\text{constr}^\text{NG cars}$ 3.6& $f_{max}^\text{GEO DHN}$ 3.3 & - \\
8& $e_\text{op}^\text{Methanol}$ 1.8& $c_{p,t}^{PV}$ 3.0& $e_\text{constr}^\text{PV}$ 3.3 & - \\
9& $e_\text{constr}^\text{Diesel trucks}$ 1.8 & $\text{endUses}_{year}^I$ 3.9& $e_\text{constr}^\text{Offshore wind}$ 3.1 & - \\
10 & -& $e_\text{op}^\text{Uranium}$ 2.9 & $\%_{max}^\text{public mob}$ 3.1 & - \\
11 & -&  $avail^\text{Wet biomass}$ 2.9 & $f_{max}^\text{Onshore wind}$ 3.0 & - \\
12 & -& $f_{max}^\text{GEO DHN}$ 2.9 & $e_\text{constr}^\text{NG cars}$ 3.0 & - \\
13 & -&  $e_\text{op}^\text{Wet biomass}$ 2.9 & $e_\text{op}^\text{Elec.}$ 2.8 & - \\
14 & -&  $e_\text{op}^\text{Gas-RE}$ 2.8 & $avail^\text{Wood}$ 2.6 & - \\
15 &-&  $c_{p,t}^\text{Offshore wind}$ 2.5 & $e_\text{op}^\text{Wet biomass}$ 2.5 & - \\
16&-&  $f_{max}^\text{Offshore wind}$ 2.5 & $e_\text{constr}^\text{Elec. wet biomass}$ 2.5 & - \\
17&-&  $e_\text{op}^\text{Methanol}$ 2.5 & $c_{p,t}^\text{Offshore wind}$ 2.5 & - \\
18&-&  $f_{max}^\text{Onshore wind}$ 2.5 & - & - \\
19&-&  $\text{endUses}_{year}^{TR}$ 2.4 & - & - \\
20&-&  $e_\text{constr}^\text{H2}$ 2.4 & - & - \\
21&-&  $\%_{max}^\text{freight train}$ 2.2 & - & - \\
22&-&  $c_{p,t}^\text{Onshore wind}$ 2.1 & - & - \\
23&-&  $\%_{max}^\text{freight boat}$ 2.1 & - & - \\
24&-&  $e_\text{constr}^\text{I waste boilers}$ 2.0 & - & - \\
25&-&  $e_\text{constr}^\text{DHN waste CHP}$ 1.9 & - & - \\
26&-&  $e_\text{op}^\text{Wood}$ 1.8 & - & - \\
27&-&  $e_\text{constr}^\text{CCGT ammonia}$ 1.8 & - & - \\
\hline
\end{tabular}
\caption{Critical parameters and Sobol indices values [\%] for several GHG emissions targets [Mt\coo-eq./y]. 
Abbreviations: electric (elec.), mobility (mob), photovoltaic (PV), nuclear (NUC), renewable gas (Gas-RE), natural gas (NG), geothermal (GEO), transport (TR), industry (I), district heating networks (DHN), combined heat and power (CHP), combined cycle gas turbine (CCGT).}
\label{table:appendix-critical-parameters}
\end{table*}

\end{appendices}

\bibliography{biblio}

\end{document}